%% file: 0_JASA-all.tex
\documentclass[12pt]{article}
\usepackage{amsmath}
\usepackage{graphicx}
\usepackage{enumerate}
\usepackage{natbib}
\usepackage{bibunits}
\defaultbibliographystyle{agsm}

\newcommand{\blind}{1}

\usepackage{3_preamble}
\usepackage{subfiles}

\begin{document}

\subfile{1_JASA-main}

\newpage

\subfile{1_JASA-supp}

\end{document}

%% file: 1_JASA-main.tex
\begin{bibunit}

\def\spacingset#1{\renewcommand{\baselinestretch}%
{#1}\small\normalsize} \spacingset{1}


\if1\blind
{
  \title{\bf Heterogeneous Treatment Effects under Network Interference: A Nonparametric Approach Based on Node Connectivity}
  \author{
    Heejong Bong \\
    Department of Statistics, Purdue University \\
    Colin B. Fogarty, Elizaveta Levina and Ji Zhu \\
    Department of Statistics, University of Michigan \\
  }
  \maketitle
} \fi

\if0\blind
{
  \bigskip
  \bigskip
  \bigskip
  \begin{center}
    {\LARGE\bf Heterogeneous Treatment Effects under Network Interference: A Nonparametric Approach Based on Node Connectivity}
\end{center}
  \medskip
} \fi

\bigskip
\begin{abstract}
In network settings, interference between units makes causal inference more challenging as outcomes may depend on the treatments received by others in the network. Typical estimands in network settings focus on treatment effects aggregated across individuals in the population. We propose a framework for estimating node-wise counterfactual means, allowing for more granular insights into the impact of network structure on treatment effect heterogeneity. We develop a doubly robust and non-parametric estimation procedure, KECENI (Kernel Estimator of Causal Effect under Network Interference), which offers consistency and asymptotic normality under network dependence. The utility of this method is demonstrated through an application to microfinance data, revealing the node-wise impact of network characteristics on treatment effects.
\end{abstract}


\noindent%
\vfill

\newpage
\spacingset{1.9} 


\input{sec/1_introduction}

\input{sec/2_methods}

\input{sec/3_theory}

\input{sec/4_numerical}
\input{sec/5_discussion}

\bigskip
\begin{center}
{\large\bf DATA AVAILABILITY}
\end{center}

The data that support the findings of this study are openly available in \emph{Data on Social Networks and Microfinance in Indian Villages} at \url{https://www.stanford.edu/~jacksonm/IndianVillagesDataFiles.zip}. The preprocessed data and code vignettes used to generate the results and figures in \cref{sec:simulation,sec:application} are openly available at \url{https://doi.org/10.6084/m9.figshare.29660984.v2} and ongoing updates will be maintained at \url{https://github.com/HeejongBong/KECENI}.


\bigskip
\begin{center}
{\large\bf SUPPLEMENTARY MATERIAL}
\end{center}

\begin{description}

\item[Supplement manuscript:] A PDF manuscript providing supplemental descriptions of proofs and extensive details regarding arguments conveyed in the main text. (.pdf file)



\end{description}

\putbib[2_ref-main]

\end{bibunit}

%% file: sec/1_introduction.tex
\section{Introduction}


Estimating causal treatment effects for network-linked units is challenging because of the phenomenon known as network interference:  an individual's outcome may depend not only on their own treatment but also on the treatments received by others in the network. Within this broad challenge, there is the question of whether, and how, causal effects vary with network connectivity. Results from case studies in education \citep{lomi2011some}, economics \citep{bramoulle2009identification}, and public health \citep{buchanan2022spillover} suggest that interference effects are affected by local network structure, reflected in local features such as centrality, density, and the clustering coefficient.  Identifying and estimating this effect heterogeneity has important implications for intervention targeting, policy evaluation, and decision making; and yet most of the literature focuses on average effects over the entire network.  To address this challenge, we propose a new framework for estimating heterogeneous treatment effects under network interference, based on node-specific counterfactual mean estimands.  

The literature on causal effect estimation under network interference is extensive; see \citet{bhadra2025causal} for a recent review. The most common target estimands are average treatment effects (ATEs), which compare counterfactual means averaged across units in the network under specified treatment assignments. Existing methods include 
inverse probability weighting (IPW) \citep{liu2016inverse,aronow2017estimating,leung2022causal} and propensity score matching \citep{forastiere2021identification}. Targeted maximum likelihood estimation \citep{van2014causal,ogburn2022causal} has been developed to provide a doubly robust approach to estimating ATEs.    
Recent work has also studied average direct effects (ADEs) and average spillover effects (ASpEs), which capture the average effect of an individual’s own treatment and of treatments received by others in the network, respectively.
For example, \citet{li2022random} studied IPW estimators for ADEs and ASpEs in randomized experiments, while \citet{leung2022graph} and \citet{khatami2024graph} developed doubly robust methods using graph neural networks.  While they result from different formulations, all these estimands are averages over the entire study population.

Our main contribution is to shift attention from population-averaged estimands to {\it node-wise} counterfactual means as the primary estimands. The key advantage of node-indexed counterfactual means is that they preserve heterogeneity induced by local network structure, rather than averaging this heterogeneity away. 
Although node-level counterfactual quantities appear in earlier work on interference, such as \citet{tchetgen2021auto}, to the best of our knowledge, their practical utility as primary targets of inference has not been systematically studied.

The central hurdle is that node-wise counterfactual means cannot be identified nonparametrically from a single observation on network-linked units without additional assumptions \citep{tchetgen2021auto}. \citet{tchetgen2021auto} address this issue through an autoregressive chain graph model, but the parametric nature of this model can be restrictive in many applications. Instead our approach begins from the observation that the difficulty is closely related to limited overlap: because the number of possible neighborhood configurations and intervention assignments is growing fast, it is unlikely that the observed data contain units whose exposure exactly matches that of a target counterfactual mean. This perspective connects the problem of estimating node-wise counterfactual means to the seemingly distinct problem of estimating a dose-response function, as studied by \citet{kennedy2016semiparametric}, and motivates our smoothing-based estimation strategy.

Once this connection is made, the high-level estimation strategy is natural. However, valid inference for our estimands requires a fundamentally different approach from that of \citet{kennedy2016semiparametric}, due to network dependence. 
In this setting, consistency and asymptotic normality of the proposed estimator does not follow directly from existing results. Instead, it requires new empirical process theory to control the relevant stochastic terms under network dependence and increasing network complexity. This theory is a key ingredient of our proofs and is necessary for establishing valid asymptotic inference for the proposed method. More broadly, the empirical process tools developed here may facilitate inference in other network settings, both associational and causal, where inference theory has been previously unavailable.

Our work is also related to the local configuration approach of \citet{auerbach2021local}, particularly in its use of smoothing. However, the two approaches differ in two important aspects. First, \citet{auerbach2021local} assume randomized treatment assignment, so that there is no confounding between treatment and outcome. In contrast, our framework allows for treatment-outcome confounding and develops a doubly robust adjustment strategy based on observed covariates. Second, the theoretical results in \citet{auerbach2021local} assume that the network consists of multiple independent components, with the number of components growing with the sample size. By contrast, our asymptotic theory applies to more general network structures, potentially consisting of a {\it single} connected component. This setting precludes treating the data as a collection of independent partitions and therefore requires a fundamentally different theoretical toolkit, including the new empirical process theory developed in this paper.

The rest of the paper is organized as follows. In \cref{sec:setting}, we introduce notation, define the target estimand, and state the identifying assumptions. In \cref{sec:keceni}, we present KECENI and summarize the three-step estimation procedure in \cref{alg:keceni_estimation}. In \cref{sec:theorems}, we introduce the new empirical process theory and establish consistency and asymptotic normality. In \cref{sec:simulation}, we compare KECENI with existing methods in simulations. Finally, in \cref{sec:application}, we illustrate its ability to capture heterogeneous spillover effects on microfinance data from India \citep{banerjee2013diffusion}.

%% file: sec/2_methods.tex
\section{Node-wise Causal Effects under Network Interference} \label{sec:setting}

\subsection{Notation and setting}

Suppose we observe data on $n$ units and each unit $i \in [n]$ is associated with a binary  intervention variable $T_i \in \{0, 1\}$ and an observed outcome $Y_i \in \reals$. 
Without restrictions on interference, each unit $i$ has $2^n$ counterfactuals $(Y_i(t_1, \dots, t_n): t_1, \dots, t_n \in \{0, 1\})$, corresponding to all possible combinations of its own intervention and those of other units.
By the \emph{consistency} assumption, the observed outcome $Y_i$ under the realized assignment $(T_1,\dots,T_n)$ is $Y_i(T_1,\dots,T_n)$.
In this paper, we adopt a \emph{superpopulation perspective}, under which each counterfactual $Y_i(t_1, \dots, t_n)$ is itself a random variable, so $Y_i$ is not a deterministic function of $(T_1, \dots, T_n)$. 

Estimating all possible $2^n$ counterfactuals from a single sample $\{(T_i, Y_i): i \in [n]\}$ is clearly infeasible, and necessitates imposing restrictions on the amount of interference from other units.  Here we can take advantage of the information on connections between units, represented by a network $\mathcal{G}$ with node set $[n]$ and edge set $\mathcal{E} \subseteq [n] \times [n]$.  We only consider undirected and unweighted networks here.   Then we make a key assumption enabling estimation of counterfactuals:  
\begin{assumption}[Network Interference] \label{assmp:network_interference}
    Conditional on $\mathcal{G}$, for $i \in [n]$, 
    \begin{equation*}
        Y_i(t_1, \dots, t_n) \stackrel{\text{a.s.}}{=} Y_i(t'_1, \dots, t'_n) ~~ \text{when} ~~ t_{j} = t'_{j} \text{ for all } j \in N_i,
    \end{equation*}
    where $N_i \equiv \{i\} \cup \{j \in [n]: (i,j) \in \mathcal{E}\}$ is the network neighborhood of $i$. 
\end{assumption}
Assumption \ref{assmp:network_interference} states that interference occurs only between units that are connected in the network $\mathcal{G}$.  
This slightly relaxed version of the standard \emph{stable unit treatment on neighborhood value assumption}, SUTNVA \citep{forastiere2021identification}, 
implies that the counterfactual $Y_i(\cdot)$ can be written as a function of the intervention variables for neighboring units, which we denote $t_{N_i}$ and write $Y_i(t_1, \dots, t_n) = Y_i(t_{N_i})$.  

\subsection{Causal Estimand: Definition, Motivation, and Examples}

We are now ready to define the \emph{node-wise counterfactual mean}, our target estimand. Let $i^*$ represent the target node in $\mathcal{G}$, and let $t^*_{N_{i^*}} \in \{0,1\}^{\abs{N_{i^*}}}$ represent a hypothetical intervention assignment to the neighborhood of $i^*$. The estimand is the expected value of the node's counterfactual $Y_{i^*}(t^*_{N_{i^*}})$, defined as  
\begin{equation*}
    \theta_{i^*}(t^*_{N_{i^*}}) \equiv 
    \Exp[Y_{i^*}(t^*_{N_{i^*}}) | \mathcal{G}]. 
\end{equation*}
Conditioning on $\mathcal{G}$ is, in essence, equivalent to assuming that $\mathcal{G}$ represents the true interference structure and is observed without noise. 
This estimand was also referred to as the unit potential outcome expectation by \citet{tchetgen2021auto}.  We refer to this as a \emph{node-wise} estimand to distinguish it from unit treatment effects, which typically refer to conditional expectations of counterfactual outcomes given specific covariates. 
To our knowledge, node-wise counterfactual means have primarily been used as conceptual devices for constructing population average counterfactual means or treatment effects, rather than being treated as inferential targets in their own right. In this work, we highlight the utility of node-wise estimands and develop a framework for their identification, estimation, and inference.

Common node-wise treatment effects, such as total, direct, and spillover effects, can be represented as differences between counterfactual means under two neighborhood interventions, $t^{(1)}_{N_{i^*}}$ and $t^{(0)}_{N_{i^*}}$. As we highlight below, these node-wise effects may be of interest in their own right, and can further serve as building blocks for estimands both familiar and new.


\begin{example}[node-wise direct effect] 
For node $i^*$, the \emph{direct effect} when its neighbors are assigned treatment $t^*_{N_{i^*} \setminus \{i^*\}}$ is given by $$\theta_{i^*}(t_{i^*} = 1, t_{N_{i^*} \setminus \{i^*\}} = t^*_{N_{i^*} \setminus \{i^*\}}) - \theta_{i^*}(t_{i^*} = 0, t_{N_{i^*} \setminus \{i^*\}} = t^*_{N_{i^*} \setminus \{i^*\}}).$$
\end{example}
\begin{example}[node-wise spillover effect]
For node $i^*$, the \emph{spillover effect} of assigning treatments $t^{(1)}_{N_{i^*} \setminus \{i^*\}}$ versus $t^{(0)}_{N_{i^*} \setminus \{i^*\}}$ to the neighbors of unit $i$ when $t_i^* = 0$ is $$\theta_{i^*}(t_{i^*} = 0, t_{N_{i^*} \setminus \{i^*\}} = t^{(1)}_{N_{i^*} \setminus \{i^*\}}) - \theta_{i^*}(t_{i^*} = 0, t_{N_{i^*} \setminus \{i^*\}} = t^{(0)}_{N_{i^*} \setminus \{i^*\}}).$$
\end{example}


\begin{example}[overall average effect] 
When comparing two hypothetical interventions $t^{(0)}$ and $t^{(1)}$, the overall \emph{average treatment effect} across the nodes in the network is $$n^{-1}\sum_{i=1}^n \left\{ \theta_{i}(t^{(1)}_{N_{i}}) - \theta_{i}(t^{(0)}_{N_{i}}) \right\}. $$
\end{example}

\begin{example}[average effects over network-based subgroups] Treatment effects may vary for node types with varying network characteristics, such as node degree: for example, highly connected individuals may be affected by an intervention differently, especially if spillover effects are present. To obtain an estimate for a particular subgroup like this, we can average the estimand over a subgroup defined by local network features.   For instance, the average treatment effect for nodes with a given degree $k$ can be obtained by averaging over the corresponding node set $\mathcal{V}_k = \{ i : \sum_j A_{ij} = k \}$, as 
\[
\frac{1}{|\mathcal{V}_k|} \sum_{i \in \mathcal{V}_k} \left\{ \theta_i(t^{(1)}_{N_i}) - \theta_i(t^{(0)}_{N_i}) \right\}. 
\]
\end{example}

\subsection{Strong Ignorability Condition for Identifiability} \label{sec:identifiability}

In observational studies under the assumption of no interference between units, a common identifying assumption for average treatment effects is \emph{strong ignorability}, conditionally on observed covariates $X_i \in \mathcal{X}$ for each unit $i$.
Strong ignorability requires that (i) $(Y_i(0), Y_i(1)) \indep T_i | X_i$; and (ii) $0 < \Pr[T_i = 1 | X_i] < 1$. 
Variants of this assumption are widely used in the network interference literature. For example, \citet{van2014causal}, \citet{tchetgen2021auto}, \citet{ogburn2022causal}, and \citet{leung2022causal} adopt the strongest version, assuming 
the counterfactuals $Y_{[n]}(\cdot)$ are independent of the treatments $T_{[n]}$ conditionally on covariates $X_{[n]}$ and the network $G$. 
In contrast, \citet{forastiere2021identification} proposed a weaker ignorability condition based on an \emph{exposure mapping} $g_i : t_{N_i} \mapsto g_i(t_{N_i})$, which combines the neighbors' interventions into a lower-dimensional summary, such as the proportion of treated neighbors.  They then assume that $Y_i(t_{N_i}) = Y_i(t'_{N_i})$ whenever $g_i(t_{N_i}) = g_i(t'_{N_i})$.
The resulting ignorability condition requires only that the exposure mapping values are independent of potential outcomes, that is, $Y_i(t_{N_i}) \indep g_i(T_{N_i}) | (X_i, \mathcal{G})$. 
We make a similar assumption without relying on an explicit exposure mapping, which is slightly stronger than that of \citet{forastiere2021identification}, 
yet still weaker than the strongest version, which is most commonly assumed in the literature. 
This assumption is sufficient for the identifiability of $\theta_{i^*}(t^*_{N_{i^*}})$ under network interference.
\begin{assumption}[Strong ignorability under network interference] \label{assmp:ignorability}
    For any node $i$ in the observed network $\mathcal{G}$,
    \(
        \left(Y_{i}(t_{N_i}): t_{N_i} \in \{0, 1\}^{\abs{N_i}}\right) \indep T_{N_{i}} | (X_{N_{i}}, \mathcal{G}),
    \)
    and for any $t_{N_{i}} \in \{0, 1\}^{\abs{N_i}}$,
    \(
        \Pr[T_{N_{i}} = t_{N_{i}} | X_{N_{i}}, \mathcal{G}] > 0, ~\mathrm{a.s.}
    \)
\end{assumption}

\begin{proposition}[Identifiability on the Observed Network] \label{thm:identifiability}
    Under \cref{assmp:network_interference,assmp:ignorability}, $\theta_{i^*}(t^*_{N_{i^*}})$ is identifiable for arbitrary user-specified neighborhood intervention vectors $t^*_{N_{i^*}}$.
\end{proposition}

\section{KECENI: Kernel Estimator of Causal Effect under Network Interference} \label{sec:keceni}

Although Assumption \ref{assmp:ignorability} identifies the target estimand $\theta_{i^*}(t^*_{N_{i^*}})$ as a functional of the observed data distribution, nonparametric estimation remains challenging when only a single realization of $(Y_{[n]}, T_{[n]}, X_{[n]})$ is observed.
A key difficulty is the combinatorially large number of neighborhood configurations and treatment assignments: there are few (in any) units with the exact same local network configuration and neighborhood intervention as the target unit $i^*$ and the specific intervention $t^*_{N_{i^*}}$. 
We address this challenge by developing a pseudo-outcome approach, which borrows strength from units with similar intervention values so that causal effects remain estimable even when exact matches are unavailable.  The pseudo-outcome approach was proposed by \citet{kennedy2017non} to handle limited overlap in estimating dose-response functions for continuous interventions; here we extend it to a network interference setting.

Define the outcome regression function and the propensity score, respectively, as
$
\mu_i(t_{N_i}, x_{N_i}) = \Exp[Y_i| T_{N_i} = t_{N_{i}}, X_{N_{i}} = x_{N_{i}}, \mathcal{G}]$,
$\pi_i(t_{N_i}| x_{N_i}) = \Pr[T_{N_{i}} = t_{N_{i}}| X_{N_{i}} = x_{N_{i}}, \mathcal{G}]$, 
and let $\bar\mu_i$ and $\bar\pi_i$ be their fixed approximations.  Define the pseudo-outcome
\begin{equation*}
\begin{aligned}
    \xi_i(\bar\mu_i, \bar\pi_i)
    \equiv \frac{Y_i - \bar\mu_i(T_{N_i}, X_{N_i})}{\bar\pi_i(T_{N_i} | X_{N_i})} 
    \int \bar\pi_i(T_{N_i}| x_{N_i}) dP_{X_{N_i}|\mathcal{G}}(x_{N_i}) 
    + \int \bar\mu_i(T_{N_i}, x_{N_i}) 
    dP_{X_{N_i}|\mathcal{G}}(x_{N_i}), 
\end{aligned}
\end{equation*}

These pseudo-outcomes have the double robustness property, similarly to those in  \citet{kennedy2017non}, as stated in the following proposition (the proof can be found in \cref{app:pf_double_robustness}). 

\begin{proposition} \label{thm:double_robustness_unadjusted}
    If $\bar\mu_i = \mu_i$ or $\bar\pi_i = \pi_i$, for any node $i$ and neighborhood intervention $t_{N_i}$, the pseudo-outcome is unbiased, in the sense that  
    $\mathbb{E}[\xi_i(\bar\mu_i, \bar\pi_i) | T_{N_i} = t_{N_i}, \mathcal{G}] = \theta_i(t_{N_i}).$
\end{proposition}



We estimate $\theta_{i^*}(t^*_{N_{i^*}})$ 
by pooling counterfactual mean values across nodes with similar neighborhoods in terms of connectivity, intervention assignments and covariate distributions, quantified by a user-specified dissimilarity metric $\Delta$. One example of such metrics is $
\Delta\{(i, t_{N_i}), (i', t'_{N_{i'}}) \} \equiv \norm{g_i(t_{N_i}) - g_{i'}(t'_{N_{i'}})}$, for $i, i' \in [n]$, $t, t' \in \{0, 1\}^n$,
where $g_i$ and $g_{i'}$ are exposure mappings that aggregate neighbors' interventions into summaries of causal relevance. We then approximate $\theta_{i^*}(t^*_{N_{i^*}})$ via kernel smoothing of the observed pseudo-outcomes $\xi_i$, leveraging their unbiasedness for $\theta_i(t_{N_i})$ (\cref{thm:double_robustness_unadjusted}).
%
%
The smoothing uses a kernel function $\kappa_\lambda(x) = \kappa(x/\lambda) \in [0, 1]$, where the bandwidth $\lambda > 0$ is a tuning parameter. 
KECENI thus addresses the lack of overlap by aggregating units with close but not identical neighborhood treatment configurations, while adjusting for confounders through a doubly robust estimator. 
The proposed procedure is summarized in \cref{alg:keceni_estimation}. 

\begin{algorithm}
\caption{KECENI}\label{alg:keceni_estimation}
\begin{enumerate}
    \item \textbf{Estimate nuisance functions} $\mu$, $\pi$, and ${P}_{X_{[n]}|\mathcal{G}}$ by appropriate machine learning or nonparametric methods.


    \item \textbf{Construct pseudo-outcomes} 
   $\hat{\xi}_i(\hat\mu_i, \hat\pi_i)$ by plugging in the estimated nuisance functions, $\hat\mu_i$, $\hat\pi_i$, and $\hat{P}_{X_{[n]}|\mathcal{G}}$. 
    

    \item \textbf{Apply kernel smoothing} over $\Delta_i \equiv \Delta\{(i, T_{N_i}), (i^*, t^*_{N_{i^*}})\}$ to estimate $\theta_{i^*}(t^*_{N_{i^*}})$ by 
    \begin{equation} \label{eq:keceni}
        \hat\theta_{i^*}(t^*_{N_{i^*}})
        \equiv \hat{D}^{-1} \tsum_{i=1}^n \hat{\xi}_i(\hat{\mu}_i, \hat{\pi}_i)
        \cdot \kappa_\lambda(\Delta_i),
    \end{equation}
    where $\hat{D} \equiv \sum_{i=1}^n \kappa_\lambda(\Delta_i)$ normalizes the kernel weights to ensure they sum to $1$. 
\end{enumerate}
\end{algorithm}

\begin{remark}[Choosing tuning parameters.] \label{rmk:cross_validation}


KECENI requires the analyst to specify a dissimilarity metric $\Delta$, a kernel function $\kappa$, and a bandwidth parameter $\lambda$. These choices determine the bias-variance tradeoff and can substantially affect both estimation and inference.
Cross-validation is the standard approach for choosing $\kappa$ and $\lambda$.  \citet{kennedy2017non} proposed several cross-validation schemes based on pseudo-outcomes, and we extend these ideas to KECENI under network interference; see \cref{app:cross_validation} for details.   While cross-validation can also be used to compare specific candidate similarity metrics, there is no universally optimal or fully automatic rule for choosing $\Delta$, not only for KECENI, but for all causal inference under network interference methods that rely on local similarities. That said, KECENI provides useful practical guidance, since the pseudo-outcome construction recasts the choice of the similarity metric as a model selection problem for nonparametric regression, standard tools such as residual diagnostics become available.  
Our simulations in \cref{app:simulation_cv} show that cross-validation successfully chooses the best-performing combination of $\Delta$, $\kappa$, and $\lambda$ from a given candidate set. In particular, the choice of $\kappa$ has little impact on predictive performance relative to $\lambda$ and $\Delta$. This is consistent with the standard intuition in kernel smoothing that the choice of bandwidth is typically  more important than kernel shape. In the empirical studies in \cref{sec:simulation,sec:application}, we use $\kappa(x) = e^{-\abs{x}}$.
\end{remark}

\begin{remark}[Additional assumptions]
For feasible estimation of $\mu_i$ and $\pi_i$, we assume that these functions depend only on local treatments, covariates, and connectivity, that is, 
$\mu_i(T_{N_i}, X_{N_i}) = \mu(T_{N_i}, X_{N_i}, \mathcal{G}_{N_i})$ and $\pi_i(T_{N_i} \mid X_{N_i}) = \pi(T_{N_i} \mid X_{N_i}, \mathcal{G}_{N_i})$,
where $\mathcal{G}_{N_i}$ is the subnetwork induced by $N_i$.
These additional assumptions go beyond  \cref{assmp:ignorability}, but they are usually justified in practice by \emph{locality} and \emph{homogeneity}. The locality principle is that, conditional on local treatments, covariates, and connectivity, distant parts of the network have negligible influence on $Y_i$. The homogeneity principle is that nodes with similar local configurations have similar outcome and treatment mechanisms, analogous to stationarity in time series.
An alternative to this assumption is allowing dependence on some kind of low-dimensional network summary $Z_i$ derived from $\mathcal{G}$, for example, a node embedding under a latent position model. One could instead assume
$\mu_i(T_{N_i}, X_{N_i}) = \mu(T_{N_i}, X_{N_i}, Z_{N_i}, \mathcal{G}_{N_i})$,
and similarly for the treatment mechanism. This preserves feasibility while allowing systematic heterogeneity across network regions.
\end{remark}

\begin{remark}[Two-hop neighborhoods] \label{rem:two_hop_neighborhoods}
In some cases,  it may be simpler to work with propensity scores conditional on covariates and the connectivity of two-hop neighborhoods, denoted by $N_i^{(2)} \equiv \cup_{j \in N_i} N_j$. For instance, this applies when the intervention assignment is independent across units, conditional on the covariates and connectivity of adjacent units, which is commonly assumed in many structural equation models under network interference such as in \citet{ogburn2022causal}. In such cases, one may want to work with the propensity score conditional on two-hop neighborhoods, 
\begin{equation} \label{eq:propensity_ind}
    \pi(T_{N_i} | X_{N_i^{(2)}}, \mathcal{G}_{N_i^{(2)}}) = \prod_{j \in N_i} \pi^\circ(T_j | X_{N_j}, \mathcal{G}_{N_j}),
\end{equation}
where $\pi^\circ$ represents the node-level propensity score. The pseudo-outcome can then be defined using covariates and connectivity from two-hop neighborhoods. The theoretical properties we establish hold regardless of how we condition on the network structure.
\end{remark}

%% file: sec/3_theory.tex
\section{Consistency and Asymptotic Normality} \label{sec:theorems}

In this section, we provide theoretical guarantees regarding the asymptotic properties and accuracy of KECENI under a set of regularity conditions.  We present both the regularity conditions and the asymptotic results under a triangular array framework, for the sequence of observed data, $\{(Y_{n,[n]}, T_{n,[n]}, X_{n,[n]}, \mathcal{G}_n): n \in \nats\}$, conditional on some fixed network sequence $\mathcal{G}_n$.
We omit the subscript $n$ unless needed for clarity.

\subsection{Empirical Process Theory under Network Dependence} \label{sec:empirical_process_theory}

We first present new empirical process theory which is critical for our analysis.   Since we use the same data to both estimate the nuisance functions and form the kernel-weighted averages, this creates nontrivial dependence in our setting of a {\it single} observed network. We believe this theory is of general interest to researchers working under network dependence.  

The key assumption which makes this theory tractable is a particular form of \emph{network dependence}, which restricts dependencies within the network and allows us to analyze each unit's outcome in isolation from non-neighboring units. 

\begin{assumption}[Network Dependence] \label{assmp:network_dependence}
    For all $n$ and all $i \in [n]$,
    \begin{equation*}
        X_i \indep X_{N_i^\cmpl}, ~\text{and}~ (T_i, Y_i) \indep (X_{N_i^\cmpl}, T_{N_i^\cmpl}, Y_{N_i^\cmpl}) | X_{N_i}, 
    \end{equation*}
   where $N_i \equiv \{i\} \cup \{j \in [n]: (i,j) \in \mathcal{E}\}$,  $N_i^\cmpl = [n] \setminus N_i$.
\end{assumption}

Under this assumption, the tuples \((Y_i, T_{N_i}, X_{N_i})\) and \((Y_j, T_{N_j}, X_{N_j})\) may be dependent only when nodes \(i\) and \(j\) are sufficiently close within the network \(\mathcal{G}\). Consequently, for each node \(i\) there exists a subset of nodes \(\tilde{N}_i\) such that  
$
 (Y_i, T_{N_i}, X_{N_i}) \indep (Y_j, T_{N_j}, X_{N_j}) \mathrm{~for~} j \notin \tilde{N}_i. 
$
The size of this subset \( K_i \equiv |\tilde{N}_i|\) represents the ``radius'' of local dependence of $(Y_i, T_{N_i}, X_{N_i})$ and is typically much smaller than \(n\); for example, $\tilde{N}_i$ can be taken as the $5$-hop neighborhood of $i$ in $\mathcal{G}$.  Let \(K_{\max} \equiv \max_{i \in [n]} K_i\). 

Although empirical process theory under network dependence has been studied, for example in \citet{van2014causal}, we cannot apply it to KECENI directly, because it mainly applies to unweighted empirical averages, and our smoothing weights may vary substantially across units as $\lambda$ shrinks.  Moreover, the existing theory is too conservative in its dependence on network complexity.  Here we develop a new combinatorial double-counting argument that improves the dependence on network complexity while explicitly incorporating weights.

Consider a weighted empirical process
$Z_n(\theta) \equiv \sum_{i=1}^n w_i f_i(\theta) / \sum_{i=1}^n w_i$,
indexed by $\theta \in \mathcal{F}$, where $\mathcal{F}$ is a functional space with metric $d$, $w_i$ are random weights, and $f_i(\theta) \indep f_j(\theta)$ for $j \notin \tilde{N}_i$. We establish \emph{stochastic equicontinuity} of this process.

\begin{theorem} \label{thm:asymptotic_equicontinuity}
    The process $Z_n$ is stochastically equicontinuous and 
    \begin{equation*}
        \Pr\left[ \sup_{d(\theta_1, \theta_2) < \delta_n} \sqrt{\frac{\sum_{i=1}^n w_i}{ K_{\max}}} \abs*{Z_n(\theta_1) - Z_n(\theta_2) - \Exp[Z_n(\theta_1) - Z_n(\theta_2)]} > \eta \right] \rightarrow 0,
    \end{equation*}
    for any sequence $\delta_n \rightarrow 0$ and $\eta >  0$, under regularity conditions, including (a) a Lipschitz-type moment condition, (b) total boundedness, and (c) a finite entropy integral for $\mathcal{F}$.
\end{theorem}
Conditions (a)--(c) are stated in \cref{app:pf_equicontinuity} and are standard in empirical process theory \citep{vaart1996weak}.  
To illustrate the rate improvement, consider the case of equal weights. The empirical process rate in \citet{van2014causal} is $K_{\max}/\sqrt{n}$, while the standard deviation of KECENI is  of order $\sqrt{K_{\max}/n}$, and is therefore too crude to yield asymptotic normality. Theorem \ref{thm:asymptotic_equicontinuity} improves this rate to $\sqrt{K_{\max}/n}$ while explicitly allowing for smoothing weights, which is key to establishing asymptotic normality and valid inference for KECENI.

\subsection{Consistency of KECENI}

We now establish consistency of our estimator. 
Let $\bar\mu$ and $\bar\pi$ be the limits of $\hat\mu$ and $\hat\pi$, in the sense defined in \cref{app:pf_consistency}. 

\begin{theorem} \label{thm:consistency}
    Let \cref{assmp:network_interference,assmp:ignorability,assmp:network_dependence} hold and assume that
    (a) $K_{\max} = O(\sqrt{n})$;
    (b) $D \equiv \mathbb{E}\!\left[\sum_{i=1}^n \kappa_\lambda(\Delta_i) \mid \mathcal{G}\right] = \Omega(\sqrt{n})$;
    (c) $\kappa$ has support $[0,1]$;
    (d) $\abs{\theta_i(T_{N_i}) - \theta_{i^*}(t^*_{N_{i^*}})} \leq L \Delta_i$ for some $L > 0$;
    (e) $Y_i$ are uniformly bounded over $i$ and $n$;
    (f) $(\hat{\mu}, \hat{\pi})$ and $(\bar{\mu}, \bar{\pi})$ belong to uniformly bounded function classes $\mathcal{F}$ with finite uniform entropy integrals, and $1/\hat{\pi}$ and $1/\bar{\pi}$ are uniformly bounded;
    (g) either $\bar{\pi} = \pi$ or $\bar{\mu} = \mu$; and
    (h) $\hat{P}_{X_{[n]} \mid \mathcal{G}}$ converges uniformly and equicontinuously to the true distribution over $\mathcal{F}$.
    Then, 
    \begin{equation} \label{eq:consistency}
        \abs{\hat\theta_{i^*}(t^*_{N_{i^*}}) - \theta_{i^*}(t^*_{N_{i^*}})}
        = O_p\left\{\sqrt{{K_{\max}}/{D}} + \lambda + r^\circ_n \cdot s^\circ_n + q_n \right\},
    \end{equation}
    where $r_n^\circ$ and $s_n^\circ$ represent the local convergence rate of $\hat\mu$ and $\hat\pi$ to $\mu$ and $\pi$ respectively, $q_n$ represent the uniform convergence rate of $\hat{P}_n$ over $\mathcal{F}$.
\end{theorem}

We defer the formal statement of these relatively mild conditions, the specification of $(r_n^\circ, s_n^\circ, q_n^\circ)$, and the proof to \cref{app:pf_consistency}. 
The convergence rate in \cref{eq:consistency} consists of four components. The first two terms correspond to variance and bias, respectively. 
As $\lambda$ increases, $D$ also increases, which in turn reduces the variance. This aligns with the typical bias-variance tradeoff in kernel smoothing: increasing $\lambda$ allows for borrowing information from more units, thereby reducing variance at the cost of increasing bias. The third term represents the product of the local convergence rates of the nuisance estimates $\hat\mu$ and $\hat\pi$. 
This term reflects the \emph{rate double robustness} \citep{jiang2022new} of KECENI, where the overall convergence rate can be fast 
even when the individual convergence rates of the nuisance estimators are slower. 
The fourth term captures the convergence rate of the covariate distribution estimator. Unfortunately, the estimation error for the covariate distribution does not have the double robustness property: for consistent estimation, the covariate distribution model must be correctly specified.

\subsection{Interpretation of consistency conditions}
Here we provide further discussion regarding the conditions of Theorem \ref{thm:consistency}. The first five conditions concern the nonparametric estimation of node-wise counterfactual means. Condition (a) rules out networks with extreme hubs: if a few nodes have very large degrees, their treatment assignments can induce strong dependence across a substantial fraction of the network, which can fundamentally undermine valid inference regardless of the estimation strategy; see Section 4.4 of \citet{ogburn2022causal}. 

Condition (b) is easy to satisfy if the maximum neighborhood size $K_{\max}$ is uniformly bounded, for example, with simple random assignment with treatment probability bounded away from 0 and 1. When $K_{\max}$ grows with $n$, however, (b) typically requires the marginal treatment probability $p$ to approach $1/2$ at an appropriate rate to avoid extremely high or low propensity scores induced by large neighborhoods. This phenomenon is inherent to interference settings with growing dependence and is not specific to our method.
More broadly, Condition (b) may be viewed as a positivity condition adapted to network settings. Recent work has proposed experimental designs that explicitly control or avoid small propensity scores under interference, for example by restricting treatment allocations or using conflict-aware randomization schemes \citep{kandiros2024conflict}. These designs are good examples of settings when (b) is expected to hold even when $K_{\max}$ grows with $n$.

Condition (c) restricts the kernel support to be bounded for technical reasons, though empirically KECENI performs well  with kernels of unbounded support. 
Condition (d) is the usual smoothness condition in nonparametric estimation, and requires the dissimilarity metric $\Delta_i$ to capture the causally relevant difference between nodes $i$ and $i^*$ through their connectivity and neighborhood treatment assignments, in the sense that it reflects the closeness between $\theta_i(T_{N_i})$ and $\theta_{i^*}(t^*_{N_{i^*}})$.
Condition (e) is another technical assumption common in related work, for example in \citet{kennedy2017non}.   While we expect KECENI to work just as well with unbounded outcomes as long as they are not heavy-tailed, there are no empirical processes results under network dependence available for this case, and obtaining those requires substantial additional work   beyond the scope of this paper.

The remaining conditions concern nuisance parameters. Condition (f) controls the size of the nuisance function classes for $\hat{\mu}$ and $\hat{\pi}$ through entropy numbers \citep[p.~83]{vaart1996weak}, which is standard in empirical process theory and also appears in \citet{kennedy2017non} and \citet{van2014causal}.    Such conditions can be weakened using sample splitting; see \citet[Chapter 27]{van2011targeted}. Condition (g) requires either the outcome regression or the propensity score estimator to converge to the truth, yielding double robustness. Condition (h) imposes suitable asymptotic requirements on $\hat{P}_{X_{[n]} \mid \mathcal{G}}$. This plays the role of a Glivenko-Cantelli or Donsker type condition for the estimated covariate distribution. 
In the absence of interference, if the covariate distribution is estimated by the empirical cdf,  (h) reduces to the usual Glivenko-Cantelli and Donsker properties of the nuisance classes, implied by (f). The same guarantee extends to the network interference setting when covariates are i.i.d.\ across nodes and the covariate distribution is estimated by the {\em empirical product  measure} \citep[Theorems 3.11 and 4.8]{arcones1993limit}:
\begin{equation} \label{eq:empirical_product_measure}
    \hat{P}_{X_{N_i} \mid \mathcal{G}} = \hat{P}^\otimes_{X_{N_i} \mid \mathcal{G}} \equiv \otimes_{j \in N_i} \hat{P}_X,
\end{equation}
where $\hat{P}_X \equiv \frac{1}{n} \sum_{i=1}^n \delta_{X_i}$ is the empirical measure of the observed covariates.

\subsection{Variance Estimation and Asymptotic Normality of KECENI}

We next establish asymptotic normality of our estimator.  Like any other estimator based on kernel smoothing, KECENI is biased at the scaling relevant for inference. 
Define smoothing bias $b_n$ and variance $\sigma_n^2$ as 
\begin{equation} \label{eq:asymptotic_variance}
\begin{aligned}
    b_n & \equiv \tilde\theta_{i^*}(t^*_{N_{i^*}}) - \theta_{i^*}(t^*_{N_{i^*}}),
    \quad \sigma_n^2 \equiv \Var\left[ \left.
        D^{-1} \tsum_{i=1}^n \psi_i
    \right| \mathcal{G} \right], \\
\end{aligned}
\end{equation}
where
$\psi_i 
\equiv \kappa_\lambda(\Delta_i) ~ (\xi_i(\bar\mu,\bar\pi) - \tilde\theta_{i^*}(t^*_{N_{i^*}})) 
$ 
is an \emph{influence function (IF)} of  $\tilde\theta_{i^*}(t^*_{N_{i^*}}) \equiv {D}^{-1} \Exp[ \sum_{i=1}^n \kappa_\lambda(\Delta_i) ~ \theta_{i}(T_{N_i}) | \mathcal{G}]$. 
The following theorem establishes asymptotic normality of our estimators, centered on the bias-adjusted mean. 

\begin{theorem} \label{thm:asymptotic_normality}
    Suppose that $\max\{r^\circ_n \, s^\circ_n, q_n\} = o( \sigma_n )$ and that $\sqrt{K_{\max}/D} = O( \sigma_n )$.
    Then, under the assumptions of \cref{thm:consistency}, we have
    $
        {\sigma_n^{-1}} (\hat\theta_{i^*}(t^*_{N_{i^*}}) - \theta_{i^*}(t^*_{N_{i^*}}) - b_n)
        \overset{d}{\rightarrow} N(0, 1).
    $
\end{theorem}

We note that the condition $\max\{r^\circ_n \, s^\circ_n, q_n\} = o( \sigma_n )$ in \cref{thm:asymptotic_normality} requires nuisance-estimation error to be asymptotically negligible at first order. Otherwise, Wald-type confidence intervals based on the naive plug-in influence-function variance estimator may exhibit anticonservative coverage. This limitation is well known in doubly robust estimation; see, for example, \citet{shook2025double}. If the nuisance-estimation contribution admits a linear approximation by local statistics, asymptotic normality can still be obtained after incorporating these first-order nuisance terms into the linearization. The corresponding limiting variance can then be consistently estimated by an empirical sandwich-type variance estimator.

Specifically, we assume that the first-order estimation errors in $\hat\mu$, $\hat\pi$, and the covariate-distribution integrals
admit linear representations as sums of local statistics, where the contribution indexed by node $i$ depends on the observed data only through $(Y_i,T_{N_i},X_{N_i})$. Substituting these representations into the first-order expansion of 
$\hat\theta_{i^*}(t^*_{N_{i^*}})-\theta_{i^*}(t^*_{N_{i^*}})$ yields
$
    \hat\theta_{i^*}(t^*_{N_{i^*}}) - \theta_{i^*}(t^*_{N_{i^*}})
    \approx \sum_{i=1}^n W_i ,
$
where 
each $W_i$ contains the direct influence-function contribution $\psi_i$, together with correction terms that propagate first-order uncertainty from nuisance function estimation. Finally, to account for network dependence, we aggregate cross-products only over pairs of nodes that may be locally dependent:
$
    \hat\sigma^2
    =
    \sum_{i,j=1}^n h(i,j)\hat W_i\hat W_j,
$
where $h(i,j) \equiv \mathbf 1\{j\in \tilde N_i\}$, and $\hat W_i$ is the plug-in estimator of $W_i$. This dependence-adjusted estimator consistently estimates the variance of $\hat\theta_{i^*}(t^*_{N_{i^*}})$ under regularity conditions. We defer the detailed specification of $\hat W_i$, together with the regularity conditions and proofs of both asymptotic normality for our estimator and consistency of the empirical sandwich variance estimator in the doubly robust regime to \cref{app:detail_inference}.

\begin{remark}
    In practice, the smoothing bias $b_n$ can affect the validity of Wald-type inference when the bandwidth is chosen to maximize prediction accuracy, for example, via cross-validation. In that regime, the resulting confidence intervals are most naturally interpreted as inference for $\tilde\theta_{i^*}(t^*_{N_{i^*}})$, which is aligned with the standard interpretation in smoothing-based procedures \citep{kennedy2017non,wasserman2006all}. If inference for the unsmoothed target is the goal, one quick solution is {\it undersmoothing}, that is, choosing a bandwidth smaller than optimal for prediction, so that the bias becomes asymptotically negligible.  While this approach yields valid inference, it can worsen prediction performance and is of limited practical value.  Another option is explicit bias estimation and debiasing as in \citet{calonico2018effect}; in the i.i.d.\ setting, related ideas were developed for the estimator of \citet{kennedy2017non} by \citet{takatsu2025debiased}. Extending such debiasing methods to our setting is difficult, because the smoothing is over the space of local neighborhood configurations, though it may be easier if the similarity metric is a function of some low-dimensional exposure mapping.
\end{remark}

\begin{remark}[Feasibility of the additional rate condition under structural equation model]
    The condition $\max\{r^\circ_n \, s^\circ_n, q_n \} = o( \sigma_n )$ is relatively mild and satisfied by common structural equation models. 
    Suppose that 
    $X_i = f_X\!\left(\vareps_{X_i}\right)$,  $T_i = f_T\!\left(s_X(X_{N_i}), \vareps_{T_i}\right)$, and $Y_i = f_Y\!\left(s_T(T_{N_i}), s_X(X_{N_i}), \vareps_{Y_i}\right)$, 
    where $s_X$ and $s_T$ are $d_X$- and $d_T$-dimensional summary mappings of $X_{N_i}$ and $T_{N_i}$. Typical choices include
    $s_X(X_{N_i}) = \bigl(X_i, \mathrm{Avg}(X_{N_i\setminus\{i\}})\bigr)$ or $\bigl(X_i, \mathrm{Sum}(X_{N_i\setminus\{i\}})\bigr)$, with analogous choices for $s_T$.
    For the exogenous noise, we assume $\vareps_{X_i}$ and $\vareps_{T_i}$ are i.i.d., while allowing $\vareps_{Y_i}$ to exhibit positive correlation across adjacent nodes. This aligns with the notion of outcome spillover in \citet{bhadra2025causal}.
    Suppose that \cref{eq:propensity_ind} is assumed and $\mu(T_{N_i},X_{N_i},\mathcal{G}_{N_i})$ and $\pi^\circ(X_{N_i},\mathcal{G}_{N_i})$ are estimated based on nonparametric kernel smoothing,
    with the empirical product measure for the covariate distribution. 
    Then the additional condition in \cref{thm:asymptotic_normality} holds when
    \begin{equation*}
        \max\left\{
            \sqrt{\frac{K_{\max}}{n}}, 
            \frac{K_{\max}}{ n^{\frac{1}{2+d_X} + \frac{1}{2+d_T+d_X}}}
        \right\}
        = o\!\left( \sqrt{\frac{K_{\max}}{n \lambda_n^{d_T}}}\right).
    \end{equation*}
    This imposes a polynomial upper bound on $K_{\max}$, which is compatible, for example, with degree distributions whose tails are well approximated by a power law with an exponential cutoff or by a lognormal tail, both widely used as descriptive models for empirical networks \citep{broido2019scale,clauset2009power,mitzenmacher2004brief}. See \cref{app:sem_example} for details of the derivation.
\end{remark}

%% file: sec/4_numerical.tex
\section{Numerical results} \label{sec:simulation}

\subsection{Node-wise Counterfactual Means} \label{sec:simulation_node_wise}

 We begin by evaluating KECENI for estimating node-wise counterfactual means in simulation studies. Since there are no existing competing methods for this specific task, we focus on demonstrating the consistency and asymptotic normality established in \cref{sec:theorems}. For space reasons, we provide only a brief description of the simulation settings here; full details are given in \cref{app:supp_simulation}.


\paragraph{Settings.}

The simulation consists of 80 repeated runs, all based on the same network \(\mathcal{G}\) with \(n = 1000\) units. We generate \(\mathcal{G}\) once from a latent variable network model, in which the probability of an edge between two nodes depends on the distance between their latent positions. The latent positions are drawn i.i.d.\ from the uniform distribution on $[-1, 1]^2$. 
Conditional on this fixed network, we generate covariates \(X_i \in \mathbb{R}^3\) i.i.d.\ from $\mathcal{N}(0,1)$; assign treatments \(T_i \in \{0,1\}\) independently according to a logistic model with predictors $\left(X_i, \mathrm{Avg}(X_{N_i \setminus \{i\}})\right)$; and finally generate outcomes \(Y_i \in \mathbb{R}\) following the two-hop neighborhood construction in \cref{rem:two_hop_neighborhoods},  from a linear model with predictors \(\big(T_i, \mathrm{Avg}(T_{N_i \setminus \{i\}}), X_i, \mathrm{Avg}(X_{N_i^{(2)} \setminus \{i\}})\big)\). 

We select the node $i^*$ with the smallest $\ell_\infty$ norm of the latent position and apply KECENI to estimate its counterfactual mean outcome, $\theta_{i^*}(t^*_{N_{i^*}})$, under a given treatment allocation \(t^*_{N_{i^*}} \in \{0, 1\}^{\abs{N_{i^*}}}\). This choice of target node \(i^*\) ensures that network edge effects are avoided. We consider two contrasting treatment scenarios: \(t^{(0)}_{N_{i^*}}\), where no neighbors are treated (\(t^{(0)}_i = 0\) for all \(i \in N_{i^*}\)), and \(t^{(1)}_{N_{i^*}}\), where all neighbors are treated (\(t^{(1)}_i = 1\) for all \(i \in N_{i^*}\)). 
The true target parameter values are \(\theta_{i^*}(t^{(0)}_{N_{i^*}}) = -2\) and \(\theta_{i^*}(t^{(1)}_{N_{i^*}}) = 2\), respectively.

We estimate the propensity scores and outcome regression fitting logistic and linear regressions with covariates used to generate the data, respectively.
We estimate the covariate distribution by the empirical product measure $\hat{P}^\otimes_{X_{[n]}|\mathcal{G}}$ as defined in \cref{eq:empirical_product_measure}.
We specify the dissimilarity metric $\Delta_i$ through a low-dimensional exposure mapping, 
\(
    \Delta_i \equiv \norm{(T_i, \mathrm{Avg}(T_{N_i \setminus \{i\}})) 
    - (t^*_{i^*}, \mathrm{Avg}(t^*_{N_{i^*} \setminus \{i^*\}}))}_1.
\)

\paragraph{Results.}

\begin{figure}[tbp]
    \centering
    \begin{subfigure}[t]{0.23\linewidth}
        \centering
        \includegraphics[width=\textwidth, trim={0 0.8cm 0 0}]{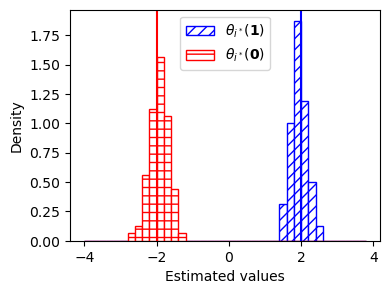}
        \subcaption{}
        \label{fig:hist_T1_T2_ite}
    \end{subfigure}
    %
    %
    \begin{subfigure}[t]{0.23\linewidth}
        \centering
        \includegraphics[width=\textwidth, trim={0 0.8cm 0 0}]{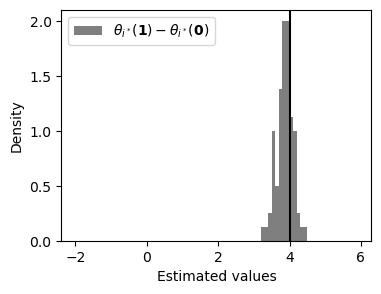}
        \subcaption{}
        \label{fig:hist_Td_ite}
    \end{subfigure}
    \begin{subfigure}[t]{0.23\linewidth}
        \centering
        \includegraphics[width=\textwidth, trim={0 0.8cm 0 0}]{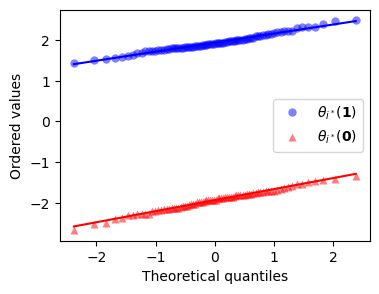}
        \subcaption{}
        \label{fig:qq_norm_T1_T2_ite}
    \end{subfigure}
    %
    %
    \begin{subfigure}[t]{0.23\linewidth}
        \centering
        \includegraphics[width=\textwidth, trim={0 0.8cm 0 0}]{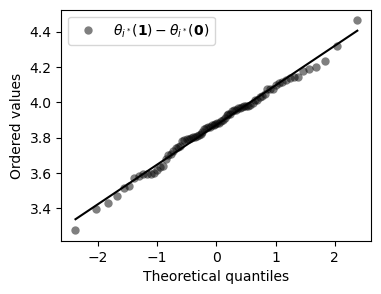}
        \subcaption{}
        \label{fig:qq_norm_Td_ite}
    \end{subfigure}

    \begin{subfigure}[t]{0.23\linewidth}
        \centering
        \includegraphics[width=\textwidth, trim={0 0.8cm 0 0}]{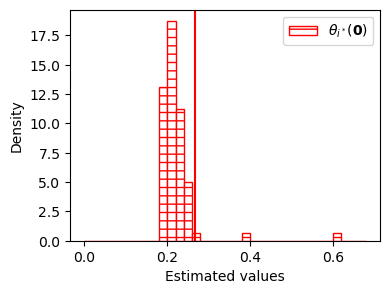}
        \subcaption{}
        \label{fig:hist_sd1s_ite}
    \end{subfigure}
    \begin{subfigure}[t]{0.23\linewidth}
        \centering
        \includegraphics[width=\textwidth, trim={0 0.8cm 0 0}]{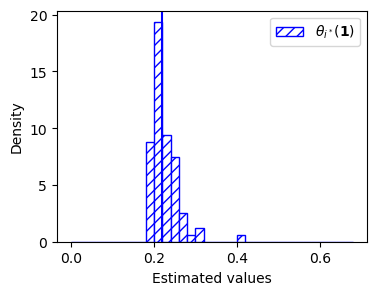}
        \subcaption{}
        \label{fig:hist_sd2s_ite}
    \end{subfigure}
    \begin{subfigure}[t]{0.23\linewidth}
        \centering
        \includegraphics[width=\textwidth, trim={0 0.8cm 0 0}]{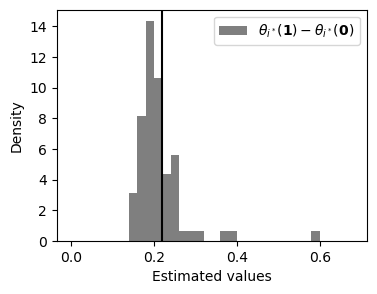}
        \subcaption{}
        \label{fig:hist_sdds_ite}
    \end{subfigure}

    \vspace{-0.1in}
    \caption{Histograms (a,b) and normal Q-Q plots (c,d) of KECENI estimates. (a,c) Node-wise counterfactual means for treatments $t^{(1)}_{N_{i^*}}$ and $t^{(0)}_{N_{i^*}}$. (b,d) Estimated treatment effect.  
    Histograms of standard error estimates (e) for $\theta_{i^*}(t^{(0)})$, (g) for $\theta_{i^*}(t^{(1)})$ and (g) for $\theta_{i^*}(t^{(1)}) - \theta_{i^*}(t^{(0)})$.
    True values are indicated by solid vertical lines. 
    }
    \label{fig:result_ite}
\end{figure}

\cref{fig:result_ite} presents the estimation and inference results from applying KECENI, averaged over the 80 replications.  
  \cref{fig:hist_T1_T2_ite} shows that the distributions of \(\hat\theta_{i^*}(t^{(1)}_{N_{i^*}})\) and \(\hat\theta_{i^*}(t^{(0)}_{N_{i^*}})\) are close to centered on the true values, but there is a slight downward bias in their estimated difference (\cref{fig:hist_Td_ite}).  This smoothing bias is expected, as shown in \cref{eq:asymptotic_variance}, especially as the target treatments are extreme cases (``all treated'' vs. ``no treated'').
The smoothing bias pulls the estimates towards the average outcome in the observed data, and their difference down.  
In \cref{app:simulation_n_node}, we show that the bias decreases as the sample size increases, confirming consistency shown in \cref{thm:consistency}.
The normal Q-Q plots in \cref{fig:qq_norm_T1_T2_ite,fig:qq_norm_Td_ite} provide empirical evidence of  asymptotic normality presented in \cref{thm:asymptotic_normality}.
\crefrange{fig:hist_sd1s_ite}{fig:hist_sdds_ite} display the distributions of the estimated standard errors from the empirical sandwich variance estimator for three target estimands: $\theta_{i^*}(t^{(0)})$, $\theta_{i^*}(t^{(1)})$, and $\theta_{i^*}(t^{(1)}) - \theta_{i^*}(t^{(0)})$. The standard error estimates for the node-wise counterfactual means $\theta_{i^*}(t^{(0)})$ and $\theta_{i^*}(t^{(1)})$ are well centered around the true standard error, estimated via Monte Carlo over the 80 simulations. The resulting 95\%-confidence intervals maintain nominal coverage: the coverages over 80 simulations are $0.913$ for $\theta_{i^*}(t^{(0)})$, $0.963$ for $\theta_{i^*}(t^{(1)})$, and $0.938$ for $\theta_{i^*}(t^{(1)}) - \theta_{i^*}(t^{(1)})$.

Supplementary results on estimation bias when network interference is ignored are available in \cref{app:simulation_nonet}.

\subsection{Double Robustness}
\label{sec:simulation_dr}

Here we demonstrate the double robustness of KECENI in estimating node-wise counterfactual means, and compare it with the \emph{G-computation} estimator \citep{robins1986new}.
Given \(\hat\mu\) and \(\hat{P}_{X_{[n]}|\mathcal{G}}\) as defined in \cref{sec:keceni}, the G-computation estimator is defined as
$\hat\theta^{G}_{i^*}(t^*_{N_{i^*}}) \equiv \int \hat\mu(t^*_{N_{i^*}}, x_{N_{i^*}^{(2)}}, \mathcal{G}_{N_{i^*}^{(2)}}) d\hat{P}_{X_{N_{i^*}^{(2)}}|\mathcal{G}}(x_{N_{i^*}^{(2)}}).$
We find that, similarly to the no-interference setting, this estimator is vulnerable to misspecification of the outcome regression model, while KECENI is doubly robust.  

\paragraph{Settings.} 

We use the same data as in the previous simulation. 
%
%
The objective is to estimate the direct treatment effect for node $i^*$. We set $t^{(0)}_{N_{i^*}}$ so that $t^{(0)}_{i^*} = 0$ and $|N_{i^*}|^{-1} \sum_{i \in N_{i^*}} T_i = 0.5$. The second treatment scenario $t^{(1)}_{N_{i^*}}$ is the same as $t^{(0)}_{N_{i^*}}$ except for $t^{(1)}_{i^*} = 1$. The true target parameter values are \(\theta_{i^*}(t^{(0)}_{N_{i^*}}) = -1\) and \(\theta_{i^*}(t^{(1)}_{N_{i^*}}) = 1\).  

We investigate the sensitivity of the G-computation and KECENI to nuisance model misspecification. For outcome regression, we use a linear model with misspecified predictors 
\(\left(T_i, \mathrm{Avg}(T_{N_i \setminus \{i\}}), X_i, \mathrm{Avg}((1-\alpha_\mu) X_{N_i^{(2)} \setminus \{i\}} + \alpha_\mu X_{N_i^{(2)} \setminus \{i\}}^2)\right)\), and similarly for the propensity score model, with $\alpha_\pi$ in place of $\alpha_\mu$. The parameters $\alpha_\mu$ and $\alpha_\pi$ quantify the extent of misspecification:   
%
the models are correctly specified when $\alpha_\pi = \alpha_\mu = 0$ and become increasingly misspecified as $\alpha_\pi$ and $\alpha_\mu$ increase towards $1$. 
We vary  $\alpha_\mu$ and $\alpha_\pi$ from 0 to 1 with a step size of 0.1, and measure the estimation accuracy by root mean square error (RMSE).   

\begin{figure}[tbp]
    \centering
    \begin{subfigure}[t]{0.3\linewidth}
        \centering
        \includegraphics[height=0.8\textwidth, trim={0 0.3cm 1.5cm 0}, clip]{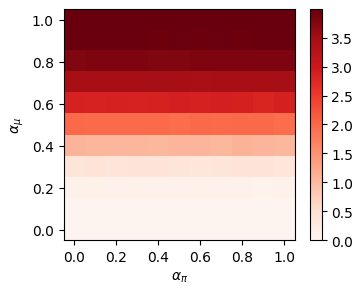}
        \subcaption{G-computation}
        \label{fig:result_alpha_G}
    \end{subfigure}
    \qquad
    \begin{subfigure}[t]{0.3\linewidth}
        \centering
        \includegraphics[height=0.8\textwidth, trim={0 0.3cm 0 0}, clip]{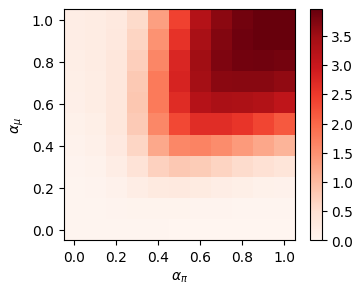}
        \subcaption{KECENI}
        \label{fig:result_alpha_DR}
    \end{subfigure}
    \vspace{-0.1in}
    \caption{RMSE for the estimated direct effect as a function of $\alpha_\pi$ and $\alpha_\mu$.}
    \label{fig:result_alpha}
\end{figure}

\paragraph{Results.} \cref{fig:result_alpha} presents heatmaps of RMSE for the estimated direct effect, $\hat\theta_{i^*}(t^{(1)}_{N_{i^*}}) - \hat\theta_{i^*}(t^{(0)}_{N_{i^*}})$, as a function of $(\alpha_\pi, \alpha_\mu)$. It shows that the G-computation is sensitive to any outcome regression misspecification, although it is not sensitive to misspecification of the propensity score model, while KECENI is robust to misspecification in either as long as the other one is not too far from correctly specified. 


\subsection{Average Treatment Effect} \label{sec:simulation_ate} 

In this simulation study, we compare KECENI with two competing methods for estimating the average treatment effect under network interference that have publicly available code, AUTOGNET \citep{tchetgen2021auto} and TMLENET \citep{ogburn2022causal}. Both rely on structural equation models: AUTOGNET assumes a parametric autoregressive model, and TMLENET uses structural equation models defined by pre-specified summary statistics.   In contrast, KECENI can work well even when the relevant parametric model or summary statistics are unknown and therefore cannot be correctly specified, because it can use nonparametric estimation based on fused Gromov-Wasserstein distances \citep{vayer2020fused}.  Again, we provide a brief overview of the settings here, with full details  given in \cref{app:supp_simulation}.

\paragraph{Settings.}

The simulation consisted of $40$ repeated runs, using the same network model $\mathcal{G}$ as that in the previous subsection with $n = 4000$ units. We use a larger number of units so that flexible nonparametric estimation based on fused Gromov-Wasserstein distances becomes feasible.
For compatibility with existing methods, we work with binary two-dimensional covariate vectors $X_i = (X_i^{(1)}, X_i^{(2)}) \in \{0, 1\}^2$ 
and binary outcome variables $Y_i \in \{0, 1\}$. Each covariate is independently sampled from the Bernoulli distribution with probability $0.5$, and treatments  $T_i \in \{0, 1\}$ are independently drawn following a logistic model with predictor $\mathrm{Avg}((X_{N_i \setminus \{i\}}^{(1)} - 0.5)(X_{N_i \setminus \{i\}}^{(2)} - 0.5))$. 
Outcomes $Y_i \in \{0,1\}$ are generated from a logistic model with predictors $\left(T_i, \mathrm{Avg}((X_{N_i \setminus \{i\}}^{(1)} - 0.5)(X_{N_i \setminus \{i\}}^{(2)} - 0.5))\right)$. 

We considered the none-treated vs.\ all-treated scenarios,  $t_i^{(0)} = 0$ or $t_i^{(1)} = 1$ for all $i$.  The true target parameter values are \( n^{-1} \sum_{i=1}^n \theta_{i}(t^{(0)}_{N_{i}}) \approx 0.406\) and \( n^{-1}\sum_{i=1}^n\theta_{i}(t^{(1)}_{N_{i}}) \approx 0.594\). 

\paragraph{Results.}

\begin{figure}[tbp]
    \centering
    \begin{subfigure}[t]{0.3\linewidth}
        \centering
        \includegraphics[width=1\textwidth]{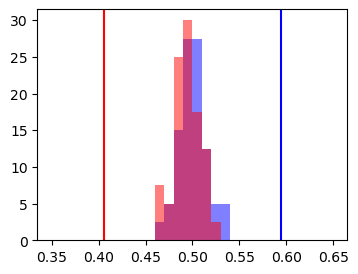}
        \subcaption{AUTOGNET}
        \label{fig:result_X_autognet}
    \end{subfigure}
    \begin{subfigure}[t]{0.3\linewidth}
        \centering
        \includegraphics[width=1\textwidth]{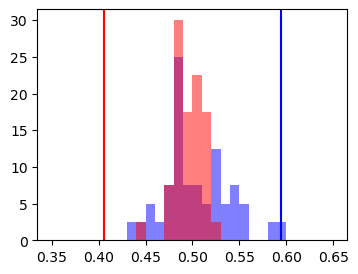}
        \subcaption{TMLENET}
        \label{fig:result_X_tmlenet}
    \end{subfigure}
    \begin{subfigure}[t]{0.3\linewidth}
        \centering
        \includegraphics[width=1\textwidth]{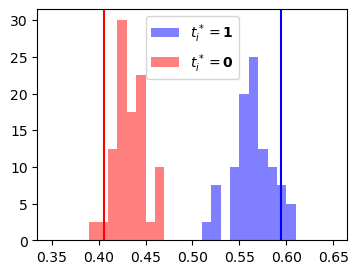}
        \subcaption{KECENI}
        \label{fig:result_X_ot}
    \end{subfigure}
    \vspace{-0.1in}
    \caption{Histogram of average counterfactual mean estimates of the three competing methods when the data generating model is misspecified.  
    Vertical lines indicate true values. }
    \label{fig:result_ate_X}
\end{figure}

We evaluate the performance of three competing methods (AUTOGNET, TMLENET and KECENI), which have no access to the true underlying model.  
In this case, a natural strategy for AUTOGNET and TMLENET is to construct structural equation models using summary statistics $\mathrm{Avg}(X_{N_i \setminus \{i\}})$. 
This results in structural equation models without the interaction term and therefore misspecify the true data-generating process, and  AUTOGNET and TMLENET both produced significantly inaccurate estimates, failing to identify the treatment effect (\cref{fig:result_X_autognet,fig:result_X_tmlenet}).

In contrast, KECENI can use a fully nonparametric approach with fused Gromov-Wasserstein distances used to measure dissimilarity in both local network structure and node attributes. In particular, we define
\begin{equation} \label{eq:delta_gw}
    \Delta_i \equiv \abs{T_i - t^*_{i^*}} 
    + W_{1,1}\left(\hat\Pr\{T_j\}_{j \in N_i \setminus \{i\}}, \hat\Pr\{t^*_j\}_{j \in N_{i^*} \setminus \{i^*\}}\right),
\end{equation}
where \(\hat\Pr\) is the empirical probability measure on the respective set, and \(W_{1,1}(\cdot, \cdot)\) denotes the Wasserstein 1-distance with respect to the \(\ell_1\) metric.
We can also nonparametrically estimate the outcome regression and propensity score using kernel smoothing estimators based on fused Gromov-Wasserstein distances defined in \cref{eq:nuisance_fgw}. 
Finally, \(\hat{P}_{X_{[n]}|\mathcal{G}}\) is taken to be the empirical product measure in \cref{eq:empirical_product_measure}. 
\cref{fig:result_X_ot} indicates that despite some smoothing bias, KECENI was able to identify the non-zero treatment effect much more successfully than AUTOGNET and TMLENET. 

If the structural equation models for AUTOGNET and TMLENET are correctly specified and include the interaction term, enabling successful identification of the causal effect (see \cref{fig:result_W_autognet,fig:result_W_tmlenet}). KECENI also improves when using this additional information (\cref{fig:result_W_linear,fig:result_W_kernel}). Additional details are provided in \cref{app:comparison_known}.

\section{Application to Indian Rural Social Networks} \label{sec:application}

In this section, we analyze the  microfinance data on rural Indian villages  from \citet{banerjee2013diffusion} to study the impact of participation in savings self-help groups (SHGs) on financial behavior.  The dataset was collected from a survey conducted in 75 villages in Karnataka, India. 
which included a village-level questionnaire on leadership, a full household census, and a detailed follow-up individual demographic survey based on stratified sampling by religion and geographic sublocation. In addition, socioeconomic interactions between the subsampled individuals were captured in a 12-dimensional multilayer network, where each layer corresponded to a different type of relationship.

In 37 of the surveyed villages, additional financial information was collected, including SHG participation and outstanding loans.
Using these villages, \citet{khatami2024graph} examined the causal effect of SHG participation on financial risk tolerance, measured by the presence of outstanding loans. We expand on their analysis by investigating not only the average direct and spillover effects across the population, but also how these effects vary with both the intervention and network characteristics of units. 
We define the interference network $\mathcal{G}$ by combining information from 7 layers (relationship types) of the surveyed network, following \citet{khatami2024graph}. 

We define the direct effect of the target node intervention and the spillover effect of the neighborhood's interventions, for a given villager $i^*$, as  
\begin{equation*}
\begin{aligned}
    \mathrm{DE}_{i^*} & \equiv \Exp[\theta_{i^*}(t_{i^*} = 1, t_{N_{i^*} \setminus \{i^*\}} = T_{N_{i^*} \setminus \{i^*\}}) | \mathcal{G}] - \Exp[\theta_{i^*}(t_{i^*} = 0, t_{N_{i^*} \setminus \{i^*\}} = T_{N_{i^*} \setminus \{i^*\}}) | \mathcal{G}], \\
    \mathrm{SpE}_{i^*} & \equiv \Exp[\theta_{i^*}(t_{i^*} = T_{i^*}, t_{N_{i^*} \setminus \{i^*\}} = \mathbf{1}) | \mathcal{G}] - \Exp[\theta_{i^*}(t_{i^*} = T_{i^*}, t_{N_{i^*} \setminus \{i^*\}} = \mathbf{0}) | \mathcal{G}], 
\end{aligned}
\end{equation*}
the same quantities \citet{khatami2024graph} studied.  
To address potential confounding, we control for six standardized demographic features $X_i \in \reals^6$ (age, education, whether native to the village, employment status, whether they work outside the village, and saving status) of the two-hop neighborhood $i \in N_{i^*}^{(2)}$, following \cref{rem:two_hop_neighborhoods}. 
We use KECENI to estimate $\mathrm{DE}_{i^*}$ and $\mathrm{SpE}_{i^*}$, assuming that covariates are i.i.d., and treatment is assigned independently across nodes given covariates, as formalized in \cref{eq:propensity_ind}.  We emphasize that these assumptions are modeling choices for the nuisance components, not structural requirements for identification or validity of our estimator, and are consistent with a commonly used structural equation model \citep{van2014causal,ogburn2022causal}. If covariates exhibit network-induced dependence or if treatment assignments are dependent within neighborhoods even after conditioning, the same overall estimation approach can still be implemented by choosing appropriate alternative models for the nuisance components, such as a Markov random field or other locally dependent assignment model. 

We estimate node-level propensity score \(\pi^\circ\) and outcome regression \(\mu\) 
by kernel smoothers on summary statistics $(x_i, \mathrm{Avg}(x_{N_i \setminus \{i\}}))$ and $(t_i, \mathrm{Avg}(t_{N_i \setminus \{i\}}), x_i, \mathrm{Avg}(x_{N_i^{(2)} \setminus \{i\}}))$, respectively.
For the dissimilarity metric, we used $\Delta_i 
\equiv \norm{(T_i, \mathrm{Avg}(T_{N_i \setminus \{i\}})) - (t^*_{i^*}, \mathrm{Avg}(t^*_{N_{i^*} \setminus \{i^*\}}))}_1$.
This type of low-dimensional exposure mapping is again a common choice in the interference literature, with most existing approaches effectively assuming that an exposure mapping is given a priori and then proceeding with identification and estimation under that mapping \citep{tchetgen2021auto,van2014causal,ogburn2022causal,forastiere2021identification,leung2022graph,khatami2024graph}. These assumptions can be evaluated similarly to any other regression assumptions, using standard diagnostic tools such as residual plots, cross-validation, and sensitivity analyses. 
However, nuisance modeling for network-dependent data is less developed than its counterpart for i.i.d.\ data, and diagnostic tools for this setting are an active area of research.
For further details, see \cref{app:supp_real_data}.

We estimate $\hat{\mathrm{DE}}_{i^*}$ and $\hat{\mathrm{SpE}}_{i^*}$ for each $i^*$ in the surveyed population, and aggregate them to obtain estimates for the average direct effect (ADE) and average spillover effect (ASpE). The ADE estimate is $0.178$ with a 95\%-confidence interval of $(0.154, 0.201)$, calculated using the empirical sandwich variance estimator. Our point estimate is smaller than $0.315$ reported by \citet{khatami2024graph} and the six baseline methods listed in their Appendix 12.2. However, our confidence interval indicates significance of the ADE, in contrast to the wider interval $(-1.570, 2.200)$ reported by \citet{khatami2024graph}.

Our estimate of the ASpE is $0.006$, with a 95\%-confidence interval of $(-0.002, 0.014)$, which aligns with the estimate of $0.050$ from \citet{khatami2024graph} and other baseline estimates.  The confidence interval includes zero, albeit only just, suggesting there is 
 no spillover effect; however, a more detailed analysis using our node-wise counterfactual framework reveals otherwise. Specifically, we investigate the spillover effect by intervening on the target node $i^*$ at a fixed value $t^*$, rather than at the observed intervention $T_{i^*}$. Define, with a slight abuse of notation, 
\begin{equation*}
    \mathrm{SpE}_{i^*}(t^*) \equiv \theta_{i^*}(t_{i^*} = t^*, t_{N_{i^*} \setminus \{i^*\}} = \mathbf{1}) - \theta_{i^*}(t_{i^*} = t^*, t_{N_{i^*} \setminus \{i^*\}} = \mathbf{0}),
\end{equation*}
for $t^* = 0, 1$, and compare the estimates $\hat{\mathrm{SpE}}_{i^*}(0)$ and $\hat{\mathrm{SpE}}_{i^*}(1)$. 
The average estimates of $\hat{\mathrm{SpE}}_{i^*}(0)$ and $\hat{\mathrm{SpE}}_{i^*}(1)$ are $-0.003$ and $0.046$, with 95\%-confidence intervals of $(-0.012, 0.005)$ and $(0.030, 0.063)$, respectively. These results indicate a significant spillover effect when the respondent received the intervention, while the ASpE approach likely did not reflect this due to the relatively small proportion of SHG participants (17\%).   

\begin{figure}[tbp]
    \centering
    \includegraphics[scale=0.75]{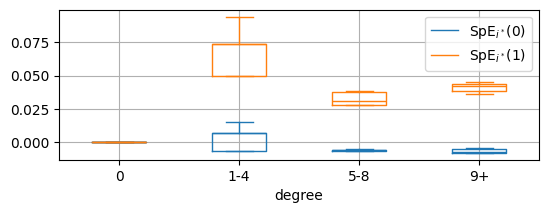}
    \caption{Box plots for the point estimates of the node-wise spillover effect, stratified by the degree of the node.  Spillover effects for treated nodes are shown in orange, and for non-treated nodes in blue. }
    \label{fig:boxplot_SpE_by_d_IndianVillage}
\end{figure}

We further investigate the heterogeneity of the spillover effect as a function of node connectivity. \cref{fig:boxplot_SpE_by_d_IndianVillage} presents boxplots of the point estimates \( \hat{\mathrm{SpE}}_{i^*}(0) \) and \( \hat{\mathrm{SpE}}_{i^*}(1) \), stratified by the degree of node \( i^* \) into four groups: isolated nodes (72 total), nodes with 1--4 connections (2832), 5--8 connections (3793), and 9 or more connections (2686). 
When node \( i^* \) does not receive an intervention (\( t^*_{i^*} = 0 \)), the spillover effect remained approximately constant and close to zero across all groups. However, when \( t^*_{i^*} = 1 \), there is substantial heterogeneity in the spillover effect. 
Excluding isolated nodes, which trivially have no spillover effects, we found that the spillover effect was generally larger for nodes with fewer connections. Specifically, \( \hat{\mathrm{SpE}}_{i^*}(1) \) was significantly higher for nodes with 1--4 connections than for those with 5--8 (\( p = 0.0014 \)) and those with 9+ (\( p = 0.026 \)). 
The difference between the two latter groups is not statistically significant (\( p = 0.134 \)).
Note that these $p$-values are not adjusted for multiple  testing and should be viewed as a data summary.  
These results suggest that less connected individuals tend to experience greater influence from their neighbors' participation in self-help groups compared to highly connected individuals,
potentially due to their greater dependence on a limited number of connections. 
Such heterogeneity in spillover effects, which our approach allows for and picks up, has implications for downstream policy decisions, especially when there is a limited intervention budget.


%% file: sec/5_discussion.tex
\section{Discussion} \label{sec:discussion}

In this paper, we have proposed a novel approach to causal inference under network interference, focused on node-wise counterfactual means.   This flexible framework accounts for heterogeneity in causal effects driven by local network characteristics, such as node degrees, and enables us to examine how treatment effects vary across subgroups of nodes defined by their connectivity. This was highlighted in our analysis of rural Indian microfinance data, where average spillover effects are not significant, but our analysis was able to detect significant spillover effects for treated nodes with few connections. In addition, our framework addresses a common practical limitation of existing approaches, which typically treat the dissimilarity metric or exposure mapping as known a priori and choose it in an ad hoc manner, by allowing data-driven selection among multiple candidate model specifications through cross-validation. This cross-validation strategy is not directly applicable to average treatment effect estimators. Because these estimators use the same set of target nodes both to define the target parameter and to estimate it, the validation set cannot be separated from the training set. We view cross-validation under node-wise counterfactual mean framework as a concrete step toward and a useful foundation for principled, data-driven causal model selection under network interference.

Lastly, we emphasize that in defining the target estimand $\theta_{i^*}(t^*_{N_{i^*}})$, the network $\mathcal{G}$ serves as a conditioning variable rather than an argument of the counterfactual. This distinction highlights that the dependency of $\theta_{i^*}(t^*_{N_{i^*}})$ on the network is associative, not causal. Consequently, our method does not imply any causal interpretation of heterogeneity observed in the Indian microfinance data analysis when intervening on the network structure. To make such claims, we would need to disentangle effects mediated through network from homophily \citep{shalizi2011homophily}, which requires either very strong assumptions, such as the availability of pre-network covariates unaffected by social factors and capable of controlling for confounding between the network and covariates due to homophily, or having multiple network observations over time. While such assumptions may be plausible in specific applications with extensive domain knowledge, our focus is on general methodology and therefore we do not pursue this direction.

%% file: 1_JASA-supp.tex
\begin{bibunit}

\def\spacingset#1{\renewcommand{\baselinestretch}%
{#1}\small\normalsize} \spacingset{1}


\newpage\appendix
\spacingset{1.9}
\counterwithin{figure}{section}
\counterwithin{table}{section}
\counterwithin{equation}{section}

\if1\blind
{
  \bigskip
  \bigskip
  \bigskip
  \begin{center}
    {\LARGE\bf Supplement to ``Heterogeneous Treatment Effects under Network Interference: A Nonparametric Approach Based on Node Connectivity''}
  \end{center}

  \begin{center}
    {\large
    Heejong Bong \\
    Department of Statistics, Purdue University \\
    Colin B. Fogarty, Elizaveta Levina and Ji Zhu \\
    Department of Statistics, University of Michigan \\
    }
  \end{center}
} \fi

\if0\blind
{
  \bigskip
  \bigskip
  \bigskip
  \begin{center}
    {\LARGE\bf Supplement to ``Heterogeneous Treatment Effects under Network Interference: A Nonparametric Approach Based on Node Connectivity''}
\end{center}
  \medskip
} \fi

\input{app/a_supp_method}

\input{app/b_proofs}

\input{app/c_supp_simulation}

\input{app/d_supp_real_data}

\putbib[2_ref-supp]

\end{bibunit}

%% file: app/a_supp_method.tex
\section{KECENI Implementation Details}

\subsection{Calibration of Hyper-parameters by Cross-validation} \label{app:cross_validation}

KECENI requires the analyst to specify the dissimilarity metric $\Delta_i$, the kernel function $\kappa$ and the bandwidth hyperaparameter $\lambda$. Different choices may induce different bias-variance tradeoffs, significantly influencing the estimates and the resulting inference. We choose these hyperparameters in a data-driven way, using a leave-neighborhood-out cross-validation approach.

Given a discrete set of candidate hyperparameter values $(\lambda, \kappa, \Delta)$, we first fit the nuisance functions $\hat{\mu}_i$, $\hat{\pi}_i$, and $d\hat{P}_{X_{[n]}|\mathcal{G}}$ using the entire dataset. With these fitted nuisance functions, we compute the pseudo-outcome $\hat{\xi}_i(\hat\mu_i, \hat\pi_i)$. This serves as an unbiased estimator of $\theta_i(T_{N_i})$, independent of the hyperparameters. Next, we compute the leave-neighborhood-out estimator for $\theta_i(T_{N_i})$ for each candidate hyperparameter set, by fitting KECENI on a dataset where the dependence neighborhood of unit $i$ is left out. In particular, for a candidate set $(\lambda^{(k)}, \kappa^{(k)}, \Delta^{(k)})$, we define
\begin{equation*}
    \hat{\theta}_i^{(k)}(T_{N_i})
    \equiv \Big( \sum_{j \in N_i^{*\cmpl}} \kappa^{(k)}_{\lambda^{(k)}}(\Delta_j^{(k)}) \Big)^{-1} 
    \sum_{j \in N_i^{*\cmpl}} \kappa^{(k)}_{\lambda^{(k)}}(\Delta_j^{(k)}) \hat{\xi}_j(\hat\mu_j, \hat\pi_j),
\end{equation*}
where $N_i^*$ is the dependence neighborhood of node $i$ under network dependence of \cref{assmp:network_dependence}.
Finally, we choose the hyperparameter set $(\lambda^{(k^*)}, \kappa^{(k^*)}, \Delta^{(k^*)})$ which minimizes the mean squared error:
\begin{equation*}
    k^* \equiv {\arg\min}_k ~\frac{1}{n} \sum_{i=1}^n (\hat{\xi}_i - \hat{\theta}_i^{(k)}(T_{N_i}))^2.
\end{equation*}

\subsection{Empirical Sandwich Variance Estimator} \label{app:detail_inference}


Efficient-influence-function-based plug-in variance estimation is a common approach to inference for AIPW estimators \citep{van2011targeted}. However, a limitation of this approach is that the resulting variance estimator is not, in general, doubly robust. Its consistency requires both nuisance models to be correctly specified, that is, $\bar\mu=\mu$ and $\bar\pi=\pi$. When only the outcome regression is correctly specified ($\bar\mu=\mu$), the variance estimator is typically conservative; when the outcome regression is misspecified ($\bar\mu\neq\mu$), valid confidence interval coverage is not guaranteed, even if the AIPW point estimator remains consistent through a correctly specified propensity score \citep{funk2011doubly,munoz2012population}. To address this limitation, \citet{shook2024double} proposed a doubly robust empirical sandwich variance estimator for settings without interference.

Here, we extend the empirical sandwich variance estimator to the network interference setting. 
The construction is motivated by stacking the estimating equation for the KECENI estimator with the estimating equations for the nuisance parameters, or with the corresponding score equations in the case of maximum likelihood estimation. To illustrate the idea, suppose that the outcome regression and propensity score are modeled parametrically, with parameters $\beta_\mu \in \reals^{p_\mu}$ and $\beta_\pi \in \reals^{p_\pi}$, respectively. We assume that these parameters are estimated by $M$-estimators $\hat\beta_\mu$ and $\hat\beta_\pi$ satisfying
\begin{equation*}
\begin{aligned}
    \phi_\mu(Y_{[n]}, T_{[n]}, X_{[n]}; \hat\beta_\mu)
    &\equiv
    \sum_{i=1}^n
    \phi_{\mu,i}(Y_i,T_{N_i},X_{N_i};\hat\beta_\mu)
    = 0, \\
    \phi_\pi(Y_{[n]}, T_{[n]}, X_{[n]}; \hat\beta_\pi)
    &\equiv
    \sum_{i=1}^n
    \phi_{\pi,i}(Y_i,T_{N_i},X_{N_i};\hat\beta_\pi)
    = 0 .
\end{aligned}
\end{equation*}
For the moment, suppose that the covariate distribution $P_{X_{[n]}\mid\mathcal G}$ is known. Then 
$(\hat\theta_{i^*}(t^*_{N_{i^*}}),\hat\beta_\mu,\hat\beta_\pi)$ can be viewed as the solution to the stacked estimating equation
$
    \phi \equiv
    \left(
        \phi_\theta,
        \phi_\mu^\top,
        \phi_\pi^\top
    \right)^\top
    = \mathbf 0 ,
$
where the estimating equation for the target parameter is
\begin{equation*}
    \phi_\theta(
        Y_{[n]},T_{[n]},X_{[n]};
        \theta,\beta_\mu,\beta_\pi
    )
    \equiv
    \sum_{i=1}^n
    \kappa_\lambda(\Delta_i)
    \left\{
        \xi_i(\mu_{\beta_\mu},\pi_{\beta_\pi})
        -
        \theta
    \right\},
\end{equation*}
and $\mu_{\beta_\mu}$ and $\pi_{\beta_\pi}$ denote the outcome regression and propensity score models indexed by $\beta_\mu$ and $\beta_\pi$, respectively.
Under standard regularity assumptions, the asymptotic covariance matrix of 
$(\hat\theta_{i^*}(t^*_{N_{i^*}}), \hat\beta_\mu, \hat\beta_\pi)$ is given by 
\begin{equation*}
    \left[
        \Exp\left\{ 
            \nabla \phi(\tilde\theta_{i^*}(t^*_{N_{i^*}}), \bar\beta_\mu, \bar\beta_\pi)
        \right\}
    \right]^{-1}
    \Var\left\{
        \phi(\tilde\theta_{i^*}(t^*_{N_{i^*}}), \bar\beta_\mu, \bar\beta_\pi)
    \right\}
    \left[
        \Exp\left\{ 
            \nabla \phi(\tilde\theta_{i^*}(t^*_{N_{i^*}}), \bar\beta_\mu, \bar\beta_\pi)
        \right\}
    \right]^{-\top},
\end{equation*}
where $\bar\beta_\mu$ and $\bar\beta_\pi$ denote the probability limits of 
$\hat\beta_\mu$ and $\hat\beta_\pi$, respectively, as determined by their estimating equations 
$\phi_\mu$ and $\phi_\pi$ \citep{stefanski2002calculus}. This covariance matrix can be estimated by a plug-in estimator, yielding the empirical sandwich variance estimator. 

We further generalize the empirical sandwich variance estimator to accommodate a broader class of nuisance estimators. These estimators need not be $M$-estimators; instead, we require only that their first-order estimation errors admit linear representations in terms of local statistics. To formulate this condition, we assume that the outcome regression and propensity score estimators satisfy
\begin{equation} \label{eq:H_nuisance}
\begin{aligned}
    \hat\mu(t_{N_j}, x_{N_j}, \mathcal{G}_{N_j}) 
    - \bar\mu(t_{N_j}, x_{N_j}, \mathcal{G}_{N_j})
    & =
    \sum_{i=1}^n H_{\bar\mu,i}(t_{N_j}, x_{N_j}, \mathcal{G}_{N_j})
    + \epsilon_{\bar\mu}(t_{N_j}, x_{N_j}, \mathcal{G}_{N_j}), \\
    \hat\pi(t_{N_j}, x_{N_j}, \mathcal{G}_{N_j}) 
    - \bar\pi(t_{N_j}, x_{N_j}, \mathcal{G}_{N_j})
    & =
    \sum_{i=1}^n H_{\bar\pi,i}(t_{N_j}, x_{N_j}, \mathcal{G}_{N_j})
    + \epsilon_{\bar\pi}(t_{N_j}, x_{N_j}, \mathcal{G}_{N_j}),
\end{aligned}
\end{equation}
where, for each evaluation point $(t_{N_j},x_{N_j},\mathcal G_{N_j})$, the random components 
$H_{\bar\mu,i}$ and $H_{\bar\pi,i}$ depend on the observed data only through the local variables 
$(Y_i,T_{N_i},X_{N_i})$. The remainder terms 
$\epsilon_{\bar\mu}$ and $\epsilon_{\bar\pi}$ are assumed to be asymptotically negligible at the scale relevant for inference.

This representation includes the usual $M$-estimation setting as a special case. For example, under standard smoothness and nonsingularity conditions, a Taylor expansion of the estimating equation gives the linear approximation
\begin{equation} \label{eq:H_M_estimator}
\begin{aligned}
    & \mu(t_{N_j}, x_{N_j}, \mathcal{G}_{N_j}; \hat\beta_\mu) 
    - \mu(t_{N_j}, x_{N_j}, \mathcal{G}_{N_j}; \bar\beta_\mu) \\
    & =
    \frac{\partial \mu}{\partial \beta_\mu}
    (t_{N_j}, x_{N_j}, \mathcal{G}_{N_j}; \bar\beta_\mu)
    \left(
        \Exp\left[
            \frac{\partial \phi_\mu}{\partial \beta_\mu}
            (Y_{[n]}, T_{[n]}, X_{[n]}; \bar\beta_\mu)
        \right]
    \right)^{-1}
    \sum_{i=1}^n
    \phi_{\mu,i}(Y_i, T_{N_i}, X_{N_i}; \bar\beta_\mu) \\
    & \quad
    + \epsilon_{\bar\mu}(t_{N_j}, x_{N_j}, \mathcal{G}_{N_j}),
\end{aligned}
\end{equation}
where $\epsilon_{\bar\mu}(t_{N_j},x_{N_j},\mathcal G_{N_j})$ denotes the second-order remainder. An analogous expansion applies to the propensity score M-estimators. 

In KECENI, the covariate distribution is also a nuisance component because it uses the marginal quantities 
$\int \hat\mu(t_{N_j}, x_{N_j}, \mathcal G_{N_j}) d\hat{P}_{X_{N_j}|\mathcal{G}}(x_{N_j})$ and $\int \hat\pi(t_{N_j}, x_{N_j}, \mathcal G_{N_j}) d\hat{P}_{X_{N_j}|\mathcal{G}}(x_{N_j})$. Therefore, variance estimation must also account for the variability induced by estimating $P_{X_{[n]}\mid\mathcal G}$. We assume that this effect admits a similar local linear representation: for any fixed $f_j\in\mathcal F_j$,
\begin{equation} \label{eq:H_covariate}
\begin{aligned}
    \int f_j(x_{N_j})\,d\hat P_{X_{N_j}\mid\mathcal G}(x_{N_j})
    -
    \int f_j(x_{N_j})\,dP_{X_{N_j}\mid\mathcal G}(x_{N_j})
    & =
    \sum_{i=1}^n H_{P_{X\mid\mathcal G},i}(f_j)
    + \epsilon_{P_{X\mid\mathcal G}}(f_j),
\end{aligned}
\end{equation}
where $H_{P_{X\mid\mathcal G},i}(f_j)$ is a local statistic, depending on the observed data only through the neighborhood of node $i$. For instance, when the covariate vectors are assumed i.i.d. and $\hat P_{X_{[n]}\mid\mathcal G}$ is taken to be the empirical product measure $\hat{P}_{X_{N_j}|\mathcal{G}}^\otimes$ as discussed in \cref{eq:empirical_product_measure},
$\int f_j(x_{N_j})\,d\hat P^\otimes_{X_{N_j}\mid\mathcal G}(x_{N_j})$ is a $V$-statistic:
\begin{equation*}
    \int f_j(x_{N_j})\,d\hat P^\otimes_{X_{N_j}\mid\mathcal G}(x_{N_j})
    =
    n^{-\abs{N_j}}
    \sum_{i_1,\dots,i_{\abs{N_j}}\in[n]}
    f_j(X_{i_1},\dots,X_{i_{\abs{N_j}}}).
\end{equation*}
Its first-order estimation error can be approximated by the Hájek projection \citep{peng2019asymptotic}. In particular, for symmetric $f_j$ and $s=\abs{N_j}$,
\begin{equation*}
    \int f_j(x_{N_j})\,d\hat P^\otimes_{X_{N_j}\mid\mathcal G}(x_{N_j})
    -
    \int f_j(x_{N_j})\,dP_{X_{N_j}\mid\mathcal G}(x_{N_j})
    \approx
    \frac{s}{n}\sum_{i=1}^n f_j^{(1)}(X_i),
\end{equation*}
where $f_j^{(1)}(x) \equiv \Exp\{f_j(x,X_2,\dots,X_s)\} - \Exp\{f_j(X_1,\dots,X_s)\}$.

Combining the preceding linear representations yields a first-order expansion of the KECENI estimation error:
$ \hat\theta_{i^*}(t^*_{N_{i^*}}) - \tilde\theta_{i^*}(t^*_{N_{i^*}}) \approx D^{-1}\sum_{i=1}^n W_i$, where $D \equiv \sum_{i=1}^n \Exp\{\kappa_\lambda(\Delta_i)\}$, and
\begin{equation*}
\begin{aligned}
    {W}_i
    & \equiv
    \kappa_\lambda(\Delta_i)
    \left\{
        \xi_i(\bar\mu, \bar\pi)
        -
        \tilde\theta_{i^*}(t^*_{N_{i^*}})
    \right\} \\
    & \quad
    + P\left[
        \sum_{j=1}^n 
        \kappa_\lambda(\Delta_j)
        \frac{\partial \xi_j}{\partial \mu_j}
        (Y_j,T_{N_j},X_{N_j};\bar\mu,\bar\pi)
        H_{\bar\mu,i}(T_{N_j},X_{N_j},\mathcal G_{N_j}) 
    \right] \\
    & \quad
    + P\left[
        \sum_{j=1}^n 
        \kappa_\lambda(\Delta_j)
        \frac{\partial \xi_j}{\partial m_j}
        (Y_j,T_{N_j},X_{N_j};\bar\mu,\bar\pi)
        \int H_{\bar\mu,i}(T_{N_j},x_{N_j},\mathcal G_{N_j})
        \,dP_{X_{N_j}\mid\mathcal G}(x_{N_j}) 
    \right] \\
    & \quad
    + P\left[
        \sum_{j=1}^n 
        \kappa_\lambda(\Delta_j)
        \frac{\partial \xi_j}{\partial \pi_j}
        (Y_j,T_{N_j},X_{N_j};\bar\mu,\bar\pi)
        H_{\bar\pi,i}(T_{N_j}\mid X_{N_j},\mathcal G_{N_j})
    \right] \\
    & \quad
    + P\left[
        \sum_{j=1}^n 
        \kappa_\lambda(\Delta_j)
        \frac{\partial \xi_j}{\partial \varpi_j}
        (Y_j,T_{N_j},X_{N_j};\bar\mu,\bar\pi)
        \int H_{\bar\pi,i}(T_{N_j}\mid x_{N_j},\mathcal G_{N_j})
        \,dP_{X_{N_j}\mid\mathcal G}(x_{N_j}) 
    \right] \\
    & \quad
    + P\left[
        \sum_{j=1}^n 
        \kappa_\lambda(\Delta_j)
        \frac{\partial \xi_j}{\partial m_j}
        (Y_j,T_{N_j},X_{N_j};\bar\mu,\bar\pi)
        H_{P_{X\mid\mathcal G},i}
        \{\bar\mu(T_{N_j},\cdot,\mathcal G_{N_j})\}
    \right] \\
    & \quad
    + P\left[
        \sum_{j=1}^n 
        \kappa_\lambda(\Delta_j)
        \frac{\partial \xi_j}{\partial \varpi_j}
        (Y_j,T_{N_j},X_{N_j};\bar\mu,\bar\pi)
        H_{P_{X\mid\mathcal G},i}
        \{\bar\pi(T_{N_j}\mid \cdot,\mathcal G_{N_j})\}
    \right].
\end{aligned}    
\end{equation*}
Here, $P$ denotes expectation over the variables indexed by $j$, while treating 
$H_{\bar\mu,i}$, $H_{\bar\pi,i}$, and $H_{P_{X\mid\mathcal G},i}$ as fixed functions. The first term in 
$W_i$ is the direct contribution of the KECENI estimating equation, while the remaining terms propagate the first-order effects of estimating the outcome regression, propensity score, and covariate distribution.

This representation is the analogue of the sandwich linearization for the target component. In the finite-dimensional $M$-estimation case, $D^{-1}W_i$ corresponds to the first component of the usual sandwich linear contribution, $[-\{\Exp(\nabla \phi)\}^{-1}\phi_i]_1$. Consequently,
$
    \bar\sigma^2
    \equiv
    \Var\left\{
        D^{-1}\sum_{i=1}^n W_i
    \right\}
$
is the target-parameter component of the sandwich variance, $[\{\Exp(\nabla \phi)\}^{-1}\Var\{\phi\}\{\Exp(\nabla \phi)\}^{-\top}]_{11}$.

Under the local linearization above, the first-order effect of nuisance estimation can be incorporated into the leading expansion, rather than required to be asymptotically negligible. Consequently, asymptotic normality can still be established even when nuisance-estimation error contributes non-negligibly to the limiting variance. This leads to the following theorem, which weakens the corresponding condition in \cref{thm:asymptotic_normality} by requiring only
\[
    \max\{\bar r_n^\circ s_n^\circ,\; r_n^\circ \bar s_n^\circ,\; q_n\}
    =
    O(\bar\sigma_n),
\]
where $\bar{r}_n^\circ$ and $\bar{s}_n^\circ$ indicate the local convergence rates of $\hat\mu$ to $\bar\mu$ and $\hat\pi$ to $\bar\pi$, respectively. The proof is given in \cref{app:pf_sandwich_asymptotic_normality}.

\begin{theorem} \label{thm:sandwich_asymptotic_normality}
    Suppose that $\bar{r}_n^\circ \bar{s}_n^\circ = o(\bar\sigma_n)$ and $\max\{ \sqrt{K_{\max}/D}, \bar{r}^\circ_n s^\circ_n, r^\circ_n \bar{s}^\circ_n, q_n \} = O( \bar\sigma_n )$. 
    Then under the assumptions of \cref{thm:consistency} and the additional conditions on linear approximations by $(H_{\bar\mu,i}, H_{\bar\pi,i}, H_{P_{X|\mathcal{G}}, i})$ as specified in  \cref{assmp:H_nuisance,assmp:H_covariate}, we have
    $
        {\bar\sigma_n^{-1}} (\hat\theta_{i^*}(t^*_{N_{i^*}}) - \theta_{i^*}(t^*_{N_{i^*}}) - b_n)
        \overset{d}{\rightarrow} N(0, 1).
    $
\end{theorem}

Although the operator $P$ integrates over the variables indexed by $j$, the resulting terms remain random because the functions held fixed inside the expectation are themselves random. For example,
\begin{equation*}
    P\left[
        \sum_{j=1}^n 
        \kappa_\lambda(\Delta_j)
        \frac{\partial \xi_j}{\partial \mu_j}
        (Y_j,T_{N_j},X_{N_j};\bar\mu,\bar\pi)
        H_{\bar\mu,i}(T_{N_j},X_{N_j},\mathcal G_{N_j}) 
    \right]
\end{equation*}
is evaluated by treating $H_{\bar\mu,i}$ as fixed, but $H_{\bar\mu,i}$ itself depends on the local data 
$(Y_i,T_{N_i},X_{N_i})$. Consequently, each $W_i$ is a local statistic, depending on the observed data only through the dependence neighborhood of node $i$. It therefore inherits the network dependence encoded by the neighborhoods $\tilde N_i$ in \cref{assmp:network_dependence}. In particular,
\begin{equation*}
    j \notin \tilde N_i
    \quad\Longrightarrow\quad
    W_i \indep W_j.
\end{equation*}
Thus, the variance of the linearized sum can be written as
\begin{equation*}
\begin{aligned}
    \Var\left\{\sum_{i=1}^n W_i\right\}
    =
    \Exp\left[
        \sum_{i,j=1}^n h(i,j)
        \left\{W_i-\Exp[W_i]\right\}
        \left\{W_j-\Exp[W_j]\right\}
    \right],
\end{aligned}
\end{equation*}
where $h(i,j) \equiv \mathbf{1}\{j \in \tilde{N}_i\}$.
We estimate this variance by replacing the population expectations, the $P$-expectation terms, and the oracle nuisance quantities with their empirical analogues. Specifically, let $\hat W_i$ denote the plug-in counterpart of $W_i$, obtained by replacing 
$(\bar\mu,\bar\pi,P_{X_{[n]}\mid\mathcal G},P,\xi_i,\tilde\theta_{i^*})$ with 
$(\hat\mu,\hat\pi,\hat P_{X_{[n]}\mid\mathcal G},P_n,\hat\xi_i,\hat\theta_{i^*})$. We then propose the following dependence-adjusted variance estimator for 
$\hat\theta_{i^*}(t^*_{N_{i^*}})$:
\begin{equation*}
    \hat\sigma^2
    \equiv
    \sum_{i,j=1}^n h(i,j)\hat W_i\hat W_j.
\end{equation*}
The following theorem presents the consistency of $\hat\sigma^2$. The proof is given in \cref{app:pf_sandwich_variance_consistency}. 

\begin{theorem} \label{thm:sandwich_variance_consistency}
    Under the assumptions of \cref{thm:sandwich_asymptotic_normality}, we have
    \begin{equation*}
        \hat\sigma_n^2 
        = \bar\sigma_n^2 
        + D^{-2} 
        \sum_{i,j \in [n]} h(i,j) 
        \Exp[W_i]
        \Exp[W_j]
        + o_p(\bar\sigma_n^2).
    \end{equation*}

    If $\bar\mu = \mu$ or $\bar\pi = \pi$, $\abs{\Exp[W_i]} = O(\Exp[\kappa_\lambda(\Delta_i)] \lambda)$, and hence the bias is on the scale of $O(\frac{K_{\max}}{D} \lambda)$. With an additional condition that $\frac{K_{\max}}{D}\lambda = o(\bar\sigma_n^2)$, $(\hat\sigma_n - \bar\sigma_n)/\bar\sigma_n \overset{p}{\rightarrow} 0$.
\end{theorem}

Alternative estimators for the variance of 
$\sum_{i=1}^n W_i$ include network heteroskedasticity-and-autocorrelation-consistent 
(HAC) variance estimators \citep{kojevnikov2021limit} and block-based bootstrap methods 
\citep{kojevnikov2021bootstrap}. These approaches may be appropriate under weaker dependence conditions than exact local dependence. In particular, they can accommodate settings in which the local-dependence assumption fails but dependence between 
$(Y_i,T_{N_i},X_{N_i})$ and $(Y_j,T_{N_j},X_{N_j})$ decays as the distance between $i$ and $j$ in 
$\mathcal G$ increases. See the definition of $\psi$-dependence in 
\citet{kojevnikov2021limit} for a formal characterization of this type of weak network dependence. 
%

%% file: app/b_proofs.tex
\section{Supplement for Theoretical Results}

\subsection{Proof of Identifiability (Proposition~\ref{thm:identifiability})}

In this section, we prove the identifiability of the target estimand $\theta_{i^*}(t^*_{N_{i^*}})$ under the given assumptions, where $i^*$ and $t^*$ are the target node and a hypothetical treatment assignment on $\mathcal{G}$, specified by the practitioner.
By G-computation, we have
\begin{equation*}
\begin{aligned}
    \theta_{i^*}(t^*_{N_{i^*}}) 
    & = \Exp[Y_{i^*}(t^*_{N_{i^*}}) | \mathcal{G}] \\
    & = \int \Exp[Y_{i^*}(t^*_{N_{i^*}}) | X_{N_{i^*}} = x_{N_{i^*}}, \mathcal{G}] dP_{X_{N_{i^*}}|\mathcal{G}}(x_{N_{i^*}}) \\
    & \overset{\mathrm{(i)}}{=} \int \Exp[Y_{i^*}(t^*_{N_{i^*}}) | T_{N_{i^*}} = t^*_{N_{i^*}}, X_{N_{i^*}} = x_{N_{i^*}}, \mathcal{G}] dP_{X_{N_{i^*}}|\mathcal{G}}(x_{N_{i^*}}) \\
    & \overset{\mathrm{(ii)}}{=} \int \Exp[Y_{i^*} | T_{N_{i^*}} = t^*_{N_{i^*}}, X_{N_{i^*}} = x_{N_{i^*}}, \mathcal{G}] dP_{X_{N_{i^*}}|\mathcal{G}}(x_{N_{i^*}}),
\end{aligned}
\end{equation*}
where step $\mathrm{(i)}$ follows from the conditional independence (unconfoundedness) condition in \cref{assmp:ignorability}, and step $\mathrm{(ii)}$ from the consistency condition (i.e., $Y_i = Y_i(T_{N_i})$). The positivity condition in \cref{assmp:ignorability} ensures that the conditional expectation is well-defined and identifiable, as the conditioning event has positive probability almost surely with respect to $dP_{X_{N_{i^*}}|\mathcal{G}}(x_{N_{i^*}})$. This proves the identifiability of the target estimand, $\theta_{i^*}(t^*_{N_{i^*}})$.

\subsection{Proof of Double-robustness (Proposition~\ref{thm:double_robustness_unadjusted})} \label{app:pf_double_robustness}

In this section, we prove the double-robustness of the pseudo-outcomes presented in \cref{sec:keceni}. The prove follows from the double-robustness of generic augmented inverse propensity weighting (AIPW) form. Let
\begin{equation*}
    \varpi_i(t_{N_i}| \mathcal{G})
    \equiv \int \pi_i(t_{N_i}| x_{N_i}, \mathcal{G}) 
    dP_{X_{N_i}|\mathcal{G}}(x_{N_i})
    ~~ \text{and}
    ~~ \bar\varpi_i(t_{N_i}| \mathcal{G})
    \equiv \int \bar\pi_i(t_{N_i}| x_{N_i}, \mathcal{G}) 
    dP_{X_{N_i}|\mathcal{G}}(x_{N_i}).
\end{equation*}
Then,
\begin{equation*}
\begin{aligned}
    & \Exp[\xi_i(\bar\mu_i,\bar\pi_i) | T_{N_i} = t_{N_i}, \mathcal{G}] \\
    & \overset{\text{(i)}}{=} \int \mathbb{E}[\xi_i(\bar\mu_i,\bar\pi_i) | T_{N_i} = t_{N_i}, X_{N_i} = x_{N_i}, \mathcal{G}]
    ~ \frac{\pi_i(t_{N_i}| x_{N_i}, \mathcal{G})}{\varpi_i(t_{N_i}| \mathcal{G})}
    ~ dP_{X_{N_i}|\mathcal{G}}(x_{N_i}) \\
    & \overset{\text{(ii)}}{=} \int \{\mu_i(t_{N_i}, x_{N_i}, \mathcal{G}) - \bar\mu_i(t_{N_i}, x_{N_i}, \mathcal{G})\}
    \frac{\pi_i(t_{N_i}| x_{N_i}, \mathcal{G}) / \varpi_i(t_{N_i}| \mathcal{G})
    }{\bar\pi_i(t_{N_i}| x_{N_i}, \mathcal{G}) / \bar\varpi_i(t_{N_i}| \mathcal{G})}
    ~ dP_{X_{N_i}|\mathcal{G}}(x_{N_i}) \\
    & \quad + \int \bar\mu_i(t_{N_i}, x_{N_i}, \mathcal{G}) 
    dP_{X_{N_i}|\mathcal{G}}(x_{N_i}) \\
    & \overset{\text{(iii)}}{=} \int \{\mu_i(t_{N_i}, x_{N_i}, \mathcal{G}) - \bar\mu_i(t_{N_i}, x_{N_i}, \mathcal{G})\}
    \left\{ \frac{\pi_i(t_{N_i}| x_{N_i}, \mathcal{G}) / \varpi_i(t_{N_i}| \mathcal{G})
    }{\bar\pi_i(t_{N_i}| x_{N_i}, \mathcal{G}) / \bar\varpi_i(t_{N_i}| \mathcal{G})} - 1 \right\}
    ~ dP_{X_{N_i}}(x_{N_i}) \\
    & \quad + \int \mu_i(t_{N_i}, x_{N_i}, \mathcal{G})
    dP_{X_{N_i} | \mathcal{G}}(x_{N_i}), \\
    & \overset{\text{(iv)}}{=} \int \{\mu_i(t_{N_i}, x_{N_i}, \mathcal{G}) - \bar\mu_i(t_{N_i}, x_{N_i}, \mathcal{G})\}
    \left\{ \frac{\pi_i(t_{N_i}| x_{N_i}, \mathcal{G}) / \varpi_i(t_{N_i}| \mathcal{G})
    }{\bar\pi_i(t_{N_i}| x_{N_i}, \mathcal{G}) / \bar\varpi_i(t_{N_i}| \mathcal{G})} - 1 \right\}
    ~ dP_{X_{N_i}}(x_{N_i}) \\
    & \quad + \theta_{i}(t_{N_i}), \\
\end{aligned}
\end{equation*}
where $\text{(i)}$ follows from the definition of conditional expectation, $\text{(ii)}$ and $\text{(iii)}$ is due to rearranging, and $\text{(iv)}$ follows from the definition of $\theta_{i}(t_{N_i})$. 
The first term in the last line vanishes under either of the following conditions: (i) when \(\bar{\mu}_i = \mu_i\), ensuring that \(\mu_i(t_{N_i}, x_{N_i}, \mathcal{G}) = \bar{\mu}_i(t_{N_i}, x_{N_i}, \mathcal{G})\) for all \(t_{N_i}\), \(x_{N_i}\), and \(i\); or (ii) when \(\bar{\pi}_i = \pi_i\), such that the ratio 
\[
\frac{\pi_i(t_{N_i} \mid x_{N_i}, \mathcal{G}) / \varpi_i(t_{N_i} \mid \mathcal{G})}{\bar{\pi}_i(t_{N_i} \mid x_{N_i}, \mathcal{G}) / \bar{\varpi}_i(t_{N_i} \mid \mathcal{G})} = 1
\] 
for all \(t_{N_i}\), \(x_{N_i}\), and \(i\).
So as long as either $\bar\mu_i = \mu_i$ or $\bar\pi_i = \pi_i$, the last line results in 
\begin{equation*}
\begin{aligned}
     \mathbb{E}[\xi_i(\bar\mu_i, \bar\pi_i) | T_{N_i} = t_{N_i}, \mathcal{G}]
     & = \theta_{i}(t_{N_i}),
\end{aligned}
\end{equation*}
which shows the desired double robustness.

\subsection{Proof of Equicontinuity (Theorem~\ref{thm:asymptotic_equicontinuity})}
\label{app:pf_equicontinuity} 

\subsubsection{Regularity Conditions} 

Here we first formulate the regularity conditions in \cref{thm:asymptotic_equicontinuity}. To reiterate the notations, the target empirical process is a weighted average process indexed by $\theta$ in a metric space $\mathcal{F}$ equipped with metric $d$, defined as
\begin{equation} \label{eq:weighted_average_process}
    Z_n(\theta) \equiv \frac{1}{W_n} \sum_{i=1}^n w_i f_i(\theta),
\end{equation}
where $w_i$ are random weights; $W_n \equiv \sum_{i=1}^n \mathbb{E}[w_i]$; given fixed $\theta$, $w_if_i(\theta) \indep w_jf_j(\theta)$ for $j \notin \tilde{N}_i$. Let $K_{\max} \equiv \max_{i \in [n]} \abs{\tilde{N}_i}$.

The regularity conditions are formulated as follows. The conditions correspond to Conditions (a), (b) and (c) in \cref{thm:asymptotic_equicontinuity} in order.

\begin{assumption}[Moment Lipschitz condition] \label{assmp:moment_lipschitz}
    There exists a universal constant $C$ so that $\sup_i \norm{(P_n - P)[f_i(\theta_1) - f_i(\theta_2)]}_\infty \leq C d(\theta_1, \theta_2)$.
\end{assumption}

\begin{assumption}[Total boundedness] \label{assmp:total_boundedness}
    $\mathcal{F}$ is totally bounded with respect to this metric $d$.
\end{assumption}

\begin{assumption}[Finite entropy integral] \label{assmp:finite_entropy_equi}
    Either of the following conditions is true:
    \begin{enumerate}[(i)]
        \item For some $M > 0$ and $\eta > 0$, $\sum_i \frac{w_i}{W_n} \leq M$ almost surely, and $\int_0^\eta \psi_2^{-1}(N(\epsilon, \mathcal{F}, d)) d\epsilon < \infty$;
        \item For some $M > 0$ and $\eta > 0$, $\frac{\sqrt{n K_{\max}}}{W_n} \leq M$, and $\int_0^\eta \psi_{4/3}^{-1}(N(\epsilon, \mathcal{F}, d)) d\epsilon < \infty$;
    \end{enumerate}
    where $\psi_q(x) = \exp(x^q)-1$, $\norm{\cdot}_{\psi_q}$ is the respective Orlicz norm, and $N(\epsilon, \mathcal{F}, d)$ is the covering number of $\mathcal{F}$ by $d$-balls of size $\epsilon$.
\end{assumption}

\subsubsection{Proof of Theorem}

Under the regularity assumptions, we prove the stochastic equicontinuity of $Z_n(\theta)$:
\begin{equation*}
    \Pr\left[ \sup_{d(\theta_1, \theta_2) < \delta_n} \sqrt{\frac{W_n}{ K_{\max}}} \abs*{(P_n - P) [Z_n(\theta_1) - Z_n(\theta_2)]} > \eta \right] \rightarrow 0.
\end{equation*}

Here $P(f)$ indicates the expectation over $(Y_{[n]}, T_{[n]}, X_{[n]})$ treating it as an independent copy of the data (or $f$ as fixed):
\begin{equation*}
    P(f) \equiv P[f(Y_{[n]}, T_{[n]}, X_{[n]})] \equiv \int f(y_{[n]}, t_{[n]}, x_{[n]}) dP_{Y_{[n]}, T_{[n]}, X_{[n]} | \mathcal{G}}(y_{[n]}, t_{[n]}, x_{[n]}).
\end{equation*}
So if $\hat{f}$ is random depending on the observed data, $P(\hat{f})$ is also random, which is not the case for $\Exp(\hat{f})$. \(P_n\) denotes the empirical version of $P$, which places probability mass \(1\) on the observed data \((Y_{[n]}, T_{[n]}, X_{[n]})\).

We use the theoretical argument in the proof of Theorem 2.2.4 and Corollary 2.2.5 in \citet{vaart1996weak}. \citet{van2014causal} rewrote these results as follows (see Appendix (A3) therein): for a sequence $(\zeta_n(\theta): n \in \nats, \theta \in \mathcal{F})$ of random processes, if (1) $\norm{\zeta_n(\theta_1) - \zeta_n(\theta_2)}_{\psi_2} \leq C \cdot d(\theta_1, \theta_2)$ for some universal constant $C$ and metric $d(\cdot, \cdot)$, (2) $\mathcal{F}$ is totally bounded with respect to this metric $d$, and (3) for some $\eta > 0$, $\int_0^\eta \psi_2^{-1}(N(\epsilon, \mathcal{F}, d)) d\epsilon < \infty$, then the process $\zeta_n$ is stochastically equicontinuous.

Plugging in $\zeta_n(\theta) \equiv \sqrt{\frac{W_n}{K_{\max}}} (P_n - P)[Z_n(\theta)] = \frac{1}{\sqrt{W_n K_{\max}}} (P_n - P)[\sum_{i=1}^n w_i f_i(\theta)]$, it suffices to show that 
\begin{equation} \label{eq:Orlicz_norm_condition}
\begin{aligned}
    & \norm*{
        \frac{1}{\sqrt{W_n K_{\max}}} (P_n - P) \left[
            \tsum_{i=1}^n w_i (f_i(\theta_1) - f_i(\theta_2)) 
        \right]
    }_{\psi_2} \leq C d(\theta_1, \theta_2),
\end{aligned}
\end{equation}
for some universal constant $C$. 
We modify Lemma 6 of \citet{van2014causal} to address the weighted average by $w_i$ and develop a double counting method on top of the original combinatorial argument to achieve a better convergence rate with respect to $K_{\max}$. The following lemma summarizes the result and proves the theorem.

\begin{lemma} \label{lem:modified_van2014_lemma}
    Assume that there exists a universal constant $C$ so that given any $i$ and $\theta$, $\norm{f_i(\theta_1) - f_i(\theta_2)}_\infty \leq C d(\theta_1, \theta_2)$, $P[w_i f_i(\theta_1) - w_i f_i(\theta_2)] = 0$, and $(w_i, f_i(\theta)) \indep (w_j, f_j(\theta))$ for $j \notin \tilde{N}_i$ with $K_{\max} \equiv \max_{i \in [n]} \abs{\tilde{N}_i}$. If $w_i \in [0, 1]$ almost surely and $\frac{\sqrt{n K_{\max}}}{W_n} \leq M$ for some universal constant $M$, there exists another universal constant $C$ such that 
    \begin{equation*}
        \norm{\tsum_{i=1}^n w_i (f_i(\theta_1) - f_i(\theta_2))}_{\psi_{4/3}} \leq C \sqrt{W_n K_{\max}} ~d(\theta_1, \theta_2).
    \end{equation*}
    Otherwise, if $\sum_i \frac{w_i}{W_n} \leq M$ almost surely for some universal constant $M$, then there exists another universal constant $C$ such that 
    \begin{equation*}
        \norm{\tsum_{i=1}^n w_i (f_i(\theta_1) - f_i(\theta_2))}_{\psi_2} \leq C \sqrt{W_n K_{\max}} ~d(\theta_1, \theta_2).
    \end{equation*}
\end{lemma}

\subsubsection{Proof of Lemma~\ref{lem:modified_van2014_lemma}}

Following the proof of Lemma 5 in \citet{van2014causal}, for an even integer $p$ and $\vec{i} \equiv (i_1, \dots, i_p) \in \{1, \dots, n\}^p$, let $R(\vec{i})$ be an indicator that has value $1$ if there exists an element $i_k$ so that $i_l \notin N_{i_k}$, $\forall l \neq k$ or $0$ otherwise. If $R(\vec{i}) = 1$, $P[\prod_{k=1}^p w_{i_k} (f_{i_k}(\theta_1) - f_{i_k}(\theta_2))] = 0$ because $\xi_i(\theta)$ has mean zero for all $i$ and $\theta$. Hence
\begin{equation*}
\begin{aligned}
    & P[\abs{\tsum_i w_i (f_i(\theta_1) - f_i(\theta_2))}^p] 
    = \tsum_{i_1, \dots, i_p} P[\tprod_{k=1}^p w_{i_k} (f_{i_k}(\theta_1) - f_{i_k}(\theta_2))] \\
    & = \tsum_{i_1,\dots,i_p} (1 - R(\vec{i})) ~ P[ w_{i_1} \cdots w_{i_p} \tprod_{k=1}^p (f_{i_k}(\theta_1) - f_{i_k}(\theta_2))] \\
    & \overset{\mathrm{(i)}}{\leq} C \tsum_{i_1, \dots, i_p} (1 - R(\vec{i})) ~ P[w_{i_1} \cdots w_{i_p}] ~d(\theta_1,\theta_2)^p,
\end{aligned}
\end{equation*}
for some universal constant $C$, where $\mathrm{(i)}$ is due to the Lipschitz condition on $f_i$. 

Next we show that $\tsum_{i_1, \dots, i_p} (1 - R(\vec{i})) ~w_{i_1} \cdots w_{i_p} \leq C \{p ~(\sum_i w_i) ~K_{\max}\}^{p/2}$.
Let an indicator vector $B(\vec{i}) \in \{0,1,2,3\}^p$ be $B_k(\vec{i}) \equiv 0$ if $N_{i_k} \cap \cup_{l=1}^{k-1} N_{i_l} = \emptyset$; $1$ if $i_k \notin N_{i_k} \cap \cup_{l=1}^{k-1} N_{i_l} \neq \emptyset$; $3$ if $i_k \in \cup_{l \in [k-1]: B_l(\vec{i}) = 0} N_{i_l}$; or $2$ otherwise, for $k = 1, \dots, p$. For $b \in \{0,1,2,3\}^p$ and $s \in \{0, 1, 2, 3\}$, we define $\mathcal{K}_s(b) \equiv \{k: b_k = s\}$ and $\kappa_s(b) = \abs{\mathcal{K}_s(b)}$. Then for a fixed $b \equiv \{b_1, \dots, b_{p-1}\} \in \{0,1,2,3\}^p$, 
\begin{equation*}
\begin{aligned}
    &\tsum_{B(\vec{i}) = b} (1 - R(\vec{i})) ~w_{i_1} \cdots w_{i_p}
    \leq \tsum_{B(\vec{i}) = b} ~w_{i_1} \cdots w_{i_p} \\
    & = \tsum_{B(i_1,\dots,i_{p-1}) = (b_1,\dots,b_{p-1})} 
    ~w_{i_1} \cdots w_{i_{p-1}}
    \left( \begin{aligned}
        & \mathbf{1}_{\{b_p = 0\}} \tsum_{i: N_i \cap (\cup_{k=1}^{p-1} N_{i_k}) = \emptyset} ~w_i \\
        & + \mathbf{1}_{\{b_p = 1\}} \tsum_{i: i \notin N_i \cap (\cup_{k=1}^{p-1} N_{i_k}) \neq \emptyset} ~w_i \\
        & + \mathbf{1}_{\{b_p \in \{2,3\} \}} \tsum_{i \in \cup_{k=1}^{p-1} N_{i_k}} ~w_i\\
    \end{aligned} \right) \\
    & \leq \tsum_{B(i_1,\dots,i_{p-1}) = (b_1,\dots,b_{p-1})} 
    ~w_{i_1} \cdots w_{i_{p-1}}
    \left( \begin{aligned}
        \mathbf{1}_{\{b_p \in \{0,1\}\}} \tsum_i ~w_i 
        + \mathbf{1}_{\{b_p \in \{2,3\}\}} ~{\sup}_{|I| = p K_{\max}} \tsum_{i \in I} w_i
    \end{aligned} \right).
\end{aligned}
\end{equation*}
Since the $p$-th term in the summand of the last line is not dependent on $i_1, \dots, i_{p-1}$, this boils down again to upper bound $\tsum_{B(i_1,\dots,i_{p-1}) = (b_1,\dots,b_{p-1})} 
~w_{i_1} \cdots w_{i_{p-1}}$. Applying the above argument recursively, we obtain
\begin{equation} \label{eq:sum_prod_w_i}
\begin{aligned}
    & \tsum_{B(\vec{i}) = b} (1 - R(\vec{i})) ~w_{i_1} \cdots w_{i_p} \\
    & \leq \prod_{k=1}^p \left( \begin{aligned}
        \mathbf{1}_{\{b_p \in \{0,1\}\}} \tsum_i ~w_i+ \mathbf{1}_{\{b_p \in \{2,3\}\}} ~{\sup}_{|I| = p K_{\max}} \tsum_{i \in I} w_i
    \end{aligned} \right) \\
    & = (\tsum_i w_i)^{\kappa_0(b) + \kappa_1(b)}
    ({\sup}_{|I| = p K_{\max}} \tsum_{i \in I} w_i)^{\kappa_2(b)+\kappa_3(b)}.
\end{aligned}
\end{equation}

For a fixed $\vec{i}$ with $B(\vec{i}) = b$, consider a permutation $\vec{i}'$ of $\vec{i}$ such that:
\begin{itemize}
    \item $i'_1, \dots, i'_{\kappa_0(b)}$ correspond to $\mathcal{K}_0(b)$;
    \item $i'_{\kappa_0(b)+1}, \dots, i'_{\kappa_0(b)+\kappa_3(b)}$ correspond to $\mathcal{K}_3(b)$;
    \item $i'_{\kappa_0(b)+\kappa_3(b)+1}, \dots, i'_{p-\kappa_1(b)}$ correspond to $\mathcal{K}_2(b)$;
    \item $i'_{p-\kappa_1(b)+1}, \dots, i'_p$ correspond to $\mathcal{K}_1(b)$,
\end{itemize}
all preserving their original order within each group. We note that on the subset of $\vec{i}$ satisfying $B(\vec{i}) = b$, the mapping $\vec{i} \mapsto \vec{i}'$ is injective; i.e., given $b$ and $\vec{i}'$, we can recover $\vec{i}$. For such $\vec{i}'$, $B(\vec{i}')$ satisfies that
\begin{equation*}
\begin{aligned}
    & B_1(\vec{i}'), \dots, B_{\kappa_0(b)}(\vec{i}') = 0; \\
    & B_{\kappa_0(b)+1}(\vec{i}'), \dots, B_{\kappa_0(b)+\kappa_3(b)} (\vec{i}') = 3; \\
    & B_{p-\kappa_1(b)+1}(\vec{i}'), \dots, B_p(\vec{i}') \in \{2, 3\}. \\
\end{aligned}
\end{equation*}
Let $\mathcal{B}'$ indicates the collection of $b' \in \{0,1,2,3\}^p$ satisfying the above conditions; namely,
\begin{equation*}
    \mathcal{B}' \equiv \left\{ b' \in \{0,1,2,3\}^p: \begin{aligned}
        & b'_1, \dots, b'_{\kappa_0(b)} = 0;
        b'_{\kappa_0(b)+1}, \dots, b'_{\kappa_0(b)+\kappa_3(b)} = 3; \\
        & b'_{p-\kappa_1(b)+1}, \dots, b'_p \in \{2, 3\}
    \end{aligned} \right\}.
\end{equation*}
We note that
\begin{equation} \label{eq:kappa(b)_kappa(b')}
    b' \in \mathcal{B}' ~\Rightarrow~ \kappa_2(b') + \kappa_3(b') \geq \kappa_1(b) + \kappa_3(b)
\end{equation}
Then
\begin{equation*}
\begin{aligned}
    & \tsum_{B(\vec{i}) = b} (1 - R(\vec{i})) ~w_{i_1} \cdots w_{i_p} 
    \overset{\mathrm{(i)}}{\leq} \tsum_{B(\vec{i}') \in \mathcal{B}'} (1 - R(\vec{i}')) ~w_{i_1} \cdots w_{i_p} \\
    & = \tsum_{b' \in \mathcal{B}'} \tsum_{B(\vec{i}') = b'} (1 - R(\vec{i}')) ~w_{i_1} \cdots w_{i_p}\\
    & \overset{\mathrm{(ii)}}{\leq} \tsum_{b' \in \mathcal{B}'} 
    (\tsum_i w_i)^{\kappa_0(b') + \kappa_1(b')} 
    ({\sup}_{|I| = p K_{\max}} \tsum_{i \in I} w_i)^{\kappa_2(b')+\kappa_3(b')} \\
    & \overset{\mathrm{(iii)}}{\leq} \tsum_{b' \in \mathcal{B}'} (\tsum_i w_i)^{\kappa_0(b) + \kappa_2(b)} ({\sup}_{\abs{I} = p K_{\max}} \tsum_{i \in I} w_i)^{\kappa_1(b) + \kappa_3(b)} 
\end{aligned}
\end{equation*}
where $\mathrm{(i)}$ follows from that the mapping $\vec{i} \mapsto \vec{i'}$ is injective and that $R(\vec{i}) = R(\vec{i}')$; and $\mathrm{(ii)}$ from \cref{eq:sum_prod_w_i}. The last inequality $\mathrm{(iii)}$ follows from \cref{eq:kappa(b)_kappa(b')} and that $\kappa_1(b') + \kappa_2(b') + \kappa_3(b') + \kappa_4(b') = \kappa_1(b) + \kappa_2(b) + \kappa_3(b) + \kappa_4(b) = p$. Since the summand in the last line does not depend on $b'$, 
\begin{equation*}
\begin{aligned}
    & \tsum_{B(\vec{i}) = b} (1 - R(\vec{i})) ~w_{i_1} \cdots w_{i_p} \\
    & \leq \abs{\mathcal{B}'} (\tsum_i w_i)^{\kappa_0(b) + \kappa_2(b)} ({\sup}_{\abs{I} = p K_{\max}} \tsum_{i \in I} w_i)^{\kappa_1(b) + \kappa_3(b)} \\
    & \leq 4^p (\tsum_i w_i)^{\kappa_0(b) + \kappa_2(b)} ({\sup}_{\abs{I} = p K_{\max}} \tsum_{i \in I} w_i)^{\kappa_1(b) + \kappa_3(b)}.
\end{aligned}
\end{equation*}
Alongside with \cref{eq:sum_prod_w_i}, we get
\begin{equation*}
\begin{aligned}
    & \tsum_{B(\vec{i}) = b} (1 - R(\vec{i})) ~w_{i_1} \cdots w_{i_p}
    \leq \min\left\{\begin{aligned}
        & (\tsum_i w_i)^{\kappa_0(b)+\kappa_1(b)} 
        ({\sup}_{|I| = p K_{\max}} \tsum_{i \in I} w_i)^{\kappa_2(b)+\kappa_3(b)}, \\
        & 4^p (\tsum_i w_i)^{\kappa_0(b) + \kappa_2(b)} 
        ({\sup}_{\abs{I} = p K_{\max}} \tsum_{i \in I} w_i)^{\kappa_1(b) + \kappa_3(b)} \\
    \end{aligned} \right\} \\
    & \leq 2^p (\tsum_i w_i)^{\kappa_0(b) + \frac{\kappa_1(b)+\kappa_2(b)}{2}} 
    ({\sup}_{\abs{I} = p K_{\max}} \tsum_{i \in I} w_i)^{\frac{\kappa_1(b) + \kappa_2(b)}{2} + \kappa_3(b)},
\end{aligned}
\end{equation*}
where the last inequality comes from $\min\{a, b\} \leq \sqrt{ab}$ for any $a, b > 0$.

For a fixed $b \in \{0,1,2,3\}^p$, suppose that there exists $\vec{i}$ such that $B(\vec{i}) = b$ and $R(\vec{i}) = 0$. Then each $k \in \mathcal{K}_0(b)$ has at least one $l \in \{k+1, \dots, p\}$ satisfying $i_l \in N_{i_k}$, and this $l$ cannot be shared with other $k' \in \mathcal{K}_0(b)$. Because every such $l$ has $B_l(b) = 3$, $\kappa_0(b) \leq \kappa_3(b)$.
Since the $p$-th term in the summand of the last line is not dependent on $i_1, \dots, i_{p-1}$, this boils down again to upper bound $\tsum_{B(i_1,\dots,i_{p-1}) = (b_1,\dots,b_{p-1})} 
~w_{i_1} \cdots w_{i_{p-1}}$. Hence $\frac{\kappa_1(b) + \kappa_2(b)}{2} + \kappa_3(b) \geq \kappa_0(b) + \frac{\kappa_1(b) + \kappa_2(b)}{2}$, and because $\kappa_0(b) + \kappa_1(b) + \kappa_2(b) + \kappa_3(b) = p$, $\frac{\kappa_1(b) + \kappa_2(b)}{2} + \kappa_3(b) \geq p/2$. These observations result in
\begin{equation*}
\begin{aligned}
    \tsum_{B(\vec{i}) = b} (1 - R(\vec{i})) ~w_{i_1} \cdots w_{i_p}
    & \leq 2^p (\tsum_i w_i)^{\kappa_0(b) + \frac{\kappa_1(b)+\kappa_2(b)}{2}} 
    ({\sup}_{\abs{I} = p K_{\max}} \tsum_{i \in I} w_i)^{\frac{\kappa_1(b) + \kappa_2(b)}{2} + \kappa_3(b)} \\
    & \leq 2^p (\tsum_i w_i)^{p/2} 
    ({\sup}_{\abs{I} = p K_{\max}} \tsum_{i \in I} w_i)^{p/2}
    \leq 2^p (\tsum_i w_i)^{p/2} (p K_{\max})^{p/2},
\end{aligned}
\end{equation*}
where the last inequality follows from that $w_i \in [0,1]$, $\forall i$. 
We note that the final upper bound also holds for $b$ with no $\vec{i}$ such that $B(\vec{i}) = b$ and $R(\vec{i}) = 0$ and hence the bound is not dependent on $b$. Because there are at most $4^p$ different $b$'s, 
\begin{equation*}
    \tsum_{i_1, \dots, i_p} (1 - R(\vec{i})) ~w_{i_1} \cdots w_{i_p} \leq 8^p (p ~(\tsum_i w_i) ~K_{\max})^{p/2}.
\end{equation*}
If $w_i \in [0, 1]$ almost surely, 
\begin{equation*}
    \norm{\tsum_i w_i}_p 
    \leq W_n + \norm{\tsum_i (w_i - \Exp[w_i])}_p.
\end{equation*}
For even $p$,
\begin{equation*}
\begin{aligned}
    \norm{\tsum_i (w_i - \Exp[w_i])}_p
    & = \tsum_{i_1, \dots, i_p} P[\tprod_{k=1}^p (w_{i_k} - \Exp[w_{i_k}])] \\
    & = \tsum_{i_1,\dots,i_p} (1 - R(\vec{i})) ~ P[ \tprod_{k=1}^p (w_{i_k} - \Exp[w_{i_k}])],
\end{aligned}
\end{equation*}
and following from the same argument as above, this term is bounded by $\sqrt{p n K_{\max}}$.
Hence
\begin{equation*}
\begin{aligned}
    & \norm{\tsum_i w_i (f_i(\theta_1) - f_i(\theta_2))}_p 
    \leq C \norm{\tsum_i w_i}_{p/2} \sqrt{p ~K_{\max}} d(\theta_1, \theta_2) \\
    & \leq C ~(\sqrt{p ~W_n K_{\max}} + p^{3/4} n^{1/4} K_{\max}^{3/4}) ~d(\theta_1,\theta_2) \\
    & \leq C p^{3/4} \sqrt{W_n K_{\max}} ~d(\theta_1, \theta_2),
\end{aligned}
\end{equation*}
as long as $\sqrt{n K_{\max}} = O(W_n)$.
By Proposition 2.5.2 of \citet{vershynin2018high}, $\norm{\tsum_i w_i (f_i(\theta_1) - f_i(\theta_2))}_{\psi_{4/3}} \leq C \sqrt{W_n K_{\max}} ~d(\theta_1,\theta_2)$ for another universal constant $C$, which proves the first part of the lemma.

If additionally $\sum_i \frac{w_i}{W_n} \leq M$ almost surely for some $M > 0$, then
\begin{equation*}
    \norm{\tsum_i w_i (f_i(\theta_1) - f_i(\theta_2))}_p \leq C \norm{\tsum_i w_i}_{p/2} \sqrt{p ~K_{\max}} d(\theta_1, \theta_2) \leq CM \sqrt{p ~W_n K_{\max}} ~d(\theta_1,\theta_2).
\end{equation*}
By Proposition 2.5.2 of \citet{vershynin2018high}, $\norm{\tsum_i w_i (f_i(\theta_1) - f_i(\theta_2))}_{\psi_2} \leq C \sqrt{W_n K_{\max}} ~d(\theta_1,\theta_2)$ for another universal constant $C$, which proves the second part of the lemma.


\subsection{Proof of Consistency (Theorem~\ref{thm:consistency})}
\label{app:pf_consistency}

\subsubsection{Regularity Conditions}

Under the network dependence (\cref{assmp:network_dependence}), we use the novel empirical process theory (\cref{sec:empirical_process_theory}) to justify using the same data for estimating both the nuisance parameter and the causal estimand. 
To allow this, we impose a restriction on the size of the parameter space, which in our case, corresponds to the functional spaces of $\hat{\mu}$ and $\hat{\pi}$, quantified by \emph{entropy numbers} \citep[p.83]{vaart1996weak}. 

\begin{assumption}[Finite Entropy Condition] \label{assmp:finite_entropy_cons}
    The estimators $\hat\mu$, $\hat\pi$ and their limits $\bar\mu$, $\bar\pi$ of all $n$ are contained in totally bounded functional classes with finite uniform entropy integrals as specified in \cref{assmp:finite_entropy_equi}. In particular, the functional space $\mathcal{F}_\mu$ of $\hat\mu$ and $\bar\mu$ is totally bounded with respect to the $\ell_\infty$ norm, and $\mathcal{F}_\pi$ of $\hat\pi$ and $\bar\pi$ is totally bounded with respect to
    \begin{equation} \label{eq:norm_pi}
        \norm{\pi}_\pi \equiv \max_{i \in [n]} 2^{\abs{N_i}} \cdot \norm{\pi(\cdot| \cdot, \mathcal{G}_{N_i})}_\infty,
    \end{equation}
    where either of the following conditions is true:
    \begin{enumerate}[(i)]
        \item For some $M > 0$ and $\eta > 0$, $\sum_i \frac{\kappa_\lambda(\Delta_i)}{D} \leq M$ almost surely,
        \begin{equation*}
            \int_0^\eta \psi_2^{-1}(N(\epsilon, \mathcal{F_\mu}, \norm{\cdot}_\infty)) d\epsilon < \infty, ~\text{and}~
            \int_0^\eta \psi_2^{-1}(N(\epsilon, \mathcal{F_\pi}, \norm{\cdot}_\pi)) d\epsilon < \infty;
        \end{equation*}
        \item For some $M > 0$ and $\eta > 0$, $\frac{\sqrt{n K_{\max}}}{D} \leq M$,
        \begin{equation*}
            \int_0^\eta \psi_{4/3}^{-1}(N(\epsilon, \mathcal{F_\mu}, \norm{\cdot}_\infty)) d\epsilon < \infty, ~\text{and}~
            \int_0^\eta \psi_{4/3}^{-1}(N(\epsilon, \mathcal{F_\pi}, \norm{\cdot}_\pi)) d\epsilon < \infty;
        \end{equation*}
    \end{enumerate}
    where $\psi_q(x) = \exp(x^q)-1$, $\norm{\cdot}_{\psi_q}$ is the respective Orlicz norm, and $N(\epsilon, \mathcal{F}, d)$ is the covering number of $\mathcal{F}$ by $d$-balls of size $\epsilon$.
\end{assumption}

In addition, we require the following uniform boundedness assumptions to control the uncertainty of each pseudo-outcome $\xi_i$. The first is the bounded outcome assumption, which was also implicitly used in \citet{kennedy2017non} and \citet{van2014causal}. 
The second is a uniform version of the \emph{positivity assumption}, commonly imposed in the causal inference literature, particularly in the context of inverse probability weighting and its extensions. 

\begin{assumption}[Uniform Boundedness] \label{assmp:uniform_boundedness}
The outcome random variables $Y_i$ are uniformly bounded over $i$ and $n$. Additionally, both the estimator $\hat{\pi}$ and its limit $\bar{\pi}$ have uniformly bounded inverses over $n$: there exists a constant $M > 0$ such that for all $n$,
\begin{equation*}
\min_{i \in [n]} \norm*{{\hat{\pi}(\cdot \mid \cdot, \mathcal{G}_{N_i})}}_\pi > M^{-1} 
\quad \text{and} \quad
\min_{i \in [n]} \norm*{{\bar{\pi}(\cdot \mid \cdot, \mathcal{G}_{N_i})}}_\pi > M^{-1},
\end{equation*}
where $\norm{\cdot}_\pi$ is defined as in \cref{assmp:finite_entropy_cons}.
\end{assumption}

\cref{assmp:finite_entropy_cons,assmp:uniform_boundedness} corresponds to Condition (f) in \cref{thm:consistency}. 
%
%
Next we assume that the dissimilarity metric $\Delta_i$, defined as a function of $(i, T_{N_i}, P_{X_{N_i}|\mathcal{G}})$, $(i^*, t^*_{N_{i^*}}, P_{X_{N_{i^*}}|\mathcal{G}})$ and $\mathcal{G}$, quantifies the causally relevant difference between nodes $i$ and $i^*$ in terms of connectivity, neighborhood intervention assignments, and covariate distributions. By \emph{causal relevance}, we mean that $\Delta_i$ appropriately captures the closeness between $\theta_i(T_{N_i})$ and $\theta_{i^*}(t^*_{N_{i^*}})$. 

\begin{assumption}[Lipschitz Continuity of Node-Wise Counterfactual Means]
\label{assmp:smoothness}
    There exists a constant $L > 0$ such that for all $n$ and $i$,
    \begin{equation}
        \abs{\theta_i(T_{N_i}) - \theta_{i^*}(t^*_{N_{i^*}})}
        \leq L \, \Delta_i.
    \end{equation}
\end{assumption}

\cref{assmp:smoothness} corresponds to Condition (d) in \cref{thm:consistency}. We introduce the following condition on the local $L_2$-convergence rate of $\hat\mu$ and $\hat\pi$ to $\mu$ and $\pi$, respectively.

\begin{assumption}[Asymptotic Property of the Outcome Regression and the Propensity Score Estimator]

\begin{equation*}
\begin{aligned}
    \sup_{(i, t_{N_i}): \Delta_i(t_{N_i}) \leq \lambda} \sqrt{\smallint \abs{(\hat\mu - \mu)(t_{N_i}, x_{N_i}, \mathcal{G}_{N_i})}^2 dP_{X_{N_i}|\mathcal{G}}(x_{N_i})} & = O_p(r_n^\circ), \\
    \sup_{(i, t_{N_i}): \Delta_i(t_{N_i}) \leq \lambda} 2^{\abs{N_i}} \sqrt{\smallint \abs{(\hat\pi - \pi)(t_{N_i}| x_{N_i}, \mathcal{G}_{N_i})}^2 dP_{X_{N_i}|\mathcal{G}}(x_{N_i})} & = O_p(s_n^\circ). \\
\end{aligned}
\end{equation*}

\end{assumption}

This condition determines the forms of $r_n^\circ$ and $s_n^\circ$ in \cref{thm:consistency}. Notably, consistency of KECENI does not require both $r_n^\circ$ and $s_n^\circ$ to converge to $0$ simultaneously. This feature reflects the double robust nature of KECENI.
Finally, we introduce the following condition on convergence of the covariate distribution estimator  $\hat{P}_{X_{[n]}|\mathcal{G}}$ to the true distribution.

\begin{assumption}[Asymptotic Property of the Covariate Distribution Estimator] \label{assmp:covariate_dist}
    For each node $i$, let $\mathcal{F}_i$ be a functional space on $\reals^{\abs{N_i} \times p}$ with the $\ell_\infty$-norm. We assume there exist a sequence $q_n$ converging to $0$ and a totally bounded subset $\mathcal{F}$ of the direct sum space
    $\oplus_{i=1}^n \mathcal{F}_i$ 
    with the norm $\norm{(f_1, \dots, f_n)} = \sum_{i=1}^n \norm{f_i}_\infty$ for each $n$ such that:
    \begin{enumerate}[(i)]
        \item $\hat{P}_{X_{[n]}|\mathcal{G}}$ converges uniformly on $\mathcal{F}$:
        \begin{equation*}
            \sup_{(f_1, \dots, f_n) \in \mathcal{F}} \abs*{\sum_{i=1}^n \int f_i(x_{N_i}) (dP_{X_{N_i}|\mathcal{G}} - d\hat{P}_{X_{N_i}|\mathcal{G}})(x_{N_i})} = O_p(q_n),
        \end{equation*}
        
        \item $\hat{P}_{X_{[n]}|\mathcal{G}}$ is equicontinuous on $\mathcal{F}$:
        \begin{equation*}
            \sup_{\norm{(f_1, \dots, f_n) - (g_1, \dots, g_n)} < \delta_n} \abs*{\sum_{i=1}^n \int (f_i - g_i)(x_{N_i}) (dP_{X_{N_i}|\mathcal{G}} - d\hat{P}_{X_{N_i}|\mathcal{G}})(x_{N_i})} = o_p(q_n),
        \end{equation*}
        as $\delta_n \to 0$, and
        
        \item $\mathcal{F}$ contains all nuisance parameter estimates and their limits:
        \begin{equation*}
            (\bar\pi(t_{N_i}| \cdot, \mathcal{G}_{N_i}): i \in [n]) \in \mathcal{F}, 
            ~\text{and}~
            (\bar\mu(t_{N_i}, \cdot, \mathcal{G}_{N_i}): i \in [n]) \in \mathcal{F},
        \end{equation*}
        \begin{equation*}
            (\hat\pi(t_{N_i}| \cdot, \mathcal{G}_{N_i}): i \in [n]) \in \mathcal{F}, 
            ~\text{and}~
            (\hat\mu(t_{N_i}, \cdot, \mathcal{G}_{N_i}): i \in [n]) \in \mathcal{F},
        \end{equation*}
        for all $t_{[n]} \in \{0, 1\}^n$, almost surely.
    \end{enumerate}
\end{assumption}

\cref{assmp:covariate_dist} corresponds to Condition (h) in \cref{thm:consistency}.

\subsubsection{Proof of Theorem}

In this proof, we provide a high-probability upper bound for the estimation error, $\abs{\hat\theta_{i^*}(t^*_{N_{i^*}}) - \theta_{i^*}(t^*_{N_{i^*}})}$. First, we define an oracle estimator:
\begin{equation*}
\begin{aligned}
    \bar\theta_{i^*}(t^*_{N_{i^*}})
    & \equiv \hat{D}^{-1} 
    \sum_{i=1}^n \kappa_\lambda(\Delta_i) \xi_i(\bar\mu, \bar\pi),
\end{aligned}
\end{equation*}
where $\hat{D} \equiv \sum_{i=1}^n \kappa_\lambda(\Delta_i)$.
Then we decompose the error into 
\begin{equation*}
    \hat{\theta}_{i^*}(t^*_{N_{i^*}}) - \theta_{i^*}(t^*_{N_{i^*}})
    = \mathcal{R}_{n,1} + \mathcal{R}_{n,2} + \mathcal{R}_{n,3},
\end{equation*}
where
\begin{equation*}
    \mathcal{R}_{n,1} = \bar\theta_{i^*}(t^*_{N_{i^*}}) - \theta_{i^*}(t^*_{N_{i^*}}),
\end{equation*}
\begin{equation*}
    \mathcal{R}_{n,2} = \hat{D}^{-1}(P_n - P) \left[ 
        \sum_{i=1}^n \kappa_\lambda(\Delta_i) (\hat\xi_i(\hat\mu, \hat\pi) - \xi_i(\bar\mu, \bar\pi)) 
    \right],
\end{equation*}
\begin{equation*}
    \mathcal{R}_{n,3} = \hat{D}^{-1} P \left[
        \sum_{i=1}^n \kappa_\lambda(\Delta_i) (\hat\xi_i(\hat\mu, \hat\pi) - \xi_i(\bar\mu, \bar\pi))
    \right].
\end{equation*}
Here $P(f)$ indicates the expectation over $(Y_{[n]}, T_{[n]}, X_{[n]})$ treating it as an independent copy of the data (or $f$ as fixed):
\begin{equation*}
    P(f) \equiv P[f(Y_{[n]}, T_{[n]}, X_{[n]})] \equiv \int f(y_{[n]}, t_{[n]}, x_{[n]}) dP_{Y_{[n]}, T_{[n]}, X_{[n]} | \mathcal{G}}(y_{[n]}, t_{[n]}, x_{[n]}).
\end{equation*}
So if $\hat{f}$ is random depending on the observed data, $P(\hat{f})$ is also random, which is not the case for $\Exp(\hat{f})$. \(P_n\) denotes the empirical version of $P$, which places probability mass \(1\) on the observed data \((Y_{[n]}, T_{[n]}, X_{[n]})\). 

Bounding the decomposed terms, we will use the following lemma. The proof is given in \cref{app:pf_consistency_D}.
\begin{lemma} \label{thm:consistency_D}
    $\abs{D - \hat{D}} = O_p(\sqrt{K_{\max} D})$, and hence $D\hat{D}^{-1} \overset{p}{\rightarrow} 1$ as long as $K_{\max} = o(D)$.
\end{lemma}

The rest of the proof proceeds as follows:
\begin{itemize}

    \item $\mathcal{R}_{n,1}$ is bounded using the Kernel smoother theory, based on Fan (1993). The variance may be calculated under a network dependence structure.
    
    \item $\mathcal{R}_{n,2}$ is bounded using the newly developed empirical process theory under a network dependence structure.
    
    \item $\mathcal{R}_{n,3}$ is bounded using the double robustness of $\xi_i(\bar\mu,\bar\pi)$, $\norm{\hat\mu - \bar\mu}_\infty = o_p(1)$ and $ \norm*{\hat\pi - \bar\pi}_\pi = o_p(1)$ and the accuracy of the covariate distribution estimate $\hat{P}_{X_{[n]}|\mathcal{G}}$.

\end{itemize}

\subsubsection{Bounding $\mathcal{R}_{n,1}$}
\label{app:pf_consistency_R_n1}

\begin{equation*}
\begin{aligned}
    & \abs{\bar\theta_{i^*}(t^*_{N_{i^*}}) - \theta_{i^*}(t^*_{N_{i^*}})}
    = \abs{\hat{D}^{-1} \tsum_{i=1}^n \kappa_\lambda(\Delta_i)\xi_i(\bar\mu, \bar\pi) - \theta_{i^*}(t^*_{N_{i^*}})} \\
    & \leq \abs{\hat{D}^{-1} \tsum_{i=1}^n \kappa_\lambda(\Delta_i)(\xi_i(\bar\mu, \bar\pi) - \theta_i(T_{N_i}))} 
    + \abs{\hat{D}^{-1}  \tsum_{i=1}^n \kappa_\lambda(\Delta_i)(\theta_i(T_{N_i}) - \theta_{i^*}(t^*_{N_{i^*}}))} \\
    & \leq \abs{\hat{D}^{-1} \tsum_{i=1}^n \kappa_\lambda(\Delta_i)(\xi_i(\bar\mu, \bar\pi) - \theta_i(T_{N_i}))} 
    + L \abs{\hat{D}^{-1}
    \tsum_{i=1}^n \kappa_\lambda(\Delta_i) \cdot \Delta_i}, \\
\end{aligned}
\end{equation*}
where the last equality follows from the smoothness result in \cref{assmp:smoothness}.
Due to the uniform boundedness of $\xi_i$, for some $M_\xi > 0$
\begin{equation*}
\begin{aligned}
    \Exp[\abs{ \kappa_\lambda(\Delta_i)(\xi_i(\bar\mu, \bar\pi) - \theta_i(T_{N_i}))}^2 | T, \mathcal{G}] 
    \leq 4 M_\xi^2 \abs{ \kappa_\lambda(\Delta_i)}^2, 
\end{aligned}
\end{equation*}
and as in the proof of \cref{thm:consistency_D},
\begin{equation*}
\begin{aligned}
    & \Exp[\abs{\tsum_{i=1}^n \kappa_\lambda(\Delta_i)(\xi_i(\bar\mu, \bar\pi) - \theta_i(T_{N_i}))}^2 | \mathcal{G}] \\
    & \leq \sum_{i=1}^n \sum_{j: j \neq i} \Exp\left[\left. \begin{aligned}
        & \abs{\kappa_\lambda(\Delta_i)(\xi_i(\bar\mu, \bar\pi) - \theta_i(T_{N_i}))} \\
        & \times \abs{\kappa_\lambda(\Delta_j)(\xi_j(\bar\mu, \bar\pi) - \theta_{j}(T_{N_j}))}
    \end{aligned} \right| \mathcal{G} \right] \\
    & \leq \sum_{i=1}^n \sum_{j: j \neq i} \left( \begin{aligned} 
        & \Exp\left[\left. 
        \abs{\kappa_\lambda(\Delta_i)(\xi_i(\bar\mu, \bar\pi) - \theta_i(T_{N_i}))}^2
        \right| \mathcal{G} \right] \\
        & + \Exp\left[\left. 
        \abs{\kappa_\lambda(\Delta_j)(\xi_j(\bar\mu, \bar\pi) - \theta_j(T_{N_j}))}^2
        \right| \mathcal{G} \right] 
    \end{aligned} \right) \\
    & \leq K_{\max} \sum_{i=1}^n \Exp\left[\left. \begin{aligned}
        \abs{\kappa_\lambda(\Delta_i)(\xi_i(\bar\mu, \bar\pi) - \theta_i(T_{N_i}))}^2 
    \end{aligned} \right| \mathcal{G} \right] 
    \leq 4 K_{\max} M_{\xi}^2 \sum_{i=1}^n \Exp\left[\left. \begin{aligned}
        \abs{\kappa_\lambda(\Delta_i)}^2 
    \end{aligned} \right| \mathcal{G} \right] 
    = O\left( K_{\max} D \right).
\end{aligned}
\end{equation*}
Hence with \cref{thm:consistency_D} the first term is upper bounded by
\begin{equation*}
\begin{aligned}
    & \Exp[\abs{\hat{D}^{-1} \tsum_{i=1}^n \kappa_\lambda(\Delta_i)(\xi_i(\bar\mu, \bar\pi) - \theta_i(T_{N_i}))}^2 | \mathcal{G}] \\
    & \leq O\left( \frac{1}{D^2} \sum_{i=1}^n \sum_{j: j \neq i} \Exp\left[\left. \begin{aligned}
        & \abs{\kappa_\lambda(\Delta_i)(\xi_i(\bar\mu, \bar\pi) - \theta_i(T_{N_i}))} \\
        & \times \abs{\kappa_\lambda(\Delta_j)(\xi_j(\bar\mu, \bar\pi) - \theta_{i^* \rightarrow k}(T;\mathcal{G}))}
    \end{aligned} \right| \mathcal{G} \right] \right) \\
    & \leq O\left( \frac{1}{D^2} \sum_{i=1}^n \sum_{j: j \neq i} \left( \begin{aligned} 
        & \Exp\left[\left. 
        \abs{\kappa_\lambda(\Delta_i)(\xi_i(\bar\mu, \bar\pi) - \theta_i(T_{N_i}))}^2
        \right| \mathcal{G} \right] \\
        & + \Exp\left[\left. 
        \abs{\kappa_\lambda(\Delta_j)(\xi_j(\bar\mu, \bar\pi) - \theta_{i^* \rightarrow k}(T;\mathcal{G}))}^2
        \right| \mathcal{G} \right] 
    \end{aligned} \right) \right) \\
    & \leq O\left( \frac{K_{\max}}{D} \right).
\end{aligned}
\end{equation*}
Similarly, due to the support of $\kappa$, the second term is upper bounded by
\begin{equation*}
\begin{aligned}
    \hat{D}^{-1} 
    \tsum_{i=1}^n \kappa_\lambda(\Delta_i) \cdot \Delta_j(T_{N_i}) 
    \leq \lambda.
\end{aligned}
\end{equation*}
In sum, by Markov's inequality,
\begin{equation*}
    \abs{\mathcal{R}_{n,1}} = O_p\left( \sqrt{\frac{K_{\max}}{D}} + L \lambda \right).
\end{equation*}

\subsubsection{Bounding $\mathcal{R}_{n,2}$}
\label{app:pf_consistency_R_n2}

Recall that
\begin{equation*}
\begin{aligned}
    \bar\varpi(t_{N_i}| \mathcal{G})
    & \equiv \int \bar\pi(t_{N_i}| x_{N_i}, \mathcal{G}_{N_i}) 
    dP_{X_{N_i}|\mathcal{G}}(x_{N_i}), ~\text{and} \\
    \bar{m}(t_{N_i}, \mathcal{G})
    & \equiv \int \bar\mu(t_{N_i}, x_{N_i}, \mathcal{G}_{N_i})
    dP_{X_{N_i}|\mathcal{G}}(x_{N_i}).
\end{aligned}
\end{equation*}
In addition, we define
\begin{equation*}
\begin{aligned}
    \hat\varpi(T_{N_i}| \mathcal{G})
    & \equiv \int \hat\pi(T_{N_i}| x_{N_i}, \mathcal{G}_{N_i}) 
    d\hat{P}_{X_{N_i}|\mathcal{G}}(x_{N_i} | \mathcal{G}), ~\text{and} \\
    \hat{m}(t_{N_i}, \mathcal{G})
    & \equiv \int \hat\mu(T_{N_i}, x_{N_i}, \mathcal{G}_{N_i}) 
    d\hat{P}_{X_{N_i}|\mathcal{G}}(x_{N_i} | \mathcal{G}).
\end{aligned}
\end{equation*}
Then, 
\begin{equation*}
\begin{aligned}
    & \hat\xi_i(\hat\mu,\hat\pi) - \xi_i(\bar\mu,\bar\pi) \\
    & = \frac{Y_j - \hat\mu(T_{N_i}, X_{N_i}, \mathcal{G}_{N_i})}
    {\hat\pi(T_{N_i}| X_{N_i}, \mathcal{G}_{N_i})}
    \hat\varpi(T_{N_i}| \mathcal{G}) 
    + \hat{m}(T_{N_i}, \mathcal{G}) \\
    & \quad - \frac{Y_j - \bar\mu(T_{N_i}, X_{N_i}, \mathcal{G}_{N_i})}
    {\bar\pi(T_{N_i}| X_{N_i}, \mathcal{G}_{N_i})}
    \bar\varpi(T_{N_i}| \mathcal{G}) 
    - \bar{m}(T_{N_i}, \mathcal{G}) \\
    & = \frac{Y_j - \bar\mu(T_{N_i}, X_{N_i}, \mathcal{G}_{N_i})}
    {\bar\pi(T_{N_i}| X_{N_i}, \mathcal{G}_{N_i})}
    \left\{ \hat\varpi(T_{N_i}| \mathcal{G}) - \bar\varpi(T_{N_i}| \mathcal{G}) \right\}\\
    & \quad + (Y_j - \bar\mu(T_{N_i}, X_{N_i}, \mathcal{G}_{N_i}))
    \frac{\hat\varpi(T_{N_i}| \mathcal{G})}{\hat\pi(T_{N_i}| X_{N_i}, \mathcal{G}_{N_i})}
    \left\{ \frac{\hat\pi(T_{N_i}| X_{N_i}, \mathcal{G}_{N_i}) - \bar\pi(T_{N_i}| X_{N_i}, \mathcal{G}_{N_i})}
    {\bar\pi(T_{N_i}| X_{N_i}, \mathcal{G}_{N_i})} \right\} \\
    & \quad + \frac{\hat\varpi(T_{N_i}| \mathcal{G})}{\hat\pi(T_{N_i}| X_{N_i}, \mathcal{G}_{N_i})}
    \left\{ \hat\mu(T_{N_i}, X_{N_i}, \mathcal{G}_{N_i}) - \bar\mu(T_{N_i}, X_{N_i}, \mathcal{G}_{N_i}) \right\} \\
    & \quad + \left\{ \hat{m}(T_{N_i}, \mathcal{G}) - \bar{m}(T_{N_i}, \mathcal{G}) \right\}. 
\end{aligned}
\end{equation*}
Due to the unifrom boundedness assumption (\cref{assmp:uniform_boundedness}),
\begin{equation*}
\begin{aligned}
    \abs*{\frac{\hat\varpi(T_{N_i}| \mathcal{G})}
    {\hat\pi(T_{N_i}| X_{N_i}, \mathcal{G}_{N_i})}}
    \leq \abs*{
        \int \frac{\hat\pi(T_{N_i}| x_{N_i}, \mathcal{G}_{N_i})}{\hat\pi(T_{N_i}| X_{N_i}, \mathcal{G}_{N_i})}
        d\hat{P}_{X_{N_i} | \mathcal{G}}(x_{N_i})
    } 
    \leq M_\pi M.
\end{aligned}
\end{equation*}
Similarly,
\begin{equation*}
\begin{aligned}
    \abs*{\frac{\hat\varpi(T_{N_i}| \mathcal{G}) - \bar\varpi(T_{N_i}| \mathcal{G})}
    {\bar\pi(T_{N_i}| X_{N_i}, \mathcal{G}_{N_i})}}
    & \leq \abs*{
        \frac{\int \hat\pi(T_{N_i}| x_{N_i}, \mathcal{G}_{N_i}) d\hat{P}_{X_{N_i}|\mathcal{G}}(x_{N_i})
        - \int \bar\pi(T_{N_i}| x_{N_i}, \mathcal{G}_{N_i}) dP_{X_{N_i}|\mathcal{G}}(x_{N_i})}{\bar\pi(T_{N_i}| X_{N_i}, \mathcal{G}_{N_i})}
    } \\
    & \leq \abs*{
        \frac{\int \{ \hat\pi(T_{N_i}| x_{N_i}, \mathcal{G}_{N_i})
        - \bar\pi(T_{N_i}| x_{N_i}, \mathcal{G}_{N_i}) \} d\hat{P}_{X_{N_i}|\mathcal{G}}(x_{N_i})}{\bar\pi(T_{N_i}| X_{N_i}, \mathcal{G}_{N_i})}
    } \\
    & \quad + \abs*{
        \frac{\int \bar\pi(T_{N_i}| x_{N_i}, \mathcal{G}_{N_i}) (d\hat{P}_{X_{N_i}|\mathcal{G}} - dP_{X_{N_i}|\mathcal{G}})(x_{N_i})}{\bar\pi(T_{N_i}| X_{N_i}, \mathcal{G}_{N_i})}
    } \\
    & \leq \norm*{\hat\pi - \bar\pi}_\pi M + o_p(1),
\end{aligned}
\end{equation*}
where the second term in the last inequality follows from the uniform convergence of $\hat{P}_{X_{[n]}|\mathcal{G}}$ over $\mathcal{F}_\pi$ (first part of \cref{assmp:covariate_dist}). Similarly,
\begin{equation*}
\begin{aligned}
    \abs*{\hat{m}(T_{N_i}, \mathcal{G}) - \bar{m}(T_{N_i}, \mathcal{G})}
    & \leq \norm*{\hat\mu - \bar\mu}_\infty + o_p(1).
\end{aligned}
\end{equation*}
Therefore, $\norm{\hat\varpi - \bar\varpi}_\pi \rightarrow 0$ and $\norm{\hat{m} - \bar{m}}_\infty \rightarrow 0$ as $n \rightarrow \infty$.
Due to the uniform boundedness of $\bar\mu$, $\hat\mu$, $\bar\pi$, $\hat\pi$ and $Y_i$,
\begin{equation*}
    \norm{\xi_i(\hat\mu,\hat\pi) - \xi_i(\bar\mu,\bar\pi)}_\infty
    = O\left(
        \norm*{\hat\pi - \bar\pi}_\pi 
        + \norm{\hat\mu - \bar\mu}_\infty
    \right).
\end{equation*}
Furthermore,
\begin{equation} \label{eq:xi_j_infty_bound}
\begin{aligned}
    & \norm*{\kappa_\lambda(\Delta_i) (\xi_i(\hat\mu, \hat\pi) - \xi_i(\bar\mu, \bar\pi)) - \Exp[\kappa_\lambda(\Delta_i) (\xi_i(\hat\mu, \hat\pi) - \xi_i(\bar\mu, \bar\pi)) | \mathcal{G}]}_p \\
    & \leq 2 \norm*{\xi_i(\hat\mu, \hat\pi) - \xi_i(\bar\mu, \bar\pi)}_\infty 
    = O\left(
        \norm*{\hat\pi - \bar\pi}_\pi 
        + \norm{\hat\mu - \bar\mu}_\infty
    \right).
\end{aligned}
\end{equation}
%
Now we prove the asymptotic negligibility of $\mathcal{R}_{n,2}$. 
Put $\theta = (\mu, \pi) \in \mathcal{F}_\mu \times \mathcal{F}_\pi$, $d(\theta_1,\theta_2) \equiv \norm*{\hat\pi - \bar\pi}_\pi 
+ \norm{\hat\mu - \bar\mu}_\infty$, and
\begin{equation*}
    Z_n(\theta) =  \frac{1}{\sqrt{\hat{D} K_{\max}}} (P_n - P) \left[
        \tsum_{i=1}^n \kappa_\lambda(\Delta_i) \xi_i(\mu, \pi)
    \right].
\end{equation*}
Note that $K_{\max}$ represents the maximum size of the local dependence of $(Y_i, T_{N_i}, X_{N_i})$. 
To $Z_n(\theta)$, we apply \cref{thm:asymptotic_equicontinuity}, the empirical process theory for weighted average process of neighborhood dependent observation.
Following the same argument as \cref{eq:xi_j_infty_bound}, $\norm{\xi_i(\mu_1,\pi_1) - \xi_i(\mu_2,\pi_2)}_\infty \leq C d(\theta_1, \theta_2)$, satisfying \cref{assmp:moment_lipschitz} of \cref{thm:asymptotic_equicontinuity}.
\cref{assmp:total_boundedness,assmp:finite_entropy_equi} follow \cref{assmp:finite_entropy_cons}.
%
Hence $\frac{1}{\sqrt{\hat{D} K_{\max}}} (P_n - P) [\sum_{i=1}^n \kappa_\lambda(\Delta_i) \xi_i(\cdot, \cdot) ]$ satisfies stochastic equicontinuity conditional on $T$.
That is, for any sequence $\delta_n \rightarrow 0$ and fixed $\eta > 0$,
\begin{equation*}
    \Pr\left[ \left. \sup \frac{1}{\sqrt{\hat{D} K_{\max}}} (P_n - P) \left[
        \tsum_{i=1}^n \kappa_\lambda(\Delta_i) (\xi_i(\mu_1, \pi_1) - \xi_i(\mu_2, \pi_2)) 
    \right] > \eta \right| T_{[n]}, \mathcal{G} \right] \rightarrow 0
\end{equation*}
as $n \rightarrow \infty$, where the supremum is over the space of $(\mu_1, \mu_2, \pi_1, \pi_2)$ satisfying $\norm{\mu_1 - \mu_2}_\infty < \delta_n$ and $\norm{\pi_1 - \pi_2}_\pi < \delta_n$. 
Then by the dominated convergence theorem, the stochastic equicontinuity holds marginally (i.e., without the condition on $T_{[n]}$), and it follows from \cref{thm:consistency_D} that
\begin{equation*}
    \mathcal{R}_{n,2} = \hat{D}^{-1} (P_n - P) \left[
        \tsum_{i=1}^n \kappa_\lambda(\Delta_i) (\xi_i(\mu_1, \pi_1) - \xi_i(\mu_2, \pi_2)) 
    | \mathcal{G} \right] = o_p\left( \sqrt{\frac{K_{\max}}{D}} \right).
\end{equation*}




\subsubsection{Bounding $\mathcal{R}_{n,3}$} \label{app:bounding_R_n3}

For any treatment $t$, network $g$ and its node $k$, conditioning on $T_{N_i} = t_{N_i}$,
\begin{equation*}
\begin{aligned}
    & P_{(Y_{i},X_{N_i})|T_{N_i}}[\hat\xi_i(\hat\mu,\hat\pi) - \xi_i(\bar\mu,\bar\pi) | t_{N_i}] \\
    & = P\left[
        \{\mu(t_{N_i}, X_{N_i}, \mathcal{G}_{N_i}) - \hat\mu(t_{N_i}, X_{N_i}, \mathcal{G}_{N_i})\}
        \left\{
            \frac{\pi(t_{N_i}| X_{N_i}, \mathcal{G}_{N_i})
            / \varpi(t_{N_i}| \mathcal{G})}
            {\hat\pi(t_{N_i}| X_{N_i}, \mathcal{G}_{N_i})
            / \hat\varpi(t_{N_i}| \mathcal{G})}
        \right\}
    \right] \\
    & \quad + \hat{m}(t_{N_i}, \mathcal{G})
    - \theta_i(t_{N_i}) \\
    & = \frac{\hat\varpi(t_{N_i}| \mathcal{G})}{\varpi(t_{N_i}| \mathcal{G})}
    P\left[
        (\mu - \hat\mu)(t_{N_i}, X_{N_i}, \mathcal{G}_{N_i})
        \frac{(\pi - \hat\pi)(t_{N_i}| X_{N_i}, \mathcal{G}_{N_i})}
        {\hat\pi(t_{N_i}| X_{N_i}, \mathcal{G}_{N_i})}
    \right] \\
    & \quad + \frac{1}{\varpi(t_{N_i}| \mathcal{G})}
    P\left[
        (\hat\pi - \pi)(t_{N_i}| X_{N_i}, \mathcal{G}_{N_i})
    \right] 
    P\left[
        (\mu - \hat\mu)(t_{N_i}, X_{N_i}, \mathcal{G}_{N_i})
    \right] \\
    & \quad + \frac{P\left[
        (\mu - \hat\mu)(t_{N_i}, X_{N_i}, \mathcal{G}_{N_i})
    \right]}{\varpi(t_{N_i}| \mathcal{G})}
    \int \hat\pi(t_{N_i}| x_{N_i}, \mathcal{G}_{N_i}) (d\hat{P}_{X_{N_i}|\mathcal{G}} - dP_{X_{N_i}|\mathcal{G}}) (x_{N_i})
     \\
    & \quad + \int \hat\mu(t_{N_i}, x_{N_i}, \mathcal{G}_{N_i}) (d\hat{P}_{X_{N_i}|\mathcal{G}} - dP_{X_{N_i}|\mathcal{G}}) (x_{N_i}),
\end{aligned}
\end{equation*}
where the first equality is due to the double-robustness of $\xi_i(\bar\mu, \bar\pi)$. 
Due to the uniform boundedness of $\pi$ and $\hat\pi$ and H\"older's inequality,
\begin{equation*}
\begin{aligned}
    & \abs*{ P \left[
        \sum_{i=1}^n \kappa_\lambda(\Delta_i) (\hat\xi_i(\hat\mu, \hat\pi) - \xi_i(\bar\mu, \bar\pi))
    \right]} \\
    & = O \left( \begin{aligned}
        & \sum_{i=1}^n \kappa_\lambda(\Delta_i) r(T_{N_i}) s(T_{N_i}) 
        \\
        & + \abs*{ \sum_{i=1}^n 
            \kappa_\lambda(\Delta_i) \int \hat\pi(T_{N_i}| x_{N_i}, \mathcal{G}_{N_i}) (d\hat{P}_{X_{N_i}|\mathcal{G}} - dP_{X_{N_i}|\mathcal{G}}) (x_{N_i}) }
         \\
        & + \abs*{ \sum_{i=1}^n
            \kappa_\lambda(\Delta_i) \int \hat\mu(T_{N_i}, x_{N_i}, \mathcal{G}_{N_i}) (d\hat{P}_{X_{N_i}|\mathcal{G}} - dP_{X_{N_i}|\mathcal{G}}) (x_{N_i}) }
    \end{aligned} \right),
\end{aligned}
\end{equation*}
where 
\begin{equation*}
    \begin{aligned}
        r(t_{N_{i}}) & 
        \equiv \sqrt{\int \abs{(\hat\mu - \mu)(t_{N_i}, x_{N_i}, \mathcal{G}_{N_i})}^2 dP_{X_{N_i}|\mathcal{G}}(x_{N_i})}, \\
        s(t_{N_{i}}) & 
        \equiv 2^{\abs{N_i}} \sqrt{\int \abs{(\hat\pi - \pi)(t_{N_i}| X_{N_i}, \mathcal{G}_{N_i})}^2 dP_{X_{N_i}|\mathcal{G}}(x_{N_i})}.
    \end{aligned}
    \end{equation*}

For the last two terms, by the equicontinuity of $\hat{P}_{X_{[n]}|\mathcal{G}}$ (the second part of \cref{assmp:covariate_dist}), 
\begin{equation*}
\begin{aligned}
    \abs*{ \sum_{i=1}^n \int \frac{1}{D}
    \kappa_\lambda(\Delta_i) (\hat\pi - \bar\pi)(T_{N_i}| x_{N_i}, \mathcal{G}_{N_i}) (d\hat{P}_{X_{N_i}|\mathcal{G}} - dP_{X_{N_i}|\mathcal{G}}) (x_{N_i}) }
    = o_p(q_n),
\end{aligned}
\end{equation*}
and by the uniform convergence of $\hat{P}_{X_{[n]}|\mathcal{G}}$ (the first part of \cref{assmp:covariate_dist}),
\begin{equation*}
\begin{aligned}
    \abs*{ \sum_{i=1}^n \int \frac{1}{D}
    \kappa_\lambda(\Delta_i) \bar\pi(T_{N_i}| x_{N_i}, \mathcal{G}_{N_i}) (d\hat{P}_{X_{N_i}|\mathcal{G}} - dP_{X_{N_i}|\mathcal{G}}) (x_{N_i}) }
    = O_p(q_n).
\end{aligned}
\end{equation*}
Hence, the last two terms have convergence rate $O_p(q_n)$. 
For the first term, because $\kappa_\lambda$ has support $[-\lambda, \lambda]$,
\begin{equation*}
\begin{aligned}
    \frac{1}{D} \sum_{i=1}^n 
    \kappa_\lambda(\Delta_i) r(T_{N_i}) s(T_{N_i}) 
    \leq \left\{ \sup_{(i, t_{N_i}): \Delta_i(t_{N_i}) \leq \lambda} r(t_{N_i}) \right\}
    \left\{ \sup_{(i, t_{N_i}): \Delta_i(t_{N_i}) \leq \lambda} s(t_{N_i}) \right\}
\end{aligned}
\end{equation*}
By this, we show
\begin{equation*}
    \abs{\mathcal{R}_{n,3}} = O_p\left(r^\circ_n \cdot s^\circ_n + q_n\right),
\end{equation*}
and prove the consistency of the KECENI estimate.

\subsubsection{Proof of Lemma~\ref{thm:consistency_D}} \label{app:pf_consistency_D}

We note that
\begin{equation*}
\begin{aligned}
    & \Var[\hat{D} | \mathcal{G}]
    = \Var\left[ \left. \sum_{i=1}^n \kappa_\lambda(\Delta_i) 
    \right| \mathcal{G} \right] \\
    & \leq \frac{1}{2} \sum_{i=1}^n \sum_{j: j \in \tilde{N}_i} \Cov\left[ \kappa_\lambda(\Delta_i),
    \kappa_\lambda(\Delta_j) | \mathcal{G} \right] 
    \leq \frac{1}{2} \sum_{i=1}^n \sum_{j: j \in \tilde{N}_i} (\Var\left[ \kappa_\lambda(\Delta_i) | \mathcal{G}] +
    \Var[\kappa_\lambda(\Delta_j) | \mathcal{G} \right]) \\
    & \leq K_{\max} \sum_{i=1}^n \Var\left[ \kappa_\lambda(\Delta_i) | \mathcal{G} \right] 
    \leq K_{\max} \sum_{i=1}^n \Exp\left[ (\kappa_\lambda(\Delta_i))^2 | \mathcal{G} \right] 
    \leq K_{\max} \sum_{i=1}^n \Exp\left[ \kappa_\lambda(\Delta_i) | \mathcal{G} \right] = K_{\max} D.
\end{aligned}
\end{equation*}
By the Chebyshev inequality, we obtain the desired result.

\subsection{Proof of Asymptotic Normality (Theorem~\ref{thm:asymptotic_normality})}
\label{app:pf_asymptotic_normality}

Recall that the estimation error is decomposed into 
\begin{equation*}
    \hat{\theta}_{i^*}(t^*_{N_{i^*}}) - \theta_{i^*}(t^*_{N_{i^*}})
    = \mathcal{R}_{n,1} + \mathcal{R}_{n,2} + \mathcal{R}_{n,3},
\end{equation*}
where
\begin{equation*}
    \mathcal{R}_{n,1} = \bar\theta_{i^*}(t^*_{N_{i^*}}) - \theta_{i^*}(t^*_{N_{i^*}}),
\end{equation*}
\begin{equation*}
    \mathcal{R}_{n,2} = \hat{D}^{-1}(P_n - P) \left[
        \sum_{i=1}^n \kappa_\lambda(\Delta_i)(\hat\xi_i(\hat\mu, \hat\pi) - \xi_i(\bar\mu, \bar\pi))
    \right],
\end{equation*}
\begin{equation*}
    \mathcal{R}_{n,3} = \hat{D}^{-1} P \left[
        \sum_{i=1}^n \kappa_\lambda(\Delta_i) (\hat\xi_i(\hat\mu, \hat\pi) - \xi_i(\bar\mu, \bar\pi))
    \right].
\end{equation*}

In the proof of $\hat\theta_{i^*}(t^*; \mathcal{G}^*)$'s consistency (\cref{app:pf_consistency}), we showed that $\mathcal{R}_{n,2}$ and $\mathcal{R}_{n,3}$ are $o_p(\sqrt{\frac{K_{\max}}{D}})$ and $O_p(r^\circ_n \cdot s^\circ_n + q_n)$, respectively. Due to the condition on $\sigma_n$, $\frac{\mathcal{R}_{n,2} + \mathcal{R}_{n,3}}{\sigma_n} \overset{p}{\rightarrow} 0$.
Now, it is sufficient to show the weak convergence of $\frac{1}{\sigma_n} (\mathcal{R}_{n,1} - b_n)$ to a Gaussian distribution.

\subsubsection{Weak convergence of $\frac{1}{\sigma_n} (\mathcal{R}_{n,1} - b_n)$}

The term ${\mathcal{R}}_{n,1}$ represents the error of the oracle estimate $\bar\theta_{i^*}(t^*_{N_{i^*}}) \equiv \hat{D}^{-1} \sum_{i=1}^n \kappa_\lambda(\Delta_i) \xi_i(\bar\mu, \bar\pi)$:
\begin{equation*}
    {\mathcal{R}}_{n,1} - b_n = {D}^{-1} \sum_{i=1}^n 
    \kappa_\lambda(\Delta_i) \left\{
        \xi_i(\bar\mu, \bar\pi) - \tilde\theta_{i^*}(t^*_{N_{i^*}}) 
    \right\}
    + \tilde{\epsilon}_{n,1},
\end{equation*}
where $\tilde\theta_{i^*}(t^*_{N_{i^*}}) \equiv {D}^{-1} \Exp\left[ \left.
    \sum_{i=1}^n \kappa_\lambda(\Delta_i) \xi_i(\bar\mu, \bar\pi)
\right| \mathcal{G} \right]$. The residual $\epsilon_{n,1}$ is the second-order error term: 
\begin{equation*}
\begin{aligned}
    \tilde\epsilon_{n,1} 
    & = (\hat{D}^{-1} - D^{-1}) \sum_{i=1}^n 
    \kappa_\lambda(\Delta_i) \left\{
        \xi_i(\bar\mu, \bar\pi) - \tilde\theta_{i^*}(t^*_{N_{i^*}}) 
    \right\}.
\end{aligned}
\end{equation*}
Due to our observation in \cref{app:pf_consistency} that $\Var[\frac{\hat{D}}{D}] = O(\frac{K_{\max}}{D})$ and that $\Var[\frac{1}{D}\kappa_\lambda(\Delta_i) \xi_i(\bar\mu, \bar\pi)] = O(\frac{K_{\max}}{D})$,
\begin{equation*}
    \tilde \epsilon_{n,1} = o_p\left(\sqrt{\frac{K_{\max}}{D}}\right).
\end{equation*}

Let $W_j$ be 
\begin{equation*}
    W_j \equiv \frac{1}{D} \left( \begin{aligned}
        \kappa_\lambda(\Delta_i) (\xi_i(\bar\mu,\bar\pi) - \tilde\theta_{i^*}(t^*_{N_{i^*}}))
        - \Exp[ \kappa_\lambda(\Delta_i) (\xi_i(\bar\mu,\bar\pi) - \tilde\theta_{i^*}(t^*_{N_{i^*}})) | \mathcal{G} ]
    \end{aligned} \right).
\end{equation*}
It is easily observed that $\Exp[W_j | \mathcal{G}] = 0$, and due to the uniform boundedness of $\xi_i$, for some $M_\xi > 0$
\begin{equation*}
\begin{aligned}
    \Exp[\abs{W_j}^3 | \mathcal{G}] 
    & \lesssim \frac{M_\xi^3}{D^3} \left\{
        \Exp[\kappa_\lambda(\Delta_j)^3| \mathcal{G}] + \Exp[\kappa_\lambda(\Delta_j)| \mathcal{G}]^3
    \right\}
    < \infty, ~\text{and}
\end{aligned}
\end{equation*}
\begin{equation*}
\begin{aligned}
    \Exp[W_j^4 | \mathcal{G}] 
    & \lesssim \frac{M_\xi^4}{D^4} \left\{
        \Exp[\kappa_\lambda(\Delta_j)^4| \mathcal{G}] + \Exp[\kappa_\lambda(\Delta_j)| \mathcal{G}]^4
    \right\} < \infty. 
\end{aligned}
\end{equation*}
More importantly, under the condition of $\sqrt{\frac{K_{\max}}{D}} = O(\sigma_n)$, 
\begin{equation*}
\begin{aligned}
    & \frac{K_{\max}^2}{\sigma_n^3} \sum_{i=1}^n \Exp [\abs{W_j}^3 | \mathcal{G}] + \sqrt{\frac{28}{\pi}} \frac{K_{\max}^{3/2}}{\sigma_n^2} \sqrt{\sum_{i=1}^n \Exp[W_j^4 | \mathcal{G}]} \\
    & \leq \frac{K_{\max}^2}{\sigma_n^3} \frac{M_\xi^3}{D^3} 
    \sum_{i=1}^n \left\{
        \Exp[\kappa_\lambda(\Delta_j)^3| \mathcal{G}] + \Exp[\kappa_\lambda(\Delta_j)| \mathcal{G}]^3
    \right\} \\
    & \quad + \sqrt{\frac{28}{\pi}} \frac{K_{\max}^{3/2}}{\sigma_n^2} \frac{M_\xi^2}{D^2} 
    \sqrt{\sum_{i=1}^n \left\{
        \Exp[\kappa_\lambda(\Delta_j)^4| \mathcal{G}] + \Exp[\kappa_\lambda(\Delta_j)| \mathcal{G}]^4
    \right\}} \\
    & \lesssim \frac{M_\xi^3 K_{\max}^2}{D^2 \sigma_n^3} 
    + \frac{M_\xi^2 K_{\max}^{3/2}}{D^{3/2} \sigma_n^2} 
    = o_p(1).
\end{aligned}
\end{equation*}
That is, the right hand side in the Eq. (3.8) of \citet{ross2011fundamentals} vanishes in probability. By Theorem 3.6 of \citet{ross2011fundamentals}, $\sum_j W_j$ satisfies the desired asymptotic normality.

\subsection{Proof of Sandwich Asymptotic Normality and Consistency of Empirical Sandwich Variance Estimate (\cref{thm:sandwich_asymptotic_normality,thm:sandwich_variance_consistency})} 
\label{app:pf_sandwich_theory}

\subsubsection{Regularity Conditions}

First, we introduce necessary additional regularity conditions, which will allow us to apply the empirical process theory developed in \cref{sec:empirical_process_theory} to each summand $\hat{W}_i \hat{W}_j$ of $\hat\sigma^2$. This requires additional conditions on the regularity of the linear approximation terms $H_{\mu,i}$, $H_{\pi, i}$ and $H_{P_{X|\mathcal{G}}, i}$. 

\begin{assumption}[Linear Approximation of Outcome Regression and Propensity Score Estimation Errors]
\label{assmp:H_nuisance}
Assume that the estimation errors of the outcome regression and propensity score have linear approximations as in \cref{eq:H_nuisance}, where for $\mu \in \mathcal{F}_\mu$ and $\pi \in \mathcal{F}_\pi$, $H_{\mu, i} \in \mathcal{F}_\mu$, $H_{\pi, i} \in \mathcal{F}_\pi$, and they satisfy the following conditions:
\begin{enumerate}[(i)]    
    \item
    \begin{equation*}
    \begin{aligned}
        \sup_{(i, t_{N_i}): \Delta_i(t_{N_i}) \leq \lambda} \sqrt{  
            P[\abs{(\hat\mu - \bar\mu)(t_{N_i}, X_{N_i}, \mathcal{G}_{N_i})}^2]
        }
        & = O_p(\bar{r}^{\circ}_n), \\
        \sup_{(i, t_{N_i}): \Delta_i(t_{N_i}) \leq \lambda} 2^{\abs{N_i}} \sqrt{
            P[\abs{(\hat\pi - \bar\pi)(t_{N_i}| X_{N_i}, \mathcal{G}_{N_i})}^2]
        }
        & = O_p(\bar{s}^{\circ}_n).
    \end{aligned}
    \end{equation*}

    \item For any $i$, $t_{N_i}$ and $x_{N_i}$,
    \begin{equation*}
        \Exp[H_{\bar\mu,i}(t_{N_j}, x_{N_j}, \mathcal{G}_{N_j})] = 0,
        ~\text{and}~
        \Exp[H_{\bar\pi,i}(t_{N_j}, x_{N_j}, \mathcal{G}_{N_j})] = 0.
    \end{equation*}

    \item  For each $i$, $H_{\bar\mu,i}$ and $H_{\bar\pi,i}$ are uniformly  bounded:
    \begin{equation*}
    \begin{aligned}
        \norm{
            {H}_{\bar\mu, i}(t_{N_j}, x_{N_j}, \mathcal{G}_{N_j})
        }_\infty
        = O\left(
            \bar{r}^H_n
        \right), ~\text{and}~
        2^{\abs{N_j}} \norm{
            {H}_{\bar\pi, i}(t_{N_j}| x_{N_j}, \mathcal{G}_{N_j})
        }_\infty
        = O\left(
            \bar{s}^H_n
        \right).
    \end{aligned}
    \end{equation*}

    \item $H_{\mu,i}$ and $H_{\pi,i}$ satisfy Lipschitz condition with respect to $\mu$ and $\pi$, respectively: There exists $C_\mu > 0$ such that for $\mu^{(1)}, \mu^{(2)} \in \mathcal{F}_\mu$ 
    \begin{equation*}
    \begin{aligned}
        \norm*{
            H_{\mu^{(1)}, i}(t_{N_j}, x_{N_j}, \mathcal{G}_{N_j})
            - H_{\mu^{(2)}, i}(t_{N_j}, x_{N_j}, \mathcal{G}_{N_j})
        }_\infty
        \leq C_\mu \bar{r}_n^H \norm{\mu^{(1)} - \mu^{(2)}}_\infty.
    \end{aligned}
    \end{equation*}
    Similarly, there exists $C_\pi > 0$ such that for $\pi^{(1)}, \pi^{(2)} \in \mathcal{F}_\pi$,
    \begin{equation*}
    \begin{aligned}
        2^{\abs{N_j}} \norm*{
            H_{\pi^{(1)}, i}(t_{N_j}| x_{N_j}, \mathcal{G}_{N_j})
            - H_{\pi^{(2)}, i}(t_{N_j}| x_{N_j}, \mathcal{G}_{N_j})
        }_\infty
        \leq C_\pi \bar{s}_n^H \norm{\pi^{(1)} - \pi^{(2)}}_\pi.
    \end{aligned}
    \end{equation*}


    \item $H_{\hat\mu,i}$ and $H_{\hat\pi,i}$ converge to $H_{\bar\mu,i}$ and $H_{\bar\pi,i}$, respectively, satisfying
    \begin{equation*}
    \begin{aligned}
        \sup_{(i, j, t_{N_j}): \Delta_j(t_{N_j}) \leq \lambda} \sqrt{ P[ \abs{
            H_{\hat\mu, i}(t_{N_j}, X_{N_j}, \mathcal{G}_{N_j})
            - H_{\bar\mu, i}(t_{N_j}, X_{N_j}, \mathcal{G}_{N_j})
        }^2 ]}
        & = O_p\left(
            \bar{r}^{\circ}_n \cdot \bar{r}^H_n
        \right), \\
        \sup_{(i, j, t_{N_j}): \Delta_j(t_{N_j}) \leq \lambda} 2^{\abs{N_j}} \sqrt{ P[ \abs{
            H_{\hat\pi, i}(t_{N_j}, X_{N_j}, \mathcal{G}_{N_j})
            - H_{\bar\pi, i}(t_{N_j}, X_{N_j}, \mathcal{G}_{N_j})
        }^2 ]}
        & = O_p\left(
            \bar{s}^{\circ}_n \cdot \bar{s}^H_n
        \right).
    \end{aligned}
    \end{equation*}
    
    \item The linear approximation errors satisfy
    \begin{equation*}
    \begin{aligned}
        \sup_{(i, t_{N_i}): \Delta_i(t_{N_i}) \leq \lambda} \sqrt{ P[\abs{
            \epsilon_{\bar\mu}(t_{N_i}, X_{N_i}, \mathcal{G}_{N_i})
        }^2]} 
        & = o_p(\bar{r}^\circ_n), \\
        \sup_{(i, t_{N_i}): \Delta_i(t_{N_i}) \leq \lambda} 2^{\abs{N_i}} \sqrt{ P[ \abs{ 
            \epsilon_{\bar\pi}(t_{N_i}, X_{N_i}, \mathcal{G}_{N_i})
        }^2]} 
        & = o_p(\bar{s}^\circ_n).
    \end{aligned}
    \end{equation*}

    \item 
    \begin{equation*}
        \sqrt{n K_{\max}} \bar{r}_n^H = O(\bar{r}_n^\circ)
        ~\text{and}~
        \sqrt{n K_{\max}} \bar{s}_n^H = O(\bar{s}_n^\circ).
    \end{equation*}
\end{enumerate}
\end{assumption}


\begin{assumption}[Linear Approximation for the  Covariate Distribution Estimation Error] \label{assmp:H_covariate}
For $q_n$ and $\mathcal{F}$ given in \cref{assmp:covariate_dist}, we assume the estimation error of the covariate distribution has a linear approximation as in \cref{eq:H_covariate} such that:
\begin{enumerate}[(i)]


    \item For any $(f_1, \dots, f_n) \in \mathcal{F}$,
    \begin{equation*}
        \sum_{j=1}^n \Exp[H_{P_{X|\mathcal{G}},i}(f_j)] = 0.
    \end{equation*}

    \item For each $i$, $H_{P_{X|\mathcal{G}},i}$ is uniformly bounded, 
    \begin{equation*}
        \sup_{(f_1,\dots,f_n) \in \mathcal{F}} \norm*{
           \tsum_{j=1}^n H_{P_{X|\mathcal{G}},i}(f_j)
        }_\infty 
        = O\left(q_n^H\right),
    \end{equation*}

    \item $H_{\hat{P}_{X|\mathcal{G}},i}$ converges to $H_{{P}_{X|\mathcal{G}},i}$, satisfying
    \begin{equation*}
    \begin{aligned}
        \sup_{i, (f_1,\dots,f_n) \in \mathcal{F}} \sqrt{ P[ \abs{
            \tsum_{j=1}^n H_{\hat{P}_{X|\mathcal{G}}, i}(f_j)
            - \tsum_{j=1}^n H_{{P}_{X|\mathcal{G}}, i}(f_j)
        }^2 ]}
        & = O_p\left(
            q_n^H
        \right).
    \end{aligned}
    \end{equation*}

    \item $\hat{P}_{X_{[n]}|\mathcal{G}}$ has a uniform linear approximation on $\mathcal{F}$:
    \begin{equation*}
        \sup_{(f_1, \dots, f_n) \in \mathcal{F}} \sqrt{P\left[\abs*{ \tsum_{j=1}^n \epsilon_{P_{X|\mathcal{G}}}(f_j) }^2 \right]} = o_p(q_n),
    \end{equation*}
    

    \item 
    \begin{equation*}
        \sqrt{n K_{\max}} q_n^H = O(q_n).
    \end{equation*}
\end{enumerate}
\end{assumption}

The uniform boundedness assumption in \cref{assmp:H_covariate}(ii) can also be justified under relatively mild conditions. For instance, in the case of the empirical product covariate distribution estimator $\hat{P}^\otimes_{X_{[n]}|\mathcal{G}}$, we have
\begin{equation*}
H_{P_{X|\mathcal{G}},i}(f_j) = \frac{s}{n} \sum_{i=1}^n f_j^{(1)}(X_i),
\end{equation*}
which scales as $o(n^{-1/2})$ for uniformly bounded $\mathcal{F}$, provided $K_{\max} = o(\sqrt{n})$. Under regularity conditions on the linear approximations for the outcome regression, propensity score, and covariate distribution estimation errors, we establish the consistency of the proposed variance estimator $\hat\sigma_n^2$.

Overall, these regularity conditions are moderate and hold in our simulation study.

\subsubsection{Decomposition of Estimation Error}

In a similar fashion with \cref{app:pf_consistency}, the variance component is decomposed into 
\begin{equation*}
    \hat{\theta}_{i^*}(t^*_{N_{i^*}}) - \tilde\theta_{i^*}(t^*_{N_{i^*}})
    = \mathcal{\tilde{R}}_{n,1} + \mathcal{\tilde{R}}_{n,2} + \mathcal{\tilde{R}}_{n,3}
    + \mathcal{\tilde{R}}_{n,4}
    + \mathcal{\tilde{R}}_{n,5}
    ,
\end{equation*}
where
\begin{equation*}
    \mathcal{\tilde{R}}_{n,1} = \hat{D}^{-1} P_n\left[ 
        \sum_{i=1}^n \kappa_\lambda(\Delta_i) \xi_i(\bar\mu, \bar\pi)
    \right]
    - {D}^{-1} P\left[ 
        \sum_{i=1}^n \kappa_\lambda(\Delta_i) \xi_i(\bar\mu, \bar\pi)
    \right],
\end{equation*}
\begin{equation*}
    \mathcal{\tilde{R}}_{n,2} = \hat{D}^{-1} P \left[
        \sum_{i=1}^n \kappa_\lambda(\Delta_i) (\xi_i(\hat\mu, \hat\pi) - \xi_i(\bar\mu, \bar\pi))
    \right],
\end{equation*}
\begin{equation*}
    \mathcal{\tilde{R}}_{n,3} = \hat{D}^{-1} P \left[
        \sum_{i=1}^n \kappa_\lambda(\Delta_i) (\hat\xi_i(\bar\mu, \bar\pi) - \xi_i(\bar\mu, \bar\pi))
    \right],
\end{equation*}
\begin{equation*}
    \mathcal{\tilde{R}}_{n,4} = \hat{D}^{-1} (P_n - P) \left[
        \sum_{i=1}^n \kappa_\lambda(\Delta_i) (\hat\xi_i(\hat\mu, \hat\pi) - \xi_i(\bar\mu, \bar\pi))
    \right],
\end{equation*}
\begin{equation*}
    \mathcal{\tilde{R}}_{n,5} = \hat{D}^{-1} P \left[
        \sum_{i=1}^n \kappa_\lambda(\Delta_i) \{(\hat\xi_i(\hat\mu, \hat\pi) - \xi_i(\hat\mu, \hat\pi))
        - (\hat\xi_i(\bar\mu, \bar\pi) - \xi_i(\bar\mu, \bar\pi))\} 
    \right].
\end{equation*}
For brevity, we use $\mu_i$, $\pi_i$, $m_i$ and $\varpi_i$ to indicate the terms in $\xi_i$: $\mu(T_{N_i}, X_{N_i}, \mathcal{G}_{N_i})$, $\pi(T_{N_i}| X_{N_i}, \mathcal{G}_{N_i})$, $m(T_{N_i}, \mathcal{G})$ and $\varpi(T_{N_i}| \mathcal{G})$.

The proof consists of two parts: (i) a linear approximation of the estimation error and (ii) the convergence of the empirical sandwich variance estimator based on the established linear approximation. Throughout the proof in this section, we use $\lesssim$ to indicate inequality with a constant factor which might be dependent only on $M_\pi$, $M_\mu$ and $M_\xi$ given by the uniform boundedness assumption (\cref{assmp:uniform_boundedness}).

\subsubsection{Linear Approximation of the Estimation Error}
\label{app:pf_sandwich_linear_approx}

\paragraph{Linear Approximation of $\tilde{\mathcal{R}}_{n,1}$.}

The term $\tilde{\mathcal{R}}_{n,1}$ represents the error of the oracle estimate $\bar\theta_{i^*}(t^*_{N_{i^*}}) \equiv \hat{D}^{-1} \sum_{i=1}^n \kappa_\lambda(\Delta_i) \xi_i(\bar\mu, \bar\pi)$:
\begin{equation*}
    \tilde{\mathcal{R}}_{n,1} = {D}^{-1} \sum_{i=1}^n 
    \kappa_\lambda(\Delta_i) \left\{
        \xi_i(\bar\mu, \bar\pi) - \tilde\theta_{i^*}(t^*_{N_{i^*}}) 
    \right\}
    + \tilde{\epsilon}_{n,1},
\end{equation*}
where $\tilde\theta_{i^*}(t^*_{N_{i^*}}) \equiv {D}^{-1} \Exp\left[ \left.
    \sum_{i=1}^n \kappa_\lambda(\Delta_i) \xi_i(\bar\mu, \bar\pi)
\right| \mathcal{G} \right]$. 
The residual $\tilde\epsilon_{n,1}$ is the second-order error term: 
\begin{equation*}
\begin{aligned}
    \tilde\epsilon_{n,1} 
    & = (\hat{D}^{-1} - D^{-1}) \sum_{i=1}^n 
    \kappa_\lambda(\Delta_i) \left\{
        \xi_i(\bar\mu, \bar\pi) - \tilde\theta_{i^*}(t^*_{N_{i^*}}) 
    \right\}.
\end{aligned}
\end{equation*}
Due to \cref{thm:consistency_D} and our observation in \cref{app:pf_consistency_R_n1} that $\frac{1}{D} \sum_{i=1}^n \kappa_\lambda(\Delta_i) \{
    \xi_i(\bar\mu, \bar\pi) - \tilde\theta_{i^*}(t^*_{N_{i^*}}) 
\} = O_p(\sqrt{\frac{K_{\max}}{D}})$,
\begin{equation*}
    \tilde \epsilon_{n,1} = o_p\left(\sqrt{\frac{K_{\max}}{D}}\right).
\end{equation*}

\paragraph{Linear Approximation of $\tilde{\mathcal{R}}_{n,2}$.}

By the same argument as in \cref{app:bounding_R_n3}, 
\begin{equation*}
    D^{-1} \abs*{P \left[
        \sum_{i=1}^n \kappa_\lambda(\Delta_i) (\xi_i(\hat\mu, \hat\pi) - \xi_i(\bar\mu, \bar\pi))
    \right]}
    = O_p(r_n^\circ \cdot s_n^\circ).
\end{equation*}
We note that if $\bar\mu = \mu$ or $\bar\pi = \pi$, then $r_n^\circ \cdot s_n^\circ \leq r_n^\circ \cdot \bar{s}_n^\circ + \bar{r}_n^\circ \cdot s_n^\circ$.
Because $\frac{\hat{D}}{D} - 1 = O_p(\sqrt{\frac{K_{\max}}{D}}) = o_p(1)$,
\begin{equation*}
    \tilde{\mathcal R}_{n,2} = D^{-1} P \left[
        \sum_{i=1}^n \kappa_\lambda(\Delta_i) (\xi_i(\hat\mu, \hat\pi) - \xi_i(\bar\mu, \bar\pi))
    \right] + o_p(r_n^\circ \cdot \bar{s}_n^\circ + \bar{r}_n^\circ \cdot s_n^\circ).
\end{equation*}
Next, the term $\tilde{\mathcal{R}}_{n,2}$ is linearly approximated by
\begin{equation*}
\begin{aligned}
    \tilde{\mathcal R}_{n,2} & = {D}^{-1} 
    \sum_{j=1}^n  
    P
    \left[ \begin{aligned}
        & \kappa_\lambda(\Delta_j) \frac{\partial \xi_j}{\partial \mu_j}(Y_j, T_{N_j}, X_{N_j}; \bar\mu, \bar\pi)
        \sum_{i=1}^n {H}_{\bar\mu, i}(T_{N_j}, X_{N_j}, \mathcal{G}_{N_j}) \\
        & + \kappa_\lambda(\Delta_j) \frac{\partial \xi_j}{\partial m_j}(Y_j, T_{N_j}, X_{N_j}; \bar\mu, \bar\pi)
        \sum_{i=1}^n \int {H}_{\bar\mu, i}(T_{N_j}, x_{N_j}, \mathcal{G}_{N_j}) d{P}_{X_{N_j}|\mathcal{G}}(x_{N_j}) \\
        & + \kappa_\lambda(\Delta_j) \frac{\partial \xi_j}{\partial \pi_j}(Y_j, T_{N_j}, X_{N_j}; \bar\mu, \bar\pi)
        \sum_{i=1}^n {H}_{\bar\pi, i}(T_{N_j}| X_{N_j}, \mathcal{G}_{N_j}) \\
        & + \kappa_\lambda(\Delta_j) \frac{\partial \xi_j}{\partial \varpi_j}(Y_j, T_{N_j}, X_{N_j}; \bar\mu, \bar\pi)
        \sum_{i=1}^n \int {H}_{\bar\pi, i}(T_{N_j}| x_{N_j}, \mathcal{G}_{N_j}) d{P}_{X_{N_j}|\mathcal{G}}(x_{N_j}) \\
        & + \kappa_\lambda(\Delta_j) \tilde\epsilon_{n,2,j}
    \end{aligned}
    \right] \\
    & \quad + o_p(r_n^\circ \cdot \bar{s}_n^\circ + \bar{r}_n^\circ \cdot s_n^\circ).
\end{aligned}
\end{equation*} 
The residual $\tilde\epsilon_{n,2,j}$ represents the second-order error: 
By the mean value theorem
\begin{equation*}
\begin{aligned}
    \tilde\epsilon_{n,2,j} 
    & = \frac{\partial \xi_j}{\partial \mu_j}(Y_j, T_{N_j}, X_{N_j}; \bar\mu, \bar\pi) 
    ~\epsilon_{\bar\mu}(T_{N_j}, X_{N_j}, \mathcal{G}_{N_j}) \\
    & \quad + \frac{\partial \xi_j}{\partial m_j} (Y_j, T_{N_j}, X_{N_j}; \bar\mu, \bar\pi)
    \int \epsilon_{\bar\mu}(T_{N_j}, x_{N_j}, \mathcal{G}_{N_j}) dP_{X_{N_j}|\mathcal{G}}(x_{N_j}) \\
    & \quad + \frac{\partial \xi_j}{\partial \pi_i} (Y_j, T_{N_j}, X_{N_j}; \bar\mu, \bar\pi)
    ~\epsilon_{\bar\pi}(T_{N_j}| X_{N_j}, \mathcal{G}_{N_j}) \\
    & \quad + \frac{\partial \xi_j}{\partial \varpi_j} (Y_j, T_{N_j}, X_{N_j}; \bar\mu, \bar\pi) \int \epsilon_{\bar\pi}(T_{N_j}| x_{N_j}, \mathcal{G}_{N_j}) dP_{X_{N_j}|\mathcal{G}}(x_{N_j}) \\
    & \quad + \inner*{
        \nabla^2 \xi_j, \{(\hat\mu_j, \hat\pi_j, \hat{m}_j, \hat\varpi_j) - (\bar\mu_j, \bar\pi_j, \bar{m}_j, \bar\varpi_j)\}^{\otimes 2}
    },
\end{aligned}
\end{equation*}
where $\nabla^2 \xi_j$ is the Hessian matrix with respect to $(\mu_j, \pi_j, m_j, \varpi_j)$, and the second derivatives are evaluated at some $(\mu'_j, \pi'_j, m'_j, \varpi'_j) \in \{\alpha \cdot (\bar\mu_j, \bar\pi_j, \bar{m}_j, \bar\varpi_j) + (1-\alpha) \cdot (\hat\mu_j, \hat\pi_j, \hat{m}_j, \hat\varpi_j): \alpha \in [0, 1]\}$. 

Because $\frac{\partial \xi_j}{\partial m_j}(Y_j, T_{N_j}, X_{N_j}; \bar\mu, \bar\pi) = 1$ and $\frac{\partial \xi_j}{\partial \mu_j}(Y_j, T_{N_j}, X_{N_j}; \bar\mu, \bar\pi) = - \frac{\bar\varpi(T_{N_j}, \mathcal{G}_{N_j})}{\bar\pi(T_{N_j}, X_{N_j}, \mathcal{G}_{N_j})}$, given $T_{N_j} = t_{N_j}$ and $\Delta_j(t_{N_j}) \leq \lambda$
\begin{equation*}
\begin{aligned}
    & \abs*{P_{Y_j, X_{N_j} | T_{N_j}}\left[ \left. \begin{aligned} 
        &\frac{\partial \xi_j}{\partial \mu_j}(Y_j, T_{N_j}, X_{N_j}; \bar\mu, \bar\pi) 
        ~\epsilon_{\bar\mu}(T_{N_j}, X_{N_j}, \mathcal{G}_{N_j}) \\
        & + \frac{\partial \xi_j}{\partial m_j} (Y_j, T_{N_j}, X_{N_j}; \bar\mu, \bar\pi)
        \int \epsilon_{\bar\mu}(T_{N_j}, x_{N_j}, \mathcal{G}_{N_j}) dP_{X_{N_j}|\mathcal{G}}(x_{N_j}) \\
    \end{aligned} \right| t_{N_j} \right] } \\
    & = \abs*{ P_{Y_j, X_{N_j} | T_{N_j}}\left[ \left. \begin{aligned} 
        - \frac{\bar\varpi(T_{N_j}, \mathcal{G}_{N_j})}{\bar\pi(T_{N_j}, X_{N_j}, \mathcal{G}_{N_j})} \epsilon_{\bar\mu}(T_{N_j}, X_{N_j}, \mathcal{G}_{N_j}) 
        + \int \epsilon_{\bar\mu}(T_{N_j}, x_{N_j}, \mathcal{G}_{N_j}) dP_{X_{N_j}|\mathcal{G}}(x_{N_j}) \\
    \end{aligned} \right| t_{N_j} \right] } \\
    & = \abs*{ P\left[ \begin{aligned} 
        - \frac{\bar\varpi(t_{N_j}, \mathcal{G}_{N_j})}{\bar\pi(t_{N_j}, X_{N_j}, \mathcal{G}_{N_j})} 
        \frac{\pi(t_{N_j}, X_{N_j}, \mathcal{G}_{N_j})}{\varpi(t_{N_j}, \mathcal{G}_{N_j})}
        \epsilon_{\bar\mu}(t_{N_j}, X_{N_j},  \mathcal{G}_{N_j}) 
        + \epsilon_{\bar\mu}(t_{N_j}, X _{N_j}, \mathcal{G}_{N_j})\\
    \end{aligned} \right] } \\
    & \leq M_\pi^2 \sqrt{{P\left[ \abs*{ \begin{aligned} 
        - \frac{\bar\varpi(t_{N_j}, \mathcal{G}_{N_j})}{\bar\pi(t_{N_j}, X_{N_j}, \mathcal{G}_{N_j})}
        + \frac{\varpi(t_{N_j}, \mathcal{G}_{N_j})}{\pi(t_{N_j}, X_{N_j}, \mathcal{G}_{N_j})} \\
    \end{aligned} }^2 \right]}
    {P\left[
        \abs*{\epsilon_{\bar\mu}(t_{N_j}, x_{N_j}, \mathcal{G}_{N_j})}^2
    \right]}} \\
    & = o_p(\bar{r}_n^\circ \cdot s_n^\circ).
\end{aligned}
\end{equation*}
Similarly, with respect to $\pi$, given $T_{N_j} = t_{N_j}$ and $\Delta_j(t_{N_j}) \leq \lambda$
\begin{equation*}
\begin{aligned}
    & \abs*{P_{Y_j, X_{N_j} | T_{N_j}}\left[ \left. \begin{aligned} 
        &\frac{\partial \xi_j}{\partial \pi_j}(Y_j, T_{N_j}, X_{N_j}; \bar\mu, \bar\pi) 
        ~\epsilon_{\bar\pi}(T_{N_j}, X_{N_j}, \mathcal{G}_{N_j}) \\
        & + \frac{\partial \xi_j}{\partial \varpi_j} (Y_j, T_{N_j}, X_{N_j}; \bar\mu, \bar\pi)
        \int \epsilon_{\bar\pi}(T_{N_j}, x_{N_j}, \mathcal{G}_{N_j}) dP_{X_{N_j}|\mathcal{G}}(x_{N_j}) \\
    \end{aligned} \right| t_{N_j} \right] } \\
    & = \abs*{ P_{Y_j, X_{N_j} | T_{N_j}}\left[ \left. \begin{aligned} 
        & - (Y_j - \bar\mu(T_{N_j}, X_{N_j}, \mathcal{G}_{N_j})) \frac{\bar\varpi(T_{N_j}, \mathcal{G}_{N_j})}{\bar\pi(T_{N_j}, X_{N_j}, \mathcal{G}_{N_j})^2} \epsilon_{\bar\pi}(T_{N_j}, X_{N_j}, \mathcal{G}_{N_j}) \\
        & + \frac{Y_j - \bar\mu(T_{N_j}, X_{N_j}, \mathcal{G}_{N_j})}{\bar\pi(T_{N_j}, X_{N_j}, \mathcal{G}_{N_j})} \int \epsilon_{\bar\pi}(T_{N_j}, x_{N_j}, \mathcal{G}_{N_j}) dP_{X_{N_j}|\mathcal{G}}(x_{N_j}) \\
    \end{aligned} \right| t_{N_j} \right] } \\
    & \leq M_\pi^3 \abs*{ P\left[ \begin{aligned} 
        (\mu - \bar\mu)(T_{N_j}, X_{N_j}, \mathcal{G}_{N_j}) 
        2^{\abs{N_j}} \epsilon_{\bar\pi}(T_{N_j}, X_{N_j}, \mathcal{G}_{N_j})\\
    \end{aligned} \right] } \\
    & \leq 2^{\abs{N_j}} \sqrt{{P\left[ \abs*{ (\mu - \bar\mu)(T_{N_j}, X_{N_j}, \mathcal{G}_{N_j}) }^2 \right]}
    {P\left[
        \abs*{\epsilon_{\bar\pi}(T_{N_j}, X_{N_j}, \mathcal{G}_{N_j})}^2
    \right]}} \\
    & = o_p(r_n^\circ \cdot \bar{s}_n^\circ).
\end{aligned}
\end{equation*} 
Moreover due to the uniform boundedness assumption (\cref{assmp:uniform_boundedness}),
\begin{equation*}
\begin{aligned}
    & \abs{ P_{Y_j| T_{N_j}, X_{N_j} } \inner*{
        \nabla^2 \xi_j, \{(\hat\mu_j, \hat\pi_j, \hat{m}_j, \hat\varpi_j) - (\bar\mu_j, \bar\pi_j, \bar{m}_j, \bar\varpi_j)\}^{\otimes 2}
    } } \\
    & = 2 \abs*{ 
        \left(
            \frac{\varpi'_j}{\pi'^2_j} (\hat\pi_j - \bar\pi_j)
            - \frac{1}{\pi'_j} (\hat\varpi_j - \bar\varpi_j)
        \right) (\hat\mu_j - \bar\mu_j)
        + \frac{\mu_j - \mu'_j}{\pi'^2_j} \left(
            \frac{\varpi'_j}{\pi'_j} (\hat\pi_j - \bar\pi_j)
            - (\hat\varpi_j - \bar\varpi_j)
        \right) (\hat\pi_j - \bar\pi_j)
    } \\
    & = O_p( \bar{r}_n^\circ \bar{s}_n^\circ + r_n^\circ \bar{s}_n^{\circ 2} ).\\
\end{aligned}
\end{equation*}
Hence by the linear approximation of the nuisance parameter estimation error (\cref{assmp:H_nuisance}(vi)),
\begin{equation*}
    {D}^{-1} \sum_{j=1}^n P[\kappa_\lambda(\Delta_j) \tilde\epsilon_{n,2,j}]
    = O_p(\bar{r}_n^\circ \bar{s}_n^\circ) + o_p(\bar{r}^\circ_n \cdot s^\circ_n + r^\circ_n \cdot \bar{s}^\circ_n).
\end{equation*}

In sum,
\begin{equation*}
\begin{aligned}
    \tilde{\mathcal{R}}_{n,2} 
    & = D^{-1} 
    \sum_{j=1}^n 
    P \left[ \begin{aligned}
        & \kappa_\lambda(\Delta_j) \frac{\partial \xi_j}{\partial \mu_j}(Y_j, T_{N_j}, X_{N_j}; \bar\mu, \bar\pi)
        \sum_{i=1}^n {H}_{\bar\mu, i}(T_{N_j}, X_{N_j}, \mathcal{G}_{N_j}) \\
        & + \kappa_\lambda(\Delta_j) \frac{\partial \xi_j}{\partial m_j}(Y_j, T_{N_j}, X_{N_j}; \bar\mu, \bar\pi)
        \sum_{i=1}^n \int {H}_{\bar\mu, i}(T_{N_j}, x_{N_j}, \mathcal{G}_{N_j}) d{P}_{X_{N_j}|\mathcal{G}}(x_{N_j}) \\
        & + \kappa_\lambda(\Delta_j) \frac{\partial \xi_j}{\partial \pi_j}(Y_j, T_{N_j}, X_{N_j}; \bar\mu, \bar\pi)
        \sum_{i=1}^n {H}_{\bar\pi, i}(T_{N_j}| X_{N_j}, \mathcal{G}_{N_j}) \\
        & + \kappa_\lambda(\Delta_j) \frac{\partial \xi_j}{\partial \varpi_j}(Y_j, T_{N_j}, X_{N_j}; \bar\mu, \bar\pi)
        \sum_{i=1}^n \int {H}_{\bar\pi, i}(T_{N_j}| x_{N_j}, \mathcal{G}_{N_j}) d{P}_{X_{N_j}|\mathcal{G}}(x_{N_j})
    \end{aligned} \right] \\
    & \quad + O_p(\bar{r}_n^\circ \bar{s}_n^\circ) + o_p(\bar{r}^\circ_n \cdot s^\circ_n + r^\circ_n \cdot \bar{s}^\circ_n).
\end{aligned}
\end{equation*}

\paragraph{Linear Approximation of $\tilde{\mathcal{R}}_{n,3}$.}

Again by the same argument as in \cref{app:bounding_R_n3}, 
\begin{equation*}
    D^{-1} \abs*{P \left[
        \sum_{i=1}^n \kappa_\lambda(\Delta_i) (\hat\xi_i(\bar\mu, \bar\pi) - \xi_i(\bar\mu, \bar\pi))
    \right]}
    = O_p(q_n).
\end{equation*}
Because $\frac{\hat{D}}{D} - 1 = O_p(\sqrt{\frac{K_{\max}}{D}}) = o_p(1)$,
\begin{equation*}
    \tilde{\mathcal R}_{n,3} = D^{-1} P \left[
        \sum_{i=1}^n \kappa_\lambda(\Delta_i) (\xi_i(\hat\mu, \hat\pi) - \xi_i(\bar\mu, \bar\pi))
    \right] + o_p(q_n).
\end{equation*}
Then by \cref{assmp:H_covariate}, the term $\tilde{\mathcal{R}}_{n,3}$ is linearly approximated by
\begin{equation*}
\begin{aligned}
    \tilde{\mathcal R}_{n,3} & = {D}^{-1} 
    \sum_{j=1}^n  
    P
    \left[ \begin{aligned}
        & \kappa_\lambda(\Delta_j) \frac{\partial \xi_j}{\partial m_j}(Y_j, T_{N_j}, X_{N_j}; \bar\mu, \bar\pi)
        \sum_{i=1}^n {H}_{P_{X|\mathcal{G}}, i}(\bar\mu(T_{N_j}, \cdot, \mathcal{G}_{N_j})) \\
        & + \kappa_\lambda(\Delta_j) \frac{\partial \xi_j}{\partial \varpi_j}(Y_j, T_{N_j}, X_{N_j}; \bar\mu, \bar\pi)
        \sum_{i=1}^n {H}_{P_{X|\mathcal{G}}, i}(\bar\pi(T_{N_j}| \cdot, \mathcal{G}_{N_j})) \\
        & + \kappa_\lambda(\Delta_j) \tilde\epsilon_{n,3,j}
    \end{aligned}
    \right]. \\
    & \quad + o_p(q_n),
\end{aligned}
\end{equation*} 
where
\begin{equation*}
\begin{aligned}
    \tilde\epsilon_{n,3,j} 
    & = \frac{\partial \xi_j}{\partial m_j}(Y_j, T_{N_j}, X_{N_j}; \bar\mu, \bar\pi) 
    ~\epsilon_{P_{X|\mathcal{G}}}(\bar\mu(T_{N_j}, \cdot, \mathcal{G}_{N_j})) \\
    & \quad + \frac{\partial \xi_j}{\partial \varpi_j}(Y_j, T_{N_j}, X_{N_j}; \bar\mu, \bar\pi) 
    ~\epsilon_{P_{X|\mathcal{G}}}(\bar\pi(T_{N_j}, \cdot, \mathcal{G}_{N_j})).  \\
\end{aligned}
\end{equation*}
Due to the uniform boundedness assumption (\cref{assmp:uniform_boundedness}), $\frac{\partial \xi_i}{\partial m_i}$ and $\frac{\partial \xi_i}{\partial \varpi_i}$ is uniformly bounded by some constant $M^{(1)}_\xi$, and the remainder satisfies \begin{equation*}
    {D}^{-1} \sum_{j=1}^n P[\kappa_\lambda(\Delta_j) \tilde\epsilon_{n,3,j}]
    = o_p(q_n),
\end{equation*}
following \cref{assmp:H_covariate}(iv). 

In sum,
\begin{equation*}
\begin{aligned}
    \tilde{\mathcal{R}}_{n,3} 
    & = D^{-1} 
    \sum_{j=1}^n 
    P \left[ \begin{aligned}
        & \kappa_\lambda(\Delta_j) \frac{\partial \xi_j}{\partial m_j}(Y_j, T_{N_j}, X_{N_j}; \bar\mu, \bar\pi)
        \sum_{i=1}^n {H}_{P_{X|\mathcal{G}}, i}(\bar\mu(T_{N_j}, \cdot, \mathcal{G}_{N_j})) \\
        & + \kappa_\lambda(\Delta_j) \frac{\partial \xi_j}{\partial \varpi_j}(Y_j, T_{N_j}, X_{N_j}; \bar\mu, \bar\pi)
        \sum_{i=1}^n {H}_{P_{X|\mathcal{G}}, i}(\bar\pi(T_{N_j}| \cdot, \mathcal{G}_{N_j})) \\
    \end{aligned} \right] \\
    & \quad + o_p(q_n).
\end{aligned}
\end{equation*}

\paragraph{Vanishing $\tilde{\mathcal{R}}_{n,4}$ and $\tilde{\mathcal{R}}_{n,5}$.}

For the rest of the terms, $\tilde{\mathcal{R}}_{n,4} = o_p(\sqrt{\frac{K_{\max}}{D}})$ due to the empirical process theory result (\cref{thm:asymptotic_equicontinuity}), and $\tilde{\mathcal{R}}_{n,5} = o_p(q_n)$ due to the equicontinuity of the covariate distribution estimate $\hat{P}_{X_{[n]}|\mathcal{G}}$ (\cref{assmp:covariate_dist}). See \cref{app:pf_consistency} for the details.

In sum, the estimation error is linearly approximated by 
\begin{equation*}
\begin{aligned}
    & \hat{\theta}_{i^*}(t^*_{N_{i^*}}) - \theta_{i^*}(t^*_{N_{i^*}}) 
    = D^{-1} \sum_{i=1}^n {W}_i + o_p(\bar\sigma_n), 
\end{aligned}
\end{equation*}
as long as $\bar{r}_n^\circ \bar{s}_n^\circ = o(\bar\sigma_n)$ and $\max\{\sqrt{\frac{K_{\max}}{D}}, \bar{r}^\circ_n s^\circ_n, r^\circ_n \bar{s}^\circ_n, q_n\} = O(\bar\sigma_n)$.
That is, $\Var[D^{-1} \sum_{i=1}^n {W}_i] / \bar\sigma_n^2 \overset{p}{\rightarrow} 1$.

\subsubsection{Asymptotic Normality}
\label{app:pf_sandwich_asymptotic_normality}

It is sufficient to show the weak convergence of $D^{-1} \sum_{i=1}^n {W}_i$ to a Gaussian distribution.
For the brevity of the rest of the proof, we denote $\kappa_\lambda(\Delta_i) \left\{
    \xi_i(\bar\mu, \bar\pi) - \tilde\theta_{i^*}(t^*_{N_{i^*}}) 
\right\}$ by ${W}_{\theta,i}$, and let ${W}_{\nu,i}$ indicate the nuisance estimate linear approximation parts so that ${W}_i = {W}_{\theta, i} + {W}_{\nu, i}$.

Due to unbiasedness of $H_{\mu,i}$, $H_{\pi,i}$ and $H_{P_{X|\mathcal{G}}}$ (\cref{assmp:H_nuisance}(ii) and \cref{assmp:H_covariate}(i)), $\Exp[{W}_i] = \Exp[W_{\theta,i}]$. Due to the uniform boundedness of $\mu$, $\pi$, $m$ and $\varpi$ (\cref{assmp:uniform_boundedness}), 
for some $M_\xi > 0$
\begin{equation*}
\begin{aligned}
    \Exp[\abs{{W}_{\theta,i} - \Exp{W}_{\theta,i}}^3 | \mathcal{G}] 
    & \lesssim M_\xi^3 \left\{
        \Exp[\kappa_\lambda(\Delta_j)^3| \mathcal{G}] + \Exp[\kappa_\lambda(\Delta_j)| \mathcal{G}]^3
    \right\}
    < \infty, ~\text{and}
\end{aligned}
\end{equation*}
\begin{equation*}
\begin{aligned}
    \Exp[\abs{{W}_{\theta,i} - \Exp{W}_{\theta,i}}^4 | \mathcal{G}] 
    & \lesssim M_\xi^4 \left\{
        \Exp[\kappa_\lambda(\Delta_j)^4| \mathcal{G}] + \Exp[\kappa_\lambda(\Delta_j)| \mathcal{G}]^4
    \right\} < \infty. 
\end{aligned}
\end{equation*}

For $W_{\nu,i}$, because $\frac{\partial \xi_j}{\partial m_j}(Y_j, T_{N_j}, X_{N_j}; \bar\mu, \bar\pi) = 1$ and $\frac{\partial \xi_j}{\partial \mu_j}(Y_j, T_{N_j}, X_{N_j}; \bar\mu, \bar\pi) = - \frac{\bar\varpi(T_{N_j}, \mathcal{G}_{N_j})}{\bar\pi(T_{N_j}, X_{N_j}, \mathcal{G}_{N_j})}$, given $T_{N_j} = t_{N_j}$ and $\Delta_j(t_{N_j}) \leq \lambda$
\begin{equation*}
\begin{aligned}
    & \abs*{P_{Y_j, X_{N_j} | T_{N_j}}\left[ \left. \begin{aligned} 
        &\frac{\partial \xi_j}{\partial \mu_j}(Y_j, T_{N_j}, X_{N_j}; \bar\mu, \bar\pi) 
        ~H_{\bar\mu,i}(T_{N_j}, X_{N_j}, \mathcal{G}_{N_j}) \\
        & + \frac{\partial \xi_j}{\partial m_j} (Y_j, T_{N_j}, X_{N_j}; \bar\mu, \bar\pi)
        \int H_{\bar\mu,i}(T_{N_j}, x_{N_j}, \mathcal{G}_{N_j}) dP_{X_{N_j}|\mathcal{G}}(x_{N_j}) \\
    \end{aligned} \right| t_{N_j} \right] } \\
    & = \abs*{ P_{Y_j, X_{N_j} | T_{N_j}}\left[ \left. \begin{aligned} 
        - \frac{\bar\varpi(T_{N_j}, \mathcal{G}_{N_j})}{\bar\pi(T_{N_j}, X_{N_j}, \mathcal{G}_{N_j})} H_{\bar\mu,i}(T_{N_j}, X_{N_j}, \mathcal{G}_{N_j}) 
        + \int H_{\bar\mu,i}(T_{N_j}, x_{N_j}, \mathcal{G}_{N_j}) dP_{X_{N_j}|\mathcal{G}}(x_{N_j}) \\
    \end{aligned} \right| t_{N_j} \right] } \\
    & = \abs*{ P\left[ \begin{aligned} 
        - \frac{\bar\varpi(t_{N_j}, \mathcal{G}_{N_j})}{\bar\pi(t_{N_j}, X_{N_j}, \mathcal{G}_{N_j})} 
        \frac{\pi(t_{N_j}, X_{N_j}, \mathcal{G}_{N_j})}{\varpi(t_{N_j}, \mathcal{G}_{N_j})}
        H_{\bar\mu,i}(t_{N_j}, X_{N_j}, \mathcal{G}_{N_j}) 
        + H_{\bar\mu,i}(t_{N_j}, X _{N_j}, \mathcal{G}_{N_j})\\
    \end{aligned} \right] } \\
    & \leq M_\pi^2 \sqrt{{P\left[ \abs*{ \begin{aligned} 
        - \frac{\bar\varpi(t_{N_j}, \mathcal{G}_{N_j})}{\bar\pi(t_{N_j}, X_{N_j}, \mathcal{G}_{N_j})}
        + \frac{\varpi(t_{N_j}, \mathcal{G}_{N_j})}{\pi(t_{N_j}, X_{N_j}, \mathcal{G}_{N_j})} \\
    \end{aligned} }^2 \right]}
    {P\left[
        \abs*{H_{\bar\mu,i}(t_{N_j}, x_{N_j}, \mathcal{G}_{N_j})}^2
    \right]}} \\
    & = O_p(\bar{r}_n^H \cdot s_n^\circ).
\end{aligned}
\end{equation*}
Similarly, with respect to $\pi$, given $T_{N_j} = t_{N_j}$ and $\Delta_j(t_{N_j}) \leq \lambda$
\begin{equation*}
\begin{aligned}
    & \abs*{P_{Y_j, X_{N_j} | T_{N_j}}\left[ \left. \begin{aligned} 
        &\frac{\partial \xi_j}{\partial \pi_j}(Y_j, T_{N_j}, X_{N_j}; \bar\mu, \bar\pi) 
        ~H_{\bar\pi,i}(T_{N_j}, X_{N_j}, \mathcal{G}_{N_j}) \\
        & + \frac{\partial \xi_j}{\partial \varpi_j} (Y_j, T_{N_j}, X_{N_j}; \bar\mu, \bar\pi)
        \int H_{\bar\pi,i}(T_{N_j}, x_{N_j}, \mathcal{G}_{N_j}) dP_{X_{N_j}|\mathcal{G}}(x_{N_j}) \\
    \end{aligned} \right| t_{N_j} \right] } \\
    & = \abs*{ P_{Y_j, X_{N_j} | T_{N_j}}\left[ \left. \begin{aligned} 
        & - (Y_j - \bar\mu(T_{N_j}, X_{N_j}, \mathcal{G}_{N_j})) \frac{\bar\varpi(T_{N_j}, \mathcal{G}_{N_j})}{\bar\pi(T_{N_j}, X_{N_j}, \mathcal{G}_{N_j})^2} H_{\bar\pi,i}(T_{N_j}, X_{N_j}, \mathcal{G}_{N_j}) \\
        & + \frac{Y_j - \bar\mu(T_{N_j}, X_{N_j}, \mathcal{G}_{N_j})}{\bar\pi(T_{N_j}, X_{N_j}, \mathcal{G}_{N_j})} \int H_{\bar\pi,i}(T_{N_j}, x_{N_j}, \mathcal{G}_{N_j}) dP_{X_{N_j}|\mathcal{G}}(x_{N_j}) \\
    \end{aligned} \right| t_{N_j} \right] } \\
    & \leq M_\pi^3 \abs*{ P\left[ \begin{aligned} 
        (\mu - \bar\mu)(t_{N_j}, X_{N_j}, \mathcal{G}_{N_j}) 
        2^{\abs{N_j}} H_{\bar\pi,i}(t_{N_j}, X_{N_j}, \mathcal{G}_{N_j})\\
    \end{aligned} \right] } \\
    & \leq 2^{\abs{N_j}} \sqrt{{P\left[ \abs*{ (\mu - \bar\mu)(t_{N_j}, X_{N_j}, \mathcal{G}_{N_j}) }^2 \right]}
    {P\left[
        \abs*{H_{\bar\pi,i}(t_{N_j}, X_{N_j}, \mathcal{G}_{N_j})}^2
    \right]}} \\
    & = O_p(r_n^\circ \cdot \bar{s}_n^H).
\end{aligned}
\end{equation*} 
Moreover, due to the uniform boundedness of $H_{P_{X|\mathcal{G}}, i}$ ($H_{P_{X|\mathcal{G}},i}$ (\cref{assmp:H_nuisance}(iii) and \cref{assmp:H_covariate}(ii)), 
\begin{equation} \label{eq:W_nu_i_bound}
    W_{\nu,i} = O_p \left(\sum_{j=1}^n P[\kappa_\lambda(\Delta_j)] (\bar{r}_n^H s_n^\circ + r_n^\circ \bar{s}_n^H + q_n^H)\right)
    = O_p\left(D (\bar{r}_n^H s_n^\circ + r_n^\circ \bar{s}_n^H + q_n^H)\right).
\end{equation}
Since $W_{\nu,i}$ is uniformly bounded, the above $O_p$ argument translates to the following moment bounds:
\begin{equation*}
\begin{aligned}
    \Exp[\abs{{W}_{\nu,i}}^3 | \mathcal{G}] 
    & \lesssim \{D (\bar{r}_n^H s_n^\circ + r_n^\circ \bar{s}_n^H + q_n^H)\}^3
    < \infty, \\
    \Exp[\abs{{W}_{\nu,i}}^4 | \mathcal{G}] 
    & \lesssim \{D (\bar{r}_n^H s_n^\circ + r_n^\circ \bar{s}_n^H + q_n^H)\}^4 < \infty. 
\end{aligned}
\end{equation*}
Because $H_{\bar\mu,i}$, $H_{\bar\pi,i}$ and $H_{P_{X|\mathcal{G}},i}$ are dependnet only through $(Y_i, T_{N_i}, X_{N_i})$, $\{{W}_{\nu,i}: i = 1,\dots, n\}$ has the same network dependency as $\{{W}_{\theta,i}: i = 1,\dots, n\}$.
Hence
\begin{equation*}
\begin{aligned}
    & \frac{K_{\max}^2}{D^3 \sigma_n^3} \sum_{i=1}^n \Exp [\abs{{W}_i}^3 | \mathcal{G}] 
    + \sqrt{\frac{28}{\pi}} \frac{K_{\max}^{3/2}}{D^2 \bar\sigma_n^2} \sqrt{\sum_{i=1}^n \Exp[{W}_i^4 | \mathcal{G}]} \\
    & \lesssim \frac{K_{\max}^2}{D^3 \bar\sigma_n^3} 
    \sum_{i=1}^n \left\{
        \Exp[\kappa_\lambda(\Delta_j)^3| \mathcal{G}] + \Exp[\kappa_\lambda(\Delta_j)| \mathcal{G}]^3
    \right\} 
    + \frac{K_{\max}^{3/2}}{D^2 \bar\sigma_n^2}
    \sqrt{\sum_{i=1}^n \left\{
        \Exp[\kappa_\lambda(\Delta_j)^4| \mathcal{G}] + \Exp[\kappa_\lambda(\Delta_j)| \mathcal{G}]^4
    \right\}} \\
    & \quad + \frac{K_{\max}^2}{D^3 \bar\sigma_n^3} 
    \sum_{i=1}^n \{D (\bar{r}_n^H s_n^\circ + r_n^\circ \bar{s}_n^H + q_n^H)\}^3
    + \sqrt{\frac{28}{\pi}} \frac{K_{\max}^{3/2}}{D^2 \bar\sigma_n^2}
    \sqrt{\sum_{i=1}^n \{D (\bar{r}_n^H s_n^\circ + r_n^\circ \bar{s}_n^H + q_n^H)\}^4}\\
    & \lesssim \frac{K_{\max}^2}{D^2 \bar\sigma_n^3} 
    + \frac{K_{\max}^{3/2}}{D^{3/2} \bar\sigma_n^2} 
    + \frac{K_{\max}^2}{\bar\sigma_n^3} 
    n (\bar{r}_n^H s_n^\circ + r_n^\circ \bar{s}_n^H + q_n^H)^3
    + \frac{K_{\max}^{3/2}}{\bar\sigma_n^2} 
    \sqrt{n} (\bar{r}_n^H s_n^\circ + r_n^\circ \bar{s}_n^H + q_n^H)^2 \\
    & \rightarrow 0,
\end{aligned}
\end{equation*}
as long as $\max\{\left(\frac{K_{\max}}{D}\right)^{3/4}, K_{\max}^{2/3} n^{1/3} (\bar{r}_n^H s_n^\circ + r_n^\circ \bar{s}_n^H + q_n^H)\} = o(\bar\sigma_n)$, which is satisfied by $\max\{ \sqrt{K_{\max}/D}, \bar{r}^\circ_n s^\circ_n, r^\circ_n \bar{s}^\circ_n, q_n \} = O( \bar\sigma_n )$, and $K_{\max} = O(\sqrt{n})$.
That is, the right hand side in the Eq. (3.8) of \citet{ross2011fundamentals} vanishes in probability. By Theorem 3.6 of \citet{ross2011fundamentals}, $D^{-1} \sum_i {W}_i$ satisfies the desired asymptotic normality.

\subsubsection{Consistency of the HAC Estimator}
\label{app:pf_sandwich_variance_consistency}

We conclude our proof by showing the convergence of the proposed HAC estimator $\hat{\sigma}_n^2$ to $\bar\sigma_n^2 \equiv \Var[D^{-1} \sum_{i=1}^n {W}_i]$. 

\paragraph{Convergence of the oracle HAC estimator.}

We first show that the oracle HAC estimator based on ${W}_i$ converges in probability to $\bar\sigma_n^2$. Because ${W}_i$'s are dependent only on $(T_{N_i}, X_{N_i}, \mathcal{G}_{N_i})$, they have network dependence according to the dependency neighborhoods $\tilde{N}_i$. We define the oracle HAC estimator as follows:
\begin{equation} \label{eq:oracle_HAC_estimator}
    \check\sigma_n^2 \equiv \frac{1}{D^2} \sum_{i,j \in [n]} h(i,j)
    {W}_i {W}_j,
\end{equation}
where $h(i,j) \equiv \mathbf{1}\{j \in \tilde{N}_i\}$.

We first consider the part of the variance due to ${W}_{\theta,i}$:
\begin{equation*}
    \frac{1}{D^2} \tsum_{i,j \in [n]} h(i,j) 
    ({W}_{\theta,i} - \Exp{W}_{\theta,i})
    ({W}_{\theta,j} - \Exp{W}_{\theta,j}).
\end{equation*}
Let's denote ${W}_{\theta,i} - \Exp{W}_{\theta,i}$ by $Z_{\theta,i}$. Because $\abs{Z_{\theta,i}} \leq 2 \kappa_\lambda(\Delta_i) M_\xi$ almost surely due to the uniform boundedness (\cref{assmp:uniform_boundedness}),
\begin{equation*}
\begin{aligned}
    & \Var\left[
        \frac{1}{D^2} \tsum_{i,j \in [n]} h(i,j) 
        ({W}_{\theta,i} - \Exp{W}_{\theta,i})
        ({W}_{\theta,j} - \Exp{W}_{\theta,j})
    \right] \\
    & = \frac{1}{D^4} \tsum_{i,j,k,l \in [n]} h(i,j,k,l) \Exp\left[
        (Z_{\theta,i} Z_{\theta,j} - \Exp Z_{\theta,i} Z_{\theta,j})(Z_{\theta,k} Z_l - \Exp Z_{\theta,k} Z_{\theta,l})
    \right] \\
    & \lesssim \frac{1}{D^4} \tsum_{i,j,k,l \in [n]} h(i,j,k,l) \norm{Z_{\theta,i} Z_{\theta,j} - \Exp Z_{\theta,i} Z_{\theta,j}}_2^2 \\
    & \lesssim \frac{1}{D^4} \tsum_{i,j,k,l \in [n]} h(i,j,k,l) 
    \{\norm{Z_{\theta,i} Z_{\theta,j}}_2^2 + (\Exp Z_{\theta,i} Z_{\theta,j})^2\} \\
    & \overset{\mathrm{(i)}}{\lesssim} \frac{1}{D^4} \tsum_{i,j,k,l \in [n]} h(i,j,k,l) \Exp[\kappa_\lambda(\Delta_i)] \\
    & \lesssim \frac{1}{D^4} D K_{\max}^3 = \frac{K_{\max}^3}{D^3} 
    \overset{\mathrm{(ii)}}{=} o_p(\sigma_n^4)
\end{aligned}
\end{equation*}
where $\mathrm{(i)}$ follows from that $\kappa_\lambda(\Delta_i), \kappa_\lambda(\Delta_j) \leq 1$ uniformly, and $\mathrm{(ii)}$ follows from that $\sqrt{\frac{K_{\max}}{D}} = O(\sigma_n)$.
Similarly, because $W_{\nu,i} = O_p\left(D (\bar{r}_n^H s_n^\circ + r_n^\circ \bar{s}_n^H + q_n^H)\right)$ as in \cref{eq:W_nu_i_bound},
\begin{equation*}
\begin{aligned}
    & \Var\left[
        \frac{1}{D^2} \tsum_{i,j \in [n]} h(i,j) 
        ({W}_{\theta,i} - \Exp{W}_{\theta,i})
        ({W}_{\nu,j} - \Exp{W}_{\nu,j})
    \right] \\
    & = \frac{1}{D^4} \tsum_{i,j,k,l \in [n]} h(i,j,k,l) \Exp\left[
        (Z_{\theta,i} Z_{\nu,j} - \Exp Z_{\theta,i} Z_{\nu,j})(Z_{\theta,k} Z_{\nu,l} - \Exp Z_{\theta,k} Z_{\nu,l})
    \right] \\
    & \lesssim \frac{1}{D^4} \tsum_{i,j,k,l \in [n]} h(i,j,k,l) \norm{Z_{\theta,i} Z_{\nu,j} - \Exp Z_{\theta,i} Z_{\nu,j}}_2^2 \\
    & \lesssim \frac{1}{D^4} \tsum_{i,j,k,l \in [n]} h(i,j,k,l) 
    \{\norm{Z_{\theta,i} Z_{\nu,j}}_2^2 + (\Exp Z_{\theta,i} Z_{\nu,j})^2\} \\
    & \overset{\mathrm{(i)}}{\lesssim} \frac{1}{D^4} \tsum_{i,j,k,l \in [n]} h(i,j,k,l) \Exp[\kappa_\lambda(\Delta_i)] D^2 (\bar{r}_n^H s_n^\circ + r_n^\circ \bar{s}_n^H + q_n^H)^2 \\
    & \lesssim \frac{1}{D^2} D K_{\max}^3 (\bar{r}_n^H s_n^\circ + r_n^\circ \bar{s}_n^H + q_n^H)^2 = \frac{K_{\max}^3}{D} (\bar{r}_n^H s_n^\circ + r_n^\circ \bar{s}_n^H + q_n^H)^2
    \overset{\mathrm{(ii)}}{=} o_p(\bar\sigma_n^4),
\end{aligned}
\end{equation*}
where $\mathrm{(i)}$ follows from that $\kappa_\lambda(\Delta_i), \kappa_\lambda(\Delta_j) \leq 1$ uniformly, and $\mathrm{(ii)}$ follows from 
that $$\max\{ \sqrt{K_{\max}/D}, \bar{r}^\circ_n s^\circ_n, r^\circ_n \bar{s}^\circ_n, q_n \} = O( \bar\sigma_n )$$ 
and that $K_{\max} = O(\sqrt{n})$.
Last, 
\begin{equation*}
\begin{aligned}
    & \Var\left[
        \frac{1}{D^2} \tsum_{i,j \in [n]} h(i,j) 
        ({W}_{\nu,i} - \Exp{W}_{\nu,i})
        ({W}_{\nu,j} - \Exp{W}_{\nu,j})
    \right] \\
    & = \frac{1}{D^4} \tsum_{i,j,k,l \in [n]} h(i,j,k,l) \Exp\left[
        (Z_{\nu,i} Z_{\nu,j} - \Exp Z_{\nu,i} Z_{\nu,j})(Z_{\nu,k} Z_{\nu,l} - \Exp Z_{\nu,k} Z_{\nu,l})
    \right] \\
    & \lesssim \frac{1}{D^4} \tsum_{i,j,k,l \in [n]} h(i,j,k,l) \norm{Z_{\nu,i} Z_{\nu,j} - \Exp Z_{\nu,i} Z_{\nu,j}}_2^2 \\
    & \lesssim \frac{1}{D^4} \tsum_{i,j,k,l \in [n]} h(i,j,k,l) 
    \{\norm{Z_{\nu,i} Z_{\nu,j}}_2^2 + (\Exp Z_{\nu,i} Z_{\nu,j})^2\} \\
    & \overset{\mathrm{(i)}}{\lesssim} \tsum_{i,j,k,l \in [n]} h(i,j,k,l) (\bar{r}_n^H s_n^\circ + r_n^\circ \bar{s}_n^H + q_n^H)^4 \\
    & \lesssim n K_{\max}^3 (\bar{r}_n^H s_n^\circ + r_n^\circ \bar{s}_n^H + q_n^H)^4
    \overset{\mathrm{(ii)}}{=} o(\bar\sigma_n^4),
\end{aligned}
\end{equation*}
where $\mathrm{(i)}$ follows from that $\kappa_\lambda(\Delta_i), \kappa_\lambda(\Delta_j) \leq 1$ uniformly; and $\mathrm{(ii)}$ follows from that $\sqrt{n K_{\max}}(\bar{r}_n^H, \bar{s}_n^H, q_n^H) = O(\bar{r}_n^\circ, \bar{s}_n^\circ, q_n^\circ)$, that $\max\{ \sqrt{K_{\max}/D}, \bar{r}^\circ_n s^\circ_n, r^\circ_n \bar{s}^\circ_n, q_n \} = O( \bar\sigma_n )$ 
and that $K_{\max} = O(\sqrt{n})$.
Hence by Chebyshev's inequality $\frac{1}{D^2} \sum_{i,j \in [n]} h(i,j) 
\{{W}_i - \Exp{W}_i\} \{{W}_j - \Exp{W}_j\}$ converges in probability to $\bar\sigma_n^2$ with $o_p(\bar\sigma_n^2)$ error.

We note that ${W}_i$ in \cref{eq:oracle_HAC_estimator} is not centered. The resulting estimator converges to 
$$
    \bar\sigma_n^2 + D^{-2} 
    \sum_{i,j \in [n]} h(i,j) 
    \Exp[{W}_i]
    \Exp[{W}_j].
$$ 
Hence if $H_n \equiv (h(i,j))_{i,j} \in \reals^{n \times n}$ is positive semidefinite, the oracle HAC estimator gives a conservative measure of KECENI's uncertainty. Due to unbiasedness of $H_{\mu,i}$, $H_{\pi,i}$ and $H_{P_{X|\mathcal{G}}}$ (\cref{assmp:H_nuisance}(ii) and \cref{assmp:H_covariate}(i)), $\Exp[{W}_i] = \Exp[\kappa_\lambda(\Delta_j) \{
    \xi_i(\bar\mu, \bar\pi) - \tilde\theta_{i^*}(t^*_{N_{i^*}}) 
\}]$. Under the double robustness condition (i.e., $\bar\mu = \mu$ or $\bar\pi = \pi$), $\Exp[\xi_i(\bar\mu, \bar\pi)|T_{N_i}] = \theta_{i}(T_{N_i})$. In this case due to the smoothness assumption (\cref{assmp:smoothness}),
$\abs{\Exp[{W}_i]} = \abs{\Exp[\kappa_\lambda(\Delta_j) \{
    \theta_i(t^*_{N_{i^*}} - \tilde\theta_{i^*}(t^*_{N_{i^*}}) 
\}]} = O(\Exp[\kappa_\lambda(\Delta_j)] \lambda)$; i.e., the scale of the conservatism is $O(\frac{K_{\max}}{D}\lambda)$. So if $\frac{K_{\max}}{D} \lambda = o(\bar\sigma_n)$, then $(\check\sigma_n - \bar\sigma_n)/\bar\sigma_n \overset{p}{\rightarrow} 0$.

\paragraph{Empirical linear approximation error.}

Now it is sufficient to show that $(\hat\sigma_n - \check\sigma_n)/\sigma_n \overset{p}{\rightarrow} 0$. Let $\hat{W}_{\theta,i}$ and $\hat{W}_{\nu,i}$ indicate the empirical counterparts of ${W}_{\nu,i}$ and ${W}_{\nu,i}$, respectively, so that $\hat{W}_i \equiv \hat{W}_i = \hat{W}_{\theta, i} + \hat{W}_{\nu, i}$. Let $\Delta_{\theta,i} \equiv \hat{W}_{\theta,i} - {W}_{\theta,i}$ and $\Delta_{\nu,i} \equiv \hat{W}_{\nu,i} - {W}_{\nu,i}$.

We first bound $\hat{W}_{\nu,i} - P[\hat{W}_{\nu,i}]$. 
For brevity, we omit $Y_j$, $T_{N_j}$, $X_{N_j}$ and $\mathcal{G}_{N_j}$ unless they are otherwise needed for clarity.
We decompose the difference by
\begin{equation} \label{eq:empirical_oracle_nuisance_decomp}
\begin{aligned}
    & \sum_{j=1}^n 
    \kappa_\lambda(\Delta_j)
    \frac{\partial \hat\xi_j}{\partial \mu_j}(\hat\mu, \hat\pi)
    {H}_{\hat\mu, i} 
    - P\left[ \sum_{j=1}^n 
        \kappa_\lambda(\Delta_j)
        \frac{\partial \hat\xi_j}{\partial \mu_j}(\hat\mu, \hat\pi)
        {H}_{\hat\mu, i}
    \right] \\
    & = \left( \begin{aligned}
        & \sum_{j=1}^n 
        \kappa_\lambda(\Delta_j)
        \frac{\partial \xi_j}{\partial \mu_j}(\bar\mu, \bar\pi)
        {H}_{\bar\mu, i} 
        - P\left[ \sum_{j=1}^n 
            \kappa_\lambda(\Delta_j)
            \frac{\partial \xi_j}{\partial \mu_j}(\bar\mu, \bar\pi)
            {H}_{\bar\mu, i}
        \right]
    \end{aligned} \right) \\
    & \quad + \left( \begin{aligned}
        (P_n - P)\left[ \sum_{j=1}^n 
        \kappa_\lambda(\Delta_j)
        \frac{\partial \hat\xi_j}{\partial \mu_j}(\hat\mu, \hat\pi)
        {H}_{\hat\mu, i}
        - \sum_{j=1}^n 
            \kappa_\lambda(\Delta_j)
            \frac{\partial \xi_j}{\partial \mu_j}(\bar\mu, \bar\pi)
            {H}_{\bar\mu, i}
        \right]
    \end{aligned} \right). \\
\end{aligned}
\end{equation}
For the first term, due to the uniform boundedness assumption in \cref{assmp:H_nuisance}(iii),
\begin{equation*}
\begin{aligned}
    & \sup_i \norm*{\begin{aligned}
        & 
        \kappa_\lambda(\Delta_j) 
        \frac{\partial \xi_j}{\partial \mu_j}(Y_j, T_{N_j}, X_{N_j}; \bar\mu, \bar\pi)
        {H}_{\bar\mu, i}(T_{N_j}, X_{N_j}, \mathcal{G}_{N_j}) \\
        & - P \left[ 
            \kappa_\lambda(\Delta_j) 
            \frac{\partial \xi_j}{\partial \mu_j}(Y_j, T_{N_j}, X_{N_j}; \bar\mu, \bar\pi)
            {H}_{\bar\mu, i}(T_{N_j}, X_{N_j}, \mathcal{G}_{N_j})
        \right] 
    \end{aligned}}_2 \\
    & \leq \sup_i \norm*{
        \kappa_\lambda(\Delta_j) 
        \frac{\partial \xi_j}{\partial \mu_j}(Y_j, T_{N_j}, X_{N_j}; \bar\mu, \bar\pi)
        {H}_{\bar\mu, i}(T_{N_j}, X_{N_j}, \mathcal{G}_{N_j})
    }_2 \\
    & \quad + \sup_i \norm*{
        P \left[ 
        \kappa_\lambda(\Delta_j) 
        \frac{\partial \xi_j}{\partial \mu_j}(Y_j, T_{N_j}, X_{N_j}; \bar\mu, \bar\pi)
        {H}_{\bar\mu, i}(T_{N_j}, X_{N_j}, \mathcal{G}_{N_j})
    \right]}_2 \\
    & = O \left(
        \norm{\kappa_\lambda(\Delta_j)}_2
        \bar{r}^H_n
    \right).
\end{aligned}
\end{equation*}
Following a similar argument with bounding $\tilde{\mathcal R}_{n,2}$ in \cref{app:pf_sandwich_linear_approx},
\begin{equation*}
\begin{aligned}
    & \sup_i \norm*{\begin{aligned}
        & \sum_{j \notin \tilde{N}_i} \kappa_\lambda(\Delta_j) 
        \frac{\partial \xi_j}{\partial \mu_j}(Y_j, T_{N_j}, X_{N_j}; \bar\mu, \bar\pi)
        {H}_{\bar\mu, i}(T_{N_j}, X_{N_j}, \mathcal{G}_{N_j}) \\
        & - \sum_{j \notin \tilde{N}_i} P \left[ 
            \kappa_\lambda(\Delta_j) 
            \frac{\partial \xi_j}{\partial \mu_j}(Y_j, T_{N_j}, X_{N_j}; \bar\mu, \bar\pi)
            {H}_{\bar\mu, i}(T_{N_j}, X_{N_j}, \mathcal{G}_{N_j})
        \right] 
    \end{aligned}}_2 \\
    & = O\left( 
        \sqrt{K_{\max} D} ~ \bar{r}^H_n
    \right), \\
\end{aligned}
\end{equation*}
and
\begin{equation*}
\begin{aligned}
    & \sup_i \norm*{\begin{aligned}
        & 
        \sum_{j \in \tilde{N}_i} \kappa_\lambda(\Delta_j) 
        \frac{\partial \xi_j}{\partial \mu_j}(Y_j, T_{N_j}, X_{N_j}; \bar\mu, \bar\pi)
        {H}_{\bar\mu, i}(T_{N_j}, X_{N_j}, \mathcal{G}_{N_j}) \\
        & - \sum_{j \in \tilde{N}_i} P \left[ 
            \kappa_\lambda(\Delta_j) 
            \frac{\partial \xi_j}{\partial \mu_j}(Y_j, T_{N_j}, X_{N_j}; \bar\mu, \bar\pi)
            {H}_{\bar\mu, i}(T_{N_j}, X_{N_j}, \mathcal{G}_{N_j})
        \right] 
    \end{aligned}}_2 \\
    & = O\left( 
        K_{\max} ~\bar{r}^H_n
    \right). \\
\end{aligned}
\end{equation*}
In sum, given $K_{\max} = o(D)$, the first term of \cref{eq:empirical_oracle_nuisance_decomp} satisfies
\begin{equation*}
\begin{aligned}
    & \sup_i \norm*{\begin{aligned}
        & \sum_{j=1}^n \kappa_\lambda(\Delta_j) 
        \frac{\partial \xi_j}{\partial \mu_j}(Y_j, T_{N_j}, X_{N_j}; \bar\mu, \bar\pi)
        {H}_{\bar\mu, i}(T_{N_j}, X_{N_j}, \mathcal{G}_{N_j}) \\
        & - \sum_{j=1}^n P \left[ 
            \kappa_\lambda(\Delta_j) 
            \frac{\partial \xi_j}{\partial \mu_j}(Y_j, T_{N_j}, X_{N_j}; \bar\mu, \bar\pi)
            {H}_{\bar\mu, i}(T_{N_j}, X_{N_j}, \mathcal{G}_{N_j})
        \right] 
    \end{aligned}}_2 \\
    & = O\left( 
        \sqrt{K_{\max} D} ~ \bar{r}_n^H
    \right). \\
\end{aligned}
\end{equation*}
For the second term of \cref{eq:empirical_oracle_nuisance_decomp}, we use the empirical process theory we developed in \cref{sec:discussion}. First we note that due to the uniform boundedness (\cref{assmp:uniform_boundedness}(iii)) and the regularity conditions on $H_{\bar\mu, i}$ (\cref{assmp:H_nuisance}), we obtain
\begin{equation*}
\begin{aligned}
    & \sup_{i,j} \norm*{
        \frac{\partial \hat\xi_j}{\partial \mu_j}(\hat\mu, \hat\pi)
        {H}_{\hat\mu, i}
        - \frac{\partial \xi_j}{\partial \mu_j}(\bar\mu, \bar\pi)
        {H}_{\bar\mu, i}
    }_\infty \\
    & = O_p \left( \begin{aligned}
        \bar{r}_n^H (\norm{\hat\mu - \bar\mu}_\infty + \norm{\hat\pi - \bar\pi}_\pi + \norm{\hat{m} - \bar{m}}_\infty + \norm{\hat\varpi - \bar\varpi}_\pi)
        + \norm*{H_{\hat\mu,i} -  H_{\bar\mu,i}}_\infty
    \end{aligned} \right) \\
    & = O_p \left( \begin{aligned}
        \bar{r}_n^H (\norm{\hat\mu - \bar\mu}_\infty + \norm{\hat\pi - \bar\pi}_\pi + \norm{\hat{m} - \bar{m}}_\infty + \norm{\hat\varpi - \bar\varpi}_\pi)
    \end{aligned} \right).
\end{aligned}
\end{equation*}
Then \cref{thm:asymptotic_equicontinuity} provides an stochastic equicontinuity of the kernel weighted empirical process, which also implies
\begin{equation*}
\begin{aligned}
    & \sup_i \norm*{ \begin{aligned}
        & P_n\left[ \sum_{j=1}^n 
        \kappa_\lambda(\Delta_j)
        \frac{\partial \hat\xi_j}{\partial \mu_j}(\hat\mu, \hat\pi)
        {H}_{\hat\mu, i}
        - \sum_{j=1}^n 
            \kappa_\lambda(\Delta_j)
            \frac{\partial \xi_j}{\partial \mu_j}(\bar\mu, \bar\pi)
            {H}_{\bar\mu, i}
        \right] \\
        & - P \left[\sum_{j=1}^n 
        \kappa_\lambda(\Delta_j)
        \frac{\partial \hat\xi_j}{\partial \mu_j}(\hat\mu, \hat\pi)
        {H}_{\hat\mu, i}
        - \sum_{j=1}^n 
            \kappa_\lambda(\Delta_j)
            \frac{\partial \xi_j}{\partial \mu_j}(\bar\mu, \bar\pi)
            {H}_{\bar\mu, i}
        \right] \\
    \end{aligned} }_2 
    = o\left(
        \sqrt{K_{\max} D} ~\bar{r}_n^H
    \right).
\end{aligned}
\end{equation*}
Summing the upper bounds for the two terms, we obtain
\begin{equation*}
\begin{aligned}
    \sup_i \norm*{\sum_{j=1}^n 
    \kappa_\lambda(\Delta_j)
    \frac{\partial \hat\xi_j}{\partial \mu_j}(\hat\mu, \hat\pi)
    {H}_{\hat\mu, i} 
    - P\left[ \sum_{j=1}^n 
        \kappa_\lambda(\Delta_j)
        \frac{\partial \hat\xi_j}{\partial \mu_j}(\hat\mu, \hat\pi)
        {H}_{\hat\mu, i}
    \right]}_2 
    = O\left(\sqrt{K_{\max} D} ~\bar{r}_n^H\right).
\end{aligned}
\end{equation*}
Similar results apply to the other nuisance estimate linear approximation terms, giving us
\begin{equation*}
    \sup_i \norm{\hat{W}_{\nu,i} - P[\hat{W}_{\nu,i}]}_2
    = O\left(\sqrt{K_{\max} D} ~(\bar{r}_n^H + \bar{s}_n^H + q_n^H)\right).
\end{equation*}

Now let's bound $P[\hat{W}_i] - W_i$. We consider the terms including $H_{\mu,i}$:
\begin{equation*}
\begin{aligned}
    & \sup_i \norm*{
        \sum_{j=1}^n P \left[
            \kappa_\lambda(\Delta_j)
            \left( \begin{aligned}
                & \frac{\partial \hat\xi_j}{\partial \mu_j}(\hat\mu, \hat\pi)
                {H}_{\hat\mu, i}
                - \frac{\partial \xi_j}{\partial \mu_j}(\bar\mu, \bar\pi)
                {H}_{\bar\mu, i} \\
                & + \frac{\partial \hat\xi_j}{\partial m_j}(\hat\mu, \hat\pi)
                \int {H}_{\hat\mu, i} d\hat{P}_{X_{N_j}|\mathcal{G}}
                - \frac{\partial \xi_j}{\partial m_j}(\bar\mu, \bar\pi)
                \int {H}_{\bar\mu, i} dP_{X_{N_j}|\mathcal{G}}
            \end{aligned} \right)
        \right]
    }_2 \\
    & \leq \sup_i \norm*{
        \sum_{j=1}^n P \left[ \begin{aligned}
            & \kappa_\lambda(\Delta_j)
            \left(
                \frac{\partial \hat\xi_j}{\partial \mu_j}(\hat\mu, \hat\pi)
                - \frac{\partial \xi_j}{\partial \mu_j}(\bar\mu, \bar\pi)
            \right) {H}_{\bar\mu, i} \\
            & + \kappa_\lambda(\Delta_j)
            \left(
                \frac{\partial \hat\xi_j}{\partial m_j}(\hat\mu, \hat\pi)
                - \frac{\partial \xi_j}{\partial m_j}(\bar\mu, \bar\pi)
            \right) \int {H}_{\bar\mu, i} dP_{X_{N_j}|\mathcal{G}}
        \end{aligned} \right]
    }_2 \\
    & \quad + \sup_i \norm*{
        \sum_{j=1}^n P \left[ \begin{aligned}
            & \kappa_\lambda(\Delta_j)
            \frac{\partial \xi_j}{\partial \mu_j}(\bar\mu, \bar\pi)
            ({H}_{\hat\mu, i} - {H}_{\bar\mu, i}) \\
            & + \kappa_\lambda(\Delta_j)
            \frac{\partial \xi_j}{\partial m_j}(\bar\mu, \bar\pi)
            \left(
                \int {H}_{\hat\mu, i} d\hat{P}_{X_{N_j}|\mathcal{G}}
                - \int {H}_{\bar\mu, i} dP_{X_{N_j}|\mathcal{G}}
            \right)
        \end{aligned} \right]
    }_2 \\
    & \quad + \sup_i \norm*{
        \sum_{j=1}^n P \left[ \begin{aligned}
            & \kappa_\lambda(\Delta_j)
            \left(
                \frac{\partial \hat\xi_j}{\partial \mu_j}(\hat\mu, \hat\pi)
                - \frac{\partial \xi_j}{\partial \mu_j}(\bar\mu, \bar\pi)
            \right) ({H}_{\hat\mu, i} - {H}_{\bar\mu, i}) \\
            & + \kappa_\lambda(\Delta_j)
            \left(
                \frac{\partial \hat\xi_j}{\partial m_j}(\hat\mu, \hat\pi)
                - \frac{\partial \xi_j}{\partial m_j}(\bar\mu, \bar\pi)
            \right) \left(
                \int {H}_{\hat\mu, i} d\hat{P}_{X_{N_j}|\mathcal{G}}
                - \int {H}_{\bar\mu, i} dP_{X_{N_j}|\mathcal{G}}
            \right) \\
        \end{aligned} \right]
    }_2. \\
\end{aligned}
\end{equation*}
For the first term, by a similar argument with \cref{eq:W_nu_i_bound}, given $T_{N_j} = t_{N_j}$ and $\Delta_j(t_{N_j}) \leq \lambda$,
\begin{equation*}
\begin{aligned}
    & \abs*{P_{Y_j, X_{N_j} | T_{N_j}} \left[ \begin{aligned}
        \left(
            \frac{\partial \hat\xi_j}{\partial \mu_j}(\hat\mu, \hat\pi)
            - \frac{\partial \xi_j}{\partial \mu_j}(\bar\mu, \bar\pi)
        \right) {H}_{\bar\mu, i}
        +
        \left(
            \frac{\partial \hat\xi_j}{\partial m_j}(\hat\mu, \hat\pi)
            - \frac{\partial \xi_j}{\partial m_j}(\bar\mu, \bar\pi)
        \right) \int {H}_{\bar\mu, i} dP_{X_{N_j}|\mathcal{G}}
    \end{aligned} \right]} \\
    & = \abs*{ P\left[ \begin{aligned} 
        \left(
            \frac{\hat\varpi(t_{N_j}, \mathcal{G}_{N_j})}{\hat\pi(t_{N_j}, X_{N_j}, \mathcal{G}_{N_j})}
            - \frac{\bar\varpi(t_{N_j}, \mathcal{G}_{N_j})}{\bar\pi(t_{N_j}, X_{N_j}, \mathcal{G}_{N_j})} 
        \right)
        \frac{\pi(t_{N_j}, X_{N_j}, \mathcal{G}_{N_j})}{\varpi(t_{N_j}, \mathcal{G}_{N_j})}
        H_{\bar\mu,j}(t_{N_j}, X_{N_j}, \mathcal{G}_{N_j}) \\
    \end{aligned} \right] } \\
    & = O_p((\bar{s}_n^\circ + q_n) \bar{r}_n^H ).
\end{aligned}
\end{equation*}
As a result,
\begin{equation*}
\begin{aligned}
    \sup_i \norm*{
        \sum_{j=1}^n P \left[ \begin{aligned}
            & \kappa_\lambda(\Delta_j)
            \left(
                \frac{\partial \hat\xi_j}{\partial \mu_j}(\hat\mu, \hat\pi)
                - \frac{\partial \xi_j}{\partial \mu_j}(\bar\mu, \bar\pi)
            \right) {H}_{\bar\mu, i} \\
            & + \kappa_\lambda(\Delta_j)
            \left(
                \frac{\partial \hat\xi_j}{\partial m_j}(\hat\mu, \hat\pi)
                - \frac{\partial \xi_j}{\partial m_j}(\bar\mu, \bar\pi)
            \right) \int {H}_{\bar\mu, i} dP_{X_{N_j}|\mathcal{G}}
        \end{aligned} \right]
    }_2
    = O( D (\bar{s}_n^\circ + q_n) \bar{r}_n^H ).
\end{aligned}
\end{equation*}
For the second term, due to the convergence of $H_{\hat\mu}$ to $H_{\bar\mu}$ (\cref{assmp:H_nuisance}), given $T_{N_j} = t_{N_j}$ and $\Delta_j(t_{N_j}) \leq \lambda$,  
\begin{equation*}
\begin{aligned}
    & \abs*{P_{Y_{j}, X_{N_j}| T_{N_j}} \left[ \left.
        \frac{\partial \xi_j}{\partial \mu_j}(\bar\mu, \bar\pi)
        ({H}_{\hat\mu, i} - {H}_{\bar\mu, i}) 
        +
        \frac{\partial \xi_j}{\partial m_j}(\bar\mu, \bar\pi)
        \int ({H}_{\hat\mu, i} - {H}_{\bar\mu, i}) dP_{X_{N_j}|\mathcal{G}}
    \right| t_{N_j} \right]} \\
    & = \abs*{ P_{Y_j, X_{N_j} | T_{N_j}}\left[ \left. \begin{aligned} 
        & - \frac{\bar\varpi(T_{N_j}, \mathcal{G}_{N_j})}{\bar\pi(T_{N_j}, X_{N_j}, \mathcal{G}_{N_j})} (H_{\hat\mu,i} - H_{\bar\mu,i})(T_{N_j}, X_{N_j}, \mathcal{G}_{N_j}) \\
        & + \int (H_{\hat\mu,i} - H_{\bar\mu,i})(T_{N_j}, x_{N_j}, \mathcal{G}_{N_j}) dP_{X_{N_j}|\mathcal{G}}(x_{N_j}) \\
    \end{aligned} \right| t_{N_j} \right] } \\
    & = \abs*{ P\left[ \begin{aligned} 
        & - \frac{\bar\varpi(t_{N_j}, \mathcal{G}_{N_j})}{\bar\pi(t_{N_j}, X_{N_j}, \mathcal{G}_{N_j})} 
        \frac{\pi(t_{N_j}, X_{N_j}, \mathcal{G}_{N_j})}{\varpi(t_{N_j}, \mathcal{G}_{N_j})}
        (H_{\hat\mu,i} - H_{\bar\mu,i})(t_{N_j}, X_{N_j}, \mathcal{G}_{N_j}) \\
        & + (H_{\hat\mu,i} - H_{\bar\mu,i})(t_{N_j}, X _{N_j}, \mathcal{G}_{N_j})\\
    \end{aligned} \right] } \\
    & \leq M_\pi^2 \sqrt{{P\left[ \abs*{ \begin{aligned} 
        - \frac{\bar\varpi(t_{N_j}, \mathcal{G}_{N_j})}{\bar\pi(t_{N_j}, X_{N_j}, \mathcal{G}_{N_j})}
        + \frac{\varpi(t_{N_j}, \mathcal{G}_{N_j})}{\pi(t_{N_j}, X_{N_j}, \mathcal{G}_{N_j})} \\
    \end{aligned} }^2 \right]}
    {P\left[
        \abs*{(H_{\hat\mu,i} - H_{\bar\mu,i})(t_{N_j}, x_{N_j}, \mathcal{G}_{N_j})}^2
    \right]}} \\
    & = O_p(s_n^\circ \bar{r}_n^\circ \bar{r}_n^H).
\end{aligned}
\end{equation*}
At the same time, given $T_{N_j} = t_{N_j}$ and $\Delta_j(t_{N_j}) \leq \lambda$
\begin{equation*}
\begin{aligned}
    \abs*{P_{Y_{j}, X_{N_j}| T_{N_j}} \left[ \left.
        \frac{\partial \xi_j}{\partial m_j}(\bar\mu, \bar\pi)
        \left(
            \int {H}_{\bar\mu, i} d\hat{P}_{X_{N_j}|\mathcal{G}}
            - \int {H}_{\bar\mu, i} dP_{X_{N_j}|\mathcal{G}}
        \right)
    \right| t_{N_j} \right]} 
    = O_p(q_n^\circ \bar{r}_n^H).
\end{aligned}
\end{equation*}
In sum,
\begin{equation*}
\begin{aligned}
    \sup_i \norm*{
        \sum_{j=1}^n P \left[ \begin{aligned}
            & \kappa_\lambda(\Delta_j)
            \frac{\partial \xi_j}{\partial \mu_j}(\bar\mu, \bar\pi)
            ({H}_{\hat\mu, i} - {H}_{\bar\mu, i}) \\
            & + \kappa_\lambda(\Delta_j)
            \frac{\partial \xi_j}{\partial m_j}(\bar\mu, \bar\pi)
            \left(
                \int {H}_{\hat\mu, i} d\hat{P}_{X_{N_j}|\mathcal{G}}
                - \int {H}_{\bar\mu, i} dP_{X_{N_j}|\mathcal{G}}
            \right)
        \end{aligned} \right]
    }_2
    = O(D (s_n^\circ \bar{r}_n^\circ + q_n^\circ) \bar{r}_n^H).
\end{aligned}
\end{equation*}
Similarly, the last term satisfies
\begin{equation*}
\begin{aligned}
    & \abs*{P_{Y_j, X_{N_j} | T_{N_j}} \left[ \begin{aligned}
        &
        \left(
            \frac{\partial \hat\xi_j}{\partial \mu_j}(\hat\mu, \hat\pi)
            - \frac{\partial \xi_j}{\partial \mu_j}(\bar\mu, \bar\pi)
        \right) ({H}_{\hat\mu, i} - {H}_{\bar\mu, i}) \\
        & +
        \left(
            \frac{\partial \hat\xi_j}{\partial m_j}(\hat\mu, \hat\pi)
            - \frac{\partial \xi_j}{\partial m_j}(\bar\mu, \bar\pi)
        \right) \left(
            \int {H}_{\hat\mu, i} d\hat{P}_{X_{N_j}|\mathcal{G}}
            - \int {H}_{\bar\mu, i} dP_{X_{N_j}|\mathcal{G}}
        \right) \\
    \end{aligned} \right]} \\
    & = \abs*{ P\left[ \begin{aligned} 
        \left(
            \frac{\hat\varpi(t_{N_j}, \mathcal{G}_{N_j})}{\hat\pi(t_{N_j}, X_{N_j}, \mathcal{G}_{N_j})}
            - \frac{\bar\varpi(t_{N_j}, \mathcal{G}_{N_j})}{\bar\pi(t_{N_j}, X_{N_j}, \mathcal{G}_{N_j})} 
        \right)
        \frac{\pi(t_{N_j}, X_{N_j}, \mathcal{G}_{N_j})}{\varpi(t_{N_j}, \mathcal{G}_{N_j})}
        (H_{\hat\mu,j} - H_{\bar\mu,j})(t_{N_j}, X_{N_j}, \mathcal{G}_{N_j}) \\
    \end{aligned} \right] } \\
    & = O_p((\bar{s}_n^\circ + q_n) \bar{r}_n^\circ \bar{r}_n^H ),
\end{aligned}
\end{equation*}
given $T_{N_j} = t_{N_j}$ and $\Delta_j(t_{N_j}) \leq \lambda$, and hence
\begin{equation*}
\begin{aligned}
    & \sup_i \norm*{
        \sum_{j=1}^n P \left[ \begin{aligned}
            & \kappa_\lambda(\Delta_j)
            \left(
                \frac{\partial \hat\xi_j}{\partial \mu_j}(\hat\mu, \hat\pi)
                - \frac{\partial \xi_j}{\partial \mu_j}(\bar\mu, \bar\pi)
            \right) ({H}_{\hat\mu, i} - {H}_{\bar\mu, i}) \\
            & + \kappa_\lambda(\Delta_j)
            \left(
                \frac{\partial \hat\xi_j}{\partial m_j}(\hat\mu, \hat\pi)
                - \frac{\partial \xi_j}{\partial m_j}(\bar\mu, \bar\pi)
            \right) \left(
                \int {H}_{\hat\mu, i} d\hat{P}_{X_{N_j}|\mathcal{G}}
                - \int {H}_{\bar\mu, i} dP_{X_{N_j}|\mathcal{G}}
            \right) \\
        \end{aligned} \right]
    }_2 \\
    & = O(D (\bar{s}_n^\circ + q_n) \bar{r}_n^\circ \bar{r}_n^H).
\end{aligned}
\end{equation*}
Summing up the previous results, 
\begin{equation*}
\begin{aligned}
    & \sup_i \norm*{ \begin{aligned}
        & \frac{1}{D} \sum_{j=1}^n P\left[ 
            \kappa_\lambda(\Delta_j)
            \left\{
            \frac{\partial \hat\xi_j}{\partial \mu_j}(\hat\mu, \hat\pi)
            {H}_{\hat\mu, i}
            +
            \frac{\partial \hat\xi_j}{\partial m_j}(\hat\mu, \hat\pi)
            \int {H}_{\hat\mu, i} d\hat{P}_{X_{N_j}|\mathcal{G}}
            \right\}
        \right]\\
        & - \frac{1}{D} \sum_{j=1}^n P\left[ 
            \kappa_\lambda(\Delta_j)
            \left\{
            \frac{\partial \xi_j}{\partial \mu_j}(\bar\mu, \bar\pi)
            {H}_{\bar\mu, i}
            +
            \frac{\partial \xi_j}{\partial m_j}(\bar\mu, \bar\pi)
            \int {H}_{\bar\mu, i} dP_{X_{N_j}|\mathcal{G}}
            \right\}
        \right]
    \end{aligned} }_2 \\
    & = O\left(
        (s_n^\circ \bar{r}_n^\circ + \bar{s}_n^\circ + q_n) \bar{r}_n^H
    \right),
\end{aligned}
\end{equation*}

With respect to $\pi$, we consider the terms including $H_{\pi,i}$:
\begin{equation*}
\begin{aligned}
    & \sup_i \norm*{
        \sum_{j=1}^n P \left[
            \kappa_\lambda(\Delta_j)
            \left( \begin{aligned}
                & \frac{\partial \hat\xi_j}{\partial \pi_j}(\hat\mu, \hat\pi)
                {H}_{\hat\pi, i}
                - \frac{\partial \xi_j}{\partial \pi_j}(\bar\mu, \bar\pi)
                {H}_{\bar\pi, i} \\
                & + \frac{\partial \hat\xi_j}{\partial \varpi_j}(\hat\mu, \hat\pi)
                \int {H}_{\hat\pi, i} d\hat{P}_{X_{N_j}|\mathcal{G}}
                - \frac{\partial \xi_j}{\partial \varpi_j}(\bar\mu, \bar\pi)
                \int {H}_{\bar\pi, i} dP_{X_{N_j}|\mathcal{G}}
            \end{aligned} \right)
        \right]
    }_2 \\
    & \leq \sup_i \norm*{
        \sum_{j=1}^n P \left[ \begin{aligned}
            & \kappa_\lambda(\Delta_j)
            \left(
                \frac{\partial \hat\xi_j}{\partial \pi_j}(\hat\mu, \hat\pi)
                - \frac{\partial \xi_j}{\partial \pi_j}(\bar\mu, \bar\pi)
            \right) {H}_{\bar\pi, i} \\
            & + \kappa_\lambda(\Delta_j)
            \left(
                \frac{\partial \hat\xi_j}{\partial \varpi_j}(\hat\mu, \hat\pi)
                - \frac{\partial \xi_j}{\partial \varpi_j}(\bar\mu, \bar\pi)
            \right) \int {H}_{\bar\pi, i} dP_{X_{N_j}|\mathcal{G}}
        \end{aligned} \right]
    }_2 \\
    & \quad + \sup_i \norm*{
        \sum_{j=1}^n P \left[ \begin{aligned}
            & \kappa_\lambda(\Delta_j)
            \frac{\partial \xi_j}{\partial \pi_j}(\bar\mu, \bar\pi)
            ({H}_{\hat\pi, i} - {H}_{\bar\pi, i}) \\
            & + \kappa_\lambda(\Delta_j)
            \frac{\partial \xi_j}{\partial \varpi_j}(\bar\mu, \bar\pi)
            \left(
                \int {H}_{\hat\pi, i} d\hat{P}_{X_{N_j}|\mathcal{G}}
                - \int {H}_{\bar\pi, i} dP_{X_{N_j}|\mathcal{G}}
            \right)
        \end{aligned} \right]
    }_2 \\
    & \quad + \sup_i \norm*{
        \sum_{j=1}^n P \left[ \begin{aligned}
            & \kappa_\lambda(\Delta_j)
            \left(
                \frac{\partial \hat\xi_j}{\partial \pi_j}(\hat\mu, \hat\pi)
                - \frac{\partial \xi_j}{\partial \pi_j}(\bar\mu, \bar\pi)
            \right) ({H}_{\hat\pi, i} - {H}_{\bar\pi, i}) \\
            & + \kappa_\lambda(\Delta_j)
            \left(
                \frac{\partial \hat\xi_j}{\partial \varpi_j}(\hat\mu, \hat\pi)
                - \frac{\partial \xi_j}{\partial \varpi_j}(\bar\mu, \bar\pi)
            \right) \left(
                \int {H}_{\hat\pi, i} d\hat{P}_{X_{N_j}|\mathcal{G}}
                - \int {H}_{\bar\pi, i} dP_{X_{N_j}|\mathcal{G}}
            \right) \\
        \end{aligned} \right]
    }_2. \\
\end{aligned}
\end{equation*}
For the first term, given $T_{N_j} = t_{N_j}$ and $\Delta_j(t_{N_j}) \leq \lambda$,
\begin{equation*}
\begin{aligned}
    & \abs*{P_{Y_j, X_{N_j} | T_{N_j}} \left[ \left.\begin{aligned}
        \left(
            \frac{\partial \hat\xi_j}{\partial \pi_j}(\hat\mu, \hat\pi)
            - \frac{\partial \xi_j}{\partial \pi_j}(\bar\mu, \bar\pi)
        \right) {H}_{\bar\pi, i}
        +
        \left(
            \frac{\partial \hat\xi_j}{\partial \varpi_j}(\hat\mu, \hat\pi)
            - \frac{\partial \xi_j}{\partial \varpi_j}(\bar\mu, \bar\pi)
        \right) \int {H}_{\bar\pi, i} dP_{X_{N_j}|\mathcal{G}}
    \end{aligned} \right| t_{N_j} \right]} \\
    & = \abs*{ P_{Y_j, X_{N_j} | T_{N_j}}\left[ \left. \begin{aligned} 
        & - (Y_j - \hat\mu(T_{N_j}, X_{N_j}, \mathcal{G}_{N_j})) \frac{\hat\varpi(T_{N_j}, \mathcal{G}_{N_j})}{\hat\pi(T_{N_j}, X_{N_j}, \mathcal{G}_{N_j})^2} H_{\bar\pi,i}(T_{N_j}, X_{N_j}, \mathcal{G}_{N_j}) \\
        & + (Y_j - \bar\mu(T_{N_j}, X_{N_j}, \mathcal{G}_{N_j})) \frac{\bar\varpi(T_{N_j}, \mathcal{G}_{N_j})}{\bar\pi(T_{N_j}, X_{N_j}, \mathcal{G}_{N_j})^2} H_{\bar\pi,i}(T_{N_j}, X_{N_j}, \mathcal{G}_{N_j}) \\
        & + \frac{Y_j - \hat\mu(T_{N_j}, X_{N_j}, \mathcal{G}_{N_j})}{\hat\pi(T_{N_j}, X_{N_j}, \mathcal{G}_{N_j})} \int H_{\bar\pi,i}(T_{N_j}, x_{N_j}, \mathcal{G}_{N_j}) dP_{X_{N_j}|\mathcal{G}}(x_{N_j}) \\
        & - \frac{Y_j - \bar\mu(T_{N_j}, X_{N_j}, \mathcal{G}_{N_j})}{\bar\pi(T_{N_j}, X_{N_j}, \mathcal{G}_{N_j})} \int H_{\bar\pi,i}(T_{N_j}, x_{N_j}, \mathcal{G}_{N_j}) dP_{X_{N_j}|\mathcal{G}}(x_{N_j}) \\
    \end{aligned} \right| t_{N_j} \right] } \\
    & = \abs*{ P_{Y_j, X_{N_j} | T_{N_j}}\left[ \left. \begin{aligned} 
        & - (\bar\mu_j - \hat\mu_j)\frac{\hat\varpi_j}{\hat\pi_j^2} H_{\bar\pi,i} 
        - (\mu_j - \bar\mu_j) \left(
            \frac{\hat\varpi_j}{\hat\pi_j^2}
            - \frac{\bar\varpi_j}{\bar\pi_j^2} 
        \right) H_{\bar\pi,i} \\
        & + (\bar\mu_j - \hat\mu_j) \frac{1}{\hat\pi_j} \int H_{\bar\pi,i} dP_{X_{N_j}|\mathcal{G}} 
        + (\mu_j - \bar\mu_j) \left(\frac{1}{\hat\pi_j} - \frac{1}{\bar\pi_j} \right) \int H_{\bar\pi,i} dP_{X_{N_j}|\mathcal{G}} \\
    \end{aligned} \right| t_{N_j} \right] } \\
    & = O_p((\bar{r}_n^\circ + r_n^\circ \bar{s}_n^\circ + r_n^\circ q_n) \bar{s}_n^H ),
\end{aligned}
\end{equation*}
and hence
\begin{equation*}
\begin{aligned}
    & \sum_{j=1}^n P \left[ \begin{aligned}
        & \kappa_\lambda(\Delta_j)
        \left(
            \frac{\partial \hat\xi_j}{\partial \pi_j}(\hat\mu, \hat\pi)
            - \frac{\partial \xi_j}{\partial \pi_j}(\bar\mu, \bar\pi)
        \right) {H}_{\bar\pi, i} \\
        & + \kappa_\lambda(\Delta_j)
        \left(
            \frac{\partial \hat\xi_j}{\partial \varpi_j}(\hat\mu, \hat\pi)
            - \frac{\partial \xi_j}{\partial \varpi_j}(\bar\mu, \bar\pi)
        \right) \int {H}_{\bar\pi, i} dP_{X_{N_j}|\mathcal{G}}(x_{N_j}) 
    \end{aligned} \right] \\
    & = O_p( D (\bar{r}_n^\circ + r_n^\circ \bar{s}_n^\circ + r_n^\circ q_n) \bar{s}_n^H ).
\end{aligned}
\end{equation*} 
Due to the convergence of $H_{\hat\pi}$ to $H_{\bar\pi}$ (\cref{assmp:H_nuisance}), given $T_{N_j} = t_{N_j}$ and $\Delta_j(t_{N_j}) \leq \lambda$,  
\begin{equation*}
\begin{aligned}
    & \abs*{P_{Y_{j}, X_{N_j}| T_{N_j}} \left[ \left.
        \frac{\partial \xi_j}{\partial \pi_j}(\bar\mu, \bar\pi)
        ({H}_{\hat\pi, i} - {H}_{\bar\pi, i}) 
        +
        \frac{\partial \xi_j}{\partial \varpi_j}(\bar\mu, \bar\pi)
        \int ({H}_{\hat\pi, i} - {H}_{\bar\pi, i}) dP_{X_{N_j}|\mathcal{G}}
    \right| t_{N_j} \right]} \\
    & = \abs*{ P_{Y_j, X_{N_j} | T_{N_j}}\left[ \left. \begin{aligned} 
        & - (Y_j - \bar\mu(T_{N_j}, X_{N_j}, \mathcal{G}_{N_j})) \frac{\bar\varpi(T_{N_j}, \mathcal{G}_{N_j})}{\bar\pi(T_{N_j}, X_{N_j}, \mathcal{G}_{N_j})^2} (H_{\hat\pi,i} - H_{\bar\pi,i})(T_{N_j}, X_{N_j}, \mathcal{G}_{N_j}) \\
        & + \frac{Y_j - \bar\mu(T_{N_j}, X_{N_j}, \mathcal{G}_{N_j})}{\bar\pi(T_{N_j}, X_{N_j}, \mathcal{G}_{N_j})} \int (H_{\hat\pi,i} - H_{\bar\pi,i})(T_{N_j}, x_{N_j}, \mathcal{G}_{N_j}) dP_{X_{N_j}|\mathcal{G}}(x_{N_j}) \\
    \end{aligned} \right| t_{N_j} \right] } \\
    & = \abs*{ P_{Y_j, X_{N_j} | T_{N_j}}\left[ \left. \begin{aligned} 
        - (\mu_j - \bar\mu_j) \frac{\bar\varpi_j}{\bar\pi_j^2} (H_{\hat\pi,i} - H_{\bar\pi,i}) 
        + \frac{\mu_j - \bar\mu_j}{\bar\pi_j} \int (H_{\hat\pi,i} - H_{\bar\pi,i}) dP_{X_{N_j}|\mathcal{G}} \\
    \end{aligned} \right| t_{N_j} \right] } \\
    & = O_p(r_n^\circ \bar{s}_n^\circ \bar{s}_n^H).
\end{aligned}
\end{equation*}
At the same time,
\begin{equation*}
\begin{aligned}
    \abs*{P_{Y_{j}, X_{N_j}| T_{N_j}} \left[ \left.
        \frac{\partial \xi_j}{\partial \varpi_j}(\bar\mu, \bar\pi)
        \left(
            \int {H}_{\bar\pi, i} d\hat{P}_{X_{N_j}|\mathcal{G}}
            - \int {H}_{\bar\pi, i} dP_{X_{N_j}|\mathcal{G}}
        \right)
    \right| t_{N_j} \right]} 
    = O_p(r_n^\circ q_n^\circ \bar{s}_n^H).
\end{aligned}
\end{equation*}
In sum,
\begin{equation*}
\begin{aligned}
    \sup_i \norm*{
        \sum_{j=1}^n P \left[ \begin{aligned}
            & \kappa_\lambda(\Delta_j)
            \frac{\partial \xi_j}{\partial \mu_j}(\bar\mu, \bar\pi)
            ({H}_{\hat\mu, i} - {H}_{\bar\mu, i}) \\
            & + \kappa_\lambda(\Delta_j)
            \frac{\partial \xi_j}{\partial m_j}(\bar\mu, \bar\pi)
            \left(
                \int {H}_{\hat\mu, i} d\hat{P}_{X_{N_j}|\mathcal{G}}
                - \int {H}_{\bar\mu, i} dP_{X_{N_j}|\mathcal{G}}
            \right)
        \end{aligned} \right]
    }_2
    = O(D (r_n^\circ \bar{s}_n^\circ + r_n^\circ q_n^\circ) \bar{s}_n^H).
\end{aligned}
\end{equation*}
Similarly
\begin{equation*}
\begin{aligned}
    & \abs*{P_{Y_j, X_{N_j} | T_{N_j}} \left[ \begin{aligned}
        &
        \left(
            \frac{\partial \hat\xi_j}{\partial \mu_j}(\hat\mu, \hat\pi)
            - \frac{\partial \xi_j}{\partial \mu_j}(\bar\mu, \bar\pi)
        \right) ({H}_{\hat\mu, i} - {H}_{\bar\mu, i}) \\
        & +
        \left(
            \frac{\partial \hat\xi_j}{\partial m_j}(\hat\mu, \hat\pi)
            - \frac{\partial \xi_j}{\partial m_j}(\bar\mu, \bar\pi)
        \right) \left(
            \int {H}_{\hat\mu, i} d\hat{P}_{X_{N_j}|\mathcal{G}}
            - \int {H}_{\bar\mu, i} dP_{X_{N_j}|\mathcal{G}}
        \right) \\
    \end{aligned} \right]} \\
    & = \abs*{ P\left[ \begin{aligned} 
        \left(
            \frac{\hat\varpi(t_{N_j}, \mathcal{G}_{N_j})}{\hat\pi(t_{N_j}, X_{N_j}, \mathcal{G}_{N_j})}
            - \frac{\bar\varpi(t_{N_j}, \mathcal{G}_{N_j})}{\bar\pi(t_{N_j}, X_{N_j}, \mathcal{G}_{N_j})} 
        \right)
        \frac{\pi(t_{N_j}, X_{N_j}, \mathcal{G}_{N_j})}{\varpi(t_{N_j}, \mathcal{G}_{N_j})}
        (H_{\hat\mu,j} - H_{\bar\mu,j})(t_{N_j}, X_{N_j}, \mathcal{G}_{N_j}) \\
    \end{aligned} \right] } \\
    & = O_p((\bar{r}_n^\circ + r_n^\circ \bar{s}_n^\circ + r_n^\circ q_n) \bar{s}_n^\circ \bar{s}_n^H ),
\end{aligned}
\end{equation*}
and hence
\begin{equation*}
\begin{aligned}
    & \sup_i \norm*{
        \sum_{j=1}^n P \left[ \begin{aligned}
            & \kappa_\lambda(\Delta_j)
            \left(
                \frac{\partial \hat\xi_j}{\partial \mu_j}(\hat\mu, \hat\pi)
                - \frac{\partial \xi_j}{\partial \mu_j}(\bar\mu, \bar\pi)
            \right) ({H}_{\hat\mu, i} - {H}_{\bar\mu, i}) \\
            & + \kappa_\lambda(\Delta_j)
            \left(
                \frac{\partial \hat\xi_j}{\partial m_j}(\hat\mu, \hat\pi)
                - \frac{\partial \xi_j}{\partial m_j}(\bar\mu, \bar\pi)
            \right) \left(
                \int {H}_{\hat\mu, i} d\hat{P}_{X_{N_j}|\mathcal{G}}
                - \int {H}_{\bar\mu, i} dP_{X_{N_j}|\mathcal{G}}
            \right) \\
        \end{aligned} \right]
    }_2 \\
    & = O(D (\bar{r}_n^\circ + r_n^\circ \bar{s}_n^\circ + r_n^\circ q_n) \bar{s}_n^\circ \bar{s}_n^H).
\end{aligned}
\end{equation*}
Summing up the previous result back to \cref{eq:empirical_oracle_nuisance_decomp}, 
\begin{equation*}
\begin{aligned}
    & \sup_i \norm*{ \begin{aligned}
        & \frac{1}{D} \sum_{j=1}^n P\left[ 
            \kappa_\lambda(\Delta_j)
            \left\{
            \frac{\partial \hat\xi_j}{\partial \pi_j}(\hat\mu, \hat\pi)
            {H}_{\hat\pi, i}
            +
            \frac{\partial \hat\xi_j}{\partial \varpi_j}(\hat\mu, \hat\pi)
            \int {H}_{\hat\pi, i} d\hat{P}_{X_{N_j}|\mathcal{G}}
            \right\}
        \right]\\
        & - \frac{1}{D} \sum_{j=1}^n P\left[ 
            \kappa_\lambda(\Delta_j)
            \left\{
            \frac{\partial \xi_j}{\partial \pi_j}(\bar\mu, \bar\pi)
            {H}_{\bar\pi, i}
            +
            \frac{\partial \xi_j}{\partial \varpi_j}(\bar\mu, \bar\pi)
            \int {H}_{\bar\pi, i} dP_{X_{N_j}|\mathcal{G}}
            \right\}
        \right]
    \end{aligned} }_2 \\
    & = O\left(
        (\bar{r}_n^\circ + r_n^\circ \bar{s}_n^\circ + r_n^\circ q_n) \bar{s}_n^H
    \right),
\end{aligned}
\end{equation*}

With respect to $P_{X|\mathcal{G}}$, we consider the terms including $H_{P_{X|\mathcal{G}},i}$:
\begin{equation*}
\begin{aligned}
    & \sup_i \norm*{
        \sum_{j=1}^n P \left[
            \kappa_\lambda(\Delta_j)
            \left( \begin{aligned}
                & \frac{\partial \hat\xi_j}{\partial m_j}(\hat\mu, \hat\pi)
                H_{\hat{P}_{X|\mathcal{G}},i}(\hat\mu_j)
                - \frac{\partial \xi_j}{\partial m_j}(\bar\mu, \bar\pi)
                H_{{P}_{X|\mathcal{G}},i}(\bar\mu_j) \\
                & + \frac{\partial \hat\xi_j}{\partial \varpi_j}(\hat\mu, \hat\pi)
                H_{\hat{P}_{X|\mathcal{G}},i}(\hat\pi_j)
                - \frac{\partial \xi_j}{\partial \varpi_j}(\bar\mu, \bar\pi)
                H_{{P}_{X|\mathcal{G}},i}(\bar\pi_j)
            \end{aligned} \right)
        \right]
    }_2 \\
    & \leq \sup_i \norm*{
        \sum_{j=1}^n P \left[ \begin{aligned}
            & \kappa_\lambda(\Delta_j)
            \left(
                \frac{\partial \hat\xi_j}{\partial m_j}(\hat\mu, \hat\pi)
                - \frac{\partial \xi_j}{\partial m_j}(\bar\mu, \bar\pi)
            \right) H_{{P}_{X|\mathcal{G}},i}(\bar\mu_j) \\
            & + \kappa_\lambda(\Delta_j)
            \left(
                \frac{\partial \hat\xi_j}{\partial \varpi_j}(\hat\mu, \hat\pi)
                - \frac{\partial \xi_j}{\partial \varpi_j}(\bar\mu, \bar\pi)
            \right) H_{{P}_{X|\mathcal{G}},i}(\bar\pi_j)
        \end{aligned} \right]
    }_2 \\
    & \quad + \sup_i \norm*{
        \sum_{j=1}^n P \left[ \begin{aligned}
            & \kappa_\lambda(\Delta_j)
            \frac{\partial \xi_j}{\partial m_j}(\bar\mu, \bar\pi)
            (H_{\hat{P}_{X|\mathcal{G}},i}(\hat\mu_j) - H_{{P}_{X|\mathcal{G}},i}(\bar\mu_j)) \\
            & + \kappa_\lambda(\Delta_j)
            \frac{\partial \xi_j}{\partial \varpi_j}(\bar\mu, \bar\pi)
            (H_{\hat{P}_{X|\mathcal{G}},i}(\hat\pi_j) - H_{{P}_{X|\mathcal{G}},i}(\bar\pi_j))
        \end{aligned} \right]
    }_2 \\
    & \quad + \sup_i \norm*{
        \sum_{j=1}^n P \left[ \begin{aligned}
            & \kappa_\lambda(\Delta_j)
            \left(
                \frac{\partial \hat\xi_j}{\partial m_j}(\hat\mu, \hat\pi)
                - \frac{\partial \xi_j}{\partial m_j}(\bar\mu, \bar\pi)
            \right) 
            (H_{\hat{P}_{X|\mathcal{G}},i}(\hat\mu_j) - H_{{P}_{X|\mathcal{G}},i}(\bar\mu_j))\\
            & + \kappa_\lambda(\Delta_j)
            \left(
                \frac{\partial \hat\xi_j}{\partial \varpi_j}(\hat\mu, \hat\pi)
                - \frac{\partial \xi_j}{\partial \varpi_j}(\bar\mu, \bar\pi)
            \right) 
            (H_{\hat{P}_{X|\mathcal{G}},i}(\hat\pi_j) - H_{{P}_{X|\mathcal{G}},i}(\bar\pi_j))
        \end{aligned} \right]
    }_2 \\
\end{aligned}
\end{equation*}
For the first term, given $T_{N_j} = t_{N_j}$ and $\Delta_j(t_{N_j}) \leq \lambda$,
\begin{equation*}
\begin{aligned}
    & \abs*{P_{Y_j, X_{N_j} | T_{N_j}} \left[ \left.\begin{aligned}
        & \left(
            \frac{\partial \hat\xi_j}{\partial m_j}(\hat\mu, \hat\pi)
            - \frac{\partial \xi_j}{\partial m_j}(\bar\mu, \bar\pi)
        \right) H_{{P}_{X|\mathcal{G}},i}(\bar\mu_j) \\
        & + \left(
            \frac{\partial \hat\xi_j}{\partial \varpi_j}(\hat\mu, \hat\pi)
            - \frac{\partial \xi_j}{\partial \varpi_j}(\bar\mu, \bar\pi)
        \right) H_{{P}_{X|\mathcal{G}},i}(\bar\pi_j)
    \end{aligned} \right| t_{N_j} \right]} \\
    & = \abs*{ P_{Y_j, X_{N_j} | T_{N_j}}\left[ \left. \begin{aligned} 
        & \left(
            \frac{Y_j - \hat\mu(T_{N_j}, X_{N_j}, \mathcal{G}_{N_j})}{\hat\pi(T_{N_j}, X_{N_j}, \mathcal{G}_{N_j})} 
            - \frac{Y_j - \bar\mu(T_{N_j}, X_{N_j}, \mathcal{G}_{N_j})}{\bar\pi(T_{N_j}, X_{N_j}, \mathcal{G}_{N_j})}
        \right)
        H_{P_{X|\mathcal{G}},i}(\bar\pi_j) \\
    \end{aligned} \right| t_{N_j} \right] } \\
    & = \abs*{ P_{Y_j, X_{N_j} | T_{N_j}}\left[ \left. \begin{aligned} 
        (\bar\mu_j - \hat\mu_j) \frac{1}{\hat\pi_j} H_{{P}_{X|\mathcal{G}},i}(\bar\pi_j)
        + (\mu_j - \bar\mu_j) \left(\frac{1}{\hat\pi_j} - \frac{1}{\bar\pi_j} \right) H_{{P}_{X|\mathcal{G}},i}(\bar\pi_j) \\
    \end{aligned} \right| t_{N_j} \right] } \\
    & = O_p((\bar{r}_n^\circ + r_n^\circ \bar{s}_n^\circ) q_n^H ),
\end{aligned}
\end{equation*}
and hence
\begin{equation*}
\begin{aligned}
    & \sup_i \norm*{
        \sum_{j=1}^n P \left[ \begin{aligned}
            & \kappa_\lambda(\Delta_j)
            \left(
                \frac{\partial \hat\xi_j}{\partial m_j}(\hat\mu, \hat\pi)
                - \frac{\partial \xi_j}{\partial m_j}(\bar\mu, \bar\pi)
            \right) H_{{P}_{X|\mathcal{G}},i}(\bar\mu_j) \\
            & + \kappa_\lambda(\Delta_j)
            \left(
                \frac{\partial \hat\xi_j}{\partial \varpi_j}(\hat\mu, \hat\pi)
                - \frac{\partial \xi_j}{\partial \varpi_j}(\bar\mu, \bar\pi)
            \right) H_{{P}_{X|\mathcal{G}},i}(\bar\pi_j)
        \end{aligned} \right]
    }_2 \\
    & = O_p( D (\bar{r}_n^\circ + r_n^\circ \bar{s}_n^\circ) q_n^H ).
\end{aligned}
\end{equation*} 
For the second term, due to the convergence of $H_{\hat\pi}$ to $H_{\bar\pi}$ (\cref{assmp:H_nuisance}), given $T_{N_j} = t_{N_j}$ and $\Delta_j(t_{N_j}) \leq \lambda$,  
\begin{equation*}
\begin{aligned}
    & \abs*{P_{Y_{j}, X_{N_j}| T_{N_j}} \left[ \left. \begin{aligned}
        & \frac{\partial \xi_j}{\partial m_j}(\bar\mu, \bar\pi)
        (H_{\hat{P}_{X|\mathcal{G}},i}(\hat\mu_j) - H_{{P}_{X|\mathcal{G}},i}(\hat\mu_j))\\
        & + \frac{\partial \xi_j}{\partial \varpi_j}(\bar\mu, \bar\pi)
        (H_{{P}_{X|\mathcal{G}},i}(\hat\pi_j) - H_{{P}_{X|\mathcal{G}},i}(\bar\pi_j))
    \end{aligned}\right| t_{N_j} \right]} \\
    & = \abs*{ P_{Y_j, X_{N_j} | T_{N_j}}\left[ \left. \begin{aligned} 
        & \frac{\partial \xi_j}{\partial m_j}(\bar\mu, \bar\pi)
        (H_{\hat{P}_{X|\mathcal{G}},i}(\hat\mu_j) - H_{{P}_{X|\mathcal{G}},i}(\bar\mu_j)) \\
        & + \frac{\partial \xi_j}{\partial m_j}(\bar\mu, \bar\pi)
        (H_{\hat{P}_{X|\mathcal{G}},i}(\hat\mu_j) - H_{{P}_{X|\mathcal{G}},i}(\bar\mu_j)) \\
        & + \frac{\partial \xi_j}{\partial \varpi_j}(\bar\mu, \bar\pi)
        (H_{\hat{P}_{X|\mathcal{G}},i}(\hat\pi_j) - H_{{P}_{X|\mathcal{G}},i}(\hat\pi_j)) \\
        & + \frac{\partial \xi_j}{\partial \varpi_j}(\bar\mu, \bar\pi)
        (H_{{P}_{X|\mathcal{G}},i}(\hat\pi_j) - H_{{P}_{X|\mathcal{G}},i}(\bar\pi_j))
    \end{aligned} \right| t_{N_j} \right] } \\
    & = O_p((\bar{r}_n^\circ + \bar{s}_n^\circ + q_n^\circ) q_n^H).
\end{aligned}
\end{equation*}
Hence,
\begin{equation*}
\begin{aligned}
    \sup_i \norm*{
        \sum_{j=1}^n P \left[ \begin{aligned}
            & \kappa_\lambda(\Delta_j)
            \frac{\partial \xi_j}{\partial m_j}(\bar\mu, \bar\pi)
            (H_{\hat{P}_{X|\mathcal{G}},i}(\hat\mu_j) - H_{{P}_{X|\mathcal{G}},i}(\bar\mu_j)) \\
            & + \kappa_\lambda(\Delta_j)
            \frac{\partial \xi_j}{\partial \varpi_j}(\bar\mu, \bar\pi)
            (H_{\hat{P}_{X|\mathcal{G}},i}(\hat\pi_j) - H_{{P}_{X|\mathcal{G}},i}(\bar\pi_j))
        \end{aligned} \right]
    }_2
    = O(D (\bar{r}_n^\circ + \bar{s}_n^\circ + q_n^\circ) q_n^H).
\end{aligned}
\end{equation*}
Similarly
\begin{equation*}
\begin{aligned}
    & \abs*{P_{Y_j, X_{N_j} | T_{N_j}} \left[ \begin{aligned}
        & \left(
            \frac{\partial \hat\xi_j}{\partial m_j}(\hat\mu, \hat\pi)
            - \frac{\partial \xi_j}{\partial m_j}(\bar\mu, \bar\pi)
        \right) 
        (H_{\hat{P}_{X|\mathcal{G}},i}(\hat\mu_j) - H_{{P}_{X|\mathcal{G}},i}(\bar\mu_j))\\
        & + \left(
            \frac{\partial \hat\xi_j}{\partial \varpi_j}(\hat\mu, \hat\pi)
            - \frac{\partial \xi_j}{\partial \varpi_j}(\bar\mu, \bar\pi)
        \right) 
        (H_{\hat{P}_{X|\mathcal{G}},i}(\hat\pi_j) - H_{{P}_{X|\mathcal{G}},i}(\bar\pi_j))
    \end{aligned} \right]} \\
    & = \abs*{ P\left[ \begin{aligned} 
        \left(
            \frac{Y_j - \hat\mu(T_{N_j}, X_{N_j}, \mathcal{G}_{N_j})}{\hat\pi(T_{N_j}, X_{N_j}, \mathcal{G}_{N_j})} 
            - \frac{Y_j - \bar\mu(T_{N_j}, X_{N_j}, \mathcal{G}_{N_j})}{\bar\pi(T_{N_j}, X_{N_j}, \mathcal{G}_{N_j})}
        \right)
        (H_{\hat{P}_{X|\mathcal{G}},i}(\hat\pi_j) - H_{{P}_{X|\mathcal{G}},i}(\bar\pi_j)) \\
    \end{aligned} \right] } \\
    & = O_p((\bar{r}_n^\circ + r_n^\circ \bar{s}_n^\circ) (\bar{s}_n^\circ + q_n^\circ) q_n^H),
\end{aligned}
\end{equation*}
and hence
\begin{equation*}
\begin{aligned}
    & \sup_i \norm*{
        \sum_{j=1}^n P \left[ \begin{aligned}
            & \kappa_\lambda(\Delta_j)
            \left(
                \frac{\partial \hat\xi_j}{\partial m_j}(\hat\mu, \hat\pi)
                - \frac{\partial \xi_j}{\partial m_j}(\bar\mu, \bar\pi)
            \right) 
            (H_{\hat{P}_{X|\mathcal{G}},i}(\hat\mu_j) - H_{{P}_{X|\mathcal{G}},i}(\bar\mu_j))\\
            & + \kappa_\lambda(\Delta_j)
            \left(
                \frac{\partial \hat\xi_j}{\partial \varpi_j}(\hat\mu, \hat\pi)
                - \frac{\partial \xi_j}{\partial \varpi_j}(\bar\mu, \bar\pi)
            \right) 
            (H_{\hat{P}_{X|\mathcal{G}},i}(\hat\pi_j) - H_{{P}_{X|\mathcal{G}},i}(\bar\pi_j))
        \end{aligned} \right]
    }_2 \\
    & = O(D (\bar{r}_n^\circ + r_n^\circ \bar{s}_n^\circ) (\bar{s}_n^\circ + q_n^\circ) q_n^H).
\end{aligned}
\end{equation*}
Summing up the previous result back to \cref{eq:empirical_oracle_nuisance_decomp}, 
\begin{equation*}
\begin{aligned}
    & \sup_i \norm*{ \begin{aligned}
        & \frac{1}{D} \sum_{j=1}^n P\left[ 
            \kappa_\lambda(\Delta_j)
            \left\{
            \frac{\partial \hat\xi_j}{\partial m_j}(\hat\mu, \hat\pi)
            H_{\hat{P}_{X|\mathcal{G}},i}(\hat\mu_j)
            +
            \frac{\partial \hat\xi_j}{\partial \varpi_j}(\hat\mu, \hat\pi)
            H_{\hat{P}_{X|\mathcal{G}},i}(\hat\pi_j)
            \right\}
        \right]\\
        & - \frac{1}{D} \sum_{j=1}^n P\left[ 
            \kappa_\lambda(\Delta_j)
            \left\{
            \frac{\partial \xi_j}{\partial m_j}(\bar\mu, \bar\pi)
            H_{{P}_{X|\mathcal{G}},i}(\bar\mu_j)
            +
            \frac{\partial \xi_j}{\partial \varpi_j}(\bar\mu, \bar\pi)
            H_{{P}_{X|\mathcal{G}},i}(\bar\pi_j)
            \right\}
        \right]
    \end{aligned} }_2 \\
    & = O\left(
        (\bar{r}_n^\circ + \bar{s}_n^\circ + r_n^\circ \bar{s}_n^\circ + q_n^\circ) q_n^H
    \right),
\end{aligned}
\end{equation*}

Summing the results for terms including $H_{\mu,i}$, $H_{\pi,i}$ and $H_{P_{X|\mathcal{G}},i}$, we obtain
\begin{equation*}
\begin{aligned}
    \sup_i \frac{1}{D} \norm{P[\hat{W}_{\nu,i}] - W_i}_2
    = O\left( \begin{aligned}
        & (s_n^\circ \bar{r}_n^\circ + \bar{s}_n^\circ + q_n) \bar{r}_n^H \\
        & + (\bar{r}_n^\circ + r_n^\circ \bar{s}_n^\circ + r_n^\circ q_n) \bar{s}_n^H \\
        & + (\bar{r}_n^\circ + \bar{s}_n^\circ + r_n^\circ \bar{s}_n^\circ + q_n^\circ) q_n^H
    \end{aligned} \right),
\end{aligned}
\end{equation*}
and in sum
\begin{equation} \label{eq:delta_nu}
    \sup_i \frac{1}{D} \norm{\Delta_{\nu,i}}_2
    = \sup_i \frac{1}{D} \norm{\hat{W}_{\nu,i} - W_i}_2
    = O\left( \begin{aligned}
        & \bar\sigma (\bar{r}_n^H + \bar{s}_n^H + q_n^H) \\
        & + \bar{s}_n^\circ \bar{r}_n^H + \bar{r}_n^\circ \bar{s}_n^H 
        + (\bar{r}_n^\circ + \bar{s}_n^\circ) q_n^H
    \end{aligned}
    \right).
\end{equation}

\paragraph{Convergence of the empirical HAC estimator.}

Now back to $\hat\sigma_n - \check\sigma_n$, the difference between the empirical and oracle variance estimates is decomposed into 
\begin{equation} \label{eq:empirical_oracle_var_decomp}
\begin{aligned}
    & \Exp[\abs{\hat\sigma_n - \check\sigma_n}]
    = \Exp\left[\abs*{
        \sum_{i,j} h(i,j) \hat{W}_i \hat{W}_j
        - \sum_{i,j} h(i,j) {W}_i {W}_j
    }\right] \\
    & \leq 2 D^{-2} \Exp\left[\abs*{
        \sum_{i,j} h(i,j) {W}_{\theta,i} \Delta_{\theta,j}
    }\right] 
    + 2 D^{-2} \Exp\left[\abs*{
        \sum_{i,j} h(i,j) {W}_{\theta,i} \Delta_{\nu,j}
    }\right] \\
    & + 2 D^{-2} \Exp\left[\abs*{
        \sum_{i,j} h(i,j) {W}_{\nu,i} \Delta_{\theta,j}
    }\right] 
    + 2 D^{-2} \Exp\left[\abs*{
        \sum_{i,j} h(i,j) {W}_{\nu,i} \Delta_{\nu,j}
    }\right] \\
    & + D^{-2} \Exp\left[\abs*{
        \sum_{i,j} h(i,j) \Delta_{\theta,i} \Delta_{\theta,j}
    }\right] 
    + 2 D^{-2} \Exp\left[\abs*{
        \sum_{i,j} h(i,j) \Delta_{\theta,i} \Delta_{\nu,j}
    }\right]
    + D^{-2} \Exp\left[\abs*{
        \sum_{i,j} h(i,j) \Delta_{\nu,i} \Delta_{\nu,j}
    }\right]. \\
\end{aligned}
\end{equation}
For the first term,
\begin{equation*}
\begin{aligned}
    & D^{-2} \Exp\left[\abs*{
        \tsum_{i,j} h(i,j) {W}_{\theta,i} \Delta_{\theta,j}
    }\right] \\
    & \leq D^{-2} \Exp\left[ \begin{aligned}
        \sum_{i,j} & h(i,j) \kappa_\lambda(\Delta_i) \kappa_\lambda(\Delta_j) 
        \abs*{
            \xi_i(\bar\mu, \bar\pi) - \tilde\theta_{i^*}(t^*_{N_{i^*}}) 
        } \\
        & \times \abs*{
            \left( \hat\xi_j(\hat\mu, \hat\pi) - \hat\theta_{i^*}(t^*_{N_{i^*}}) \right) -
            \left( \xi_j(\bar\mu, \bar\pi) - \tilde\theta_{i^*}(t^*_{N_{i^*}}) \right)
        }
    \end{aligned} \right] \\
    & \overset{(\mathrm{i})}{\leq} D^{-2} \sqrt{\Exp\left[ \begin{aligned}
        \sum_{i,j} h(i,j) \kappa_\lambda(\Delta_i)^2
        \left\{
            \xi_i(\bar\mu, \bar\pi) - \tilde\theta_{i^*}(t^*_{N_{i^*}}) 
        \right\}^2
    \end{aligned} \right]} \\
    & \quad \times \sqrt{\Exp\left[ \begin{aligned}
        \sum_{i,j} h(i,j) \kappa_\lambda(\Delta_j)^2
        \left\{
            \left( \hat\xi_j(\hat\mu, \hat\pi) - \hat\theta_{i^*}(t^*_{N_{i^*}}) \right) -
            \left( \xi_j(\bar\mu, \bar\pi) - \tilde\theta_{i^*}(t^*_{N_{i^*}}) \right)
        \right\}^2
    \end{aligned} \right]} \\
    & \overset{(\mathrm{ii})}{\lesssim} D^{-2} M_\xi \sqrt{\Exp[
        \tsum_{i,j} h(i,j) \kappa_\lambda(\Delta_i)^2
    ]} \\
    & \quad \times \sqrt{\Exp\left[ \begin{aligned}
        \sum_{i,j} h(i,j) \kappa_\lambda(\Delta_j)^2 
        \left\{
            \left( \hat\xi_j(\hat\mu, \hat\pi) - \hat\theta_{i^*}(t^*_{N_{i^*}}) \right) -
            \left( \xi_j(\bar\mu, \bar\pi) - \tilde\theta_{i^*}(t^*_{N_{i^*}}) \right)
        \right\}^2
    \end{aligned} \right]}, \\
\end{aligned}
\end{equation*}
where $(\mathrm{i})$ is due to Hölder's inequality, and $(\mathrm{ii})$ is due to the uniform boundedness (\cref{assmp:uniform_boundedness}).
By the Cauchy-Schwarz inequality,
\begin{equation*}
\begin{aligned}
    & \sqrt{\Exp[ 
        \tsum_{i,j} h(i,j) \kappa_\lambda(\Delta_j)^2 
        \{
            ( \hat\xi_j(\hat\mu, \hat\pi) - \hat\theta_{i^*}(t^*_{N_{i^*}}) ) -
            ( \xi_j(\bar\mu, \bar\pi) - \tilde\theta_{i^*}(t^*_{N_{i^*}}) )
        \}^2
    ]}\\
    & \lesssim \sqrt{\Exp[
        \tsum_{i,j} h(i,j) \kappa_\lambda(\Delta_j)^2 
        ( 
            \hat\xi_j(\hat\mu, \hat\pi) - \xi_j(\bar\mu, \bar\pi) 
        )^2
    ]} \\
    & \quad + \sqrt{\Exp[
        \tsum_{i,j} h(i,j) \kappa_\lambda(\Delta_j)^2 
        ( 
            \hat\theta_{i^*}(t^*_{N_{i^*}}) - \tilde\theta_{i^*}(t^*_{N_{i^*}}) 
        )^2
    ]}.
\end{aligned}
\end{equation*}
We note that due to the uniform boundedness in \cref{assmp:uniform_boundedness,assmp:H_nuisance,assmp:H_covariate},
\begin{equation*}
\begin{aligned}
    & \Exp[
        \tsum_{i,j} h(i,j) \kappa_\lambda(\Delta_j)^2 
        ( 
            \hat\xi_j(\hat\mu, \hat\pi) - \xi_j(\bar\mu, \bar\pi) 
        )^2
    ] \\
    & = \sum_{t_{[n]}}
    \sum_{i,j} h(i,j) \kappa_\lambda(\Delta_j(t_{N_j}))^2
    \Exp[( 
        \hat\xi_j(\hat\mu, \hat\pi) - \xi_j(\bar\mu, \bar\pi) 
    )^2|T_{[n]} = t_{[n]}]
    \varpi(t_{[n]}| \mathcal{G}) \\
    & \lesssim M_\xi^2 \sum_{t_{[n]}}
    \sum_{i,j} h(i,j) \kappa_\lambda(\Delta_j(t_{N_j}))^2
    (\bar{r}_n^\circ + \bar{s}_n^\circ + q_n)^2
    \varpi(t_{[n]}| \mathcal{G}) \\
    & \lesssim M_\xi^2 ~\Exp[
        \tsum_{i,j} h(i,j) \kappa_\lambda(\Delta_j)^2
    ] 
    (\bar{r}_n^\circ + \bar{s}_n^\circ + q_n)^2,
\end{aligned}
\end{equation*}
and the same bound applies to $\Exp[
    \tsum_{i,j} h(i,j) \kappa_\lambda(\Delta_j)^2  
    ( 
        \hat\theta_{i^*}(t^*_{N_{i^*}}) - \tilde\theta_{i^*}(t^*_{N_{i^*}}) 
    )^2
]$.
That is,
\begin{equation} \label{eq:delta_theta}
\begin{aligned}
    & \sqrt{\Exp[ 
        \tsum_{i,j} h(i,j) \kappa_\lambda(\Delta_j)^2 
        \{
            ( \hat\xi_j(\hat\mu, \hat\pi) - \hat\theta_{i^*}(t^*_{N_{i^*}}) ) -
            ( \xi_j(\bar\mu, \bar\pi) - \tilde\theta_{i^*}(t^*_{N_{i^*}}) )
        \}^2
    ]}\\
    & \lesssim ~\sqrt{\Exp[
        \tsum_{i,j} h(i,j) \kappa_\lambda(\Delta_j)^2
    ] }
    (\bar{r}_n^\circ + \bar{s}_n^\circ + q_n).
\end{aligned}
\end{equation}
In sum,
\begin{equation*}
\begin{aligned}
    & D^{-2} \Exp\left[\abs*{
        \tsum_{i,j} h(i,j) {W}_{\theta,i} \Delta_{\theta,j}
    }\right] \\
    & \lesssim D^{-2} \Exp[
        \tsum_{i,j} h(i,j) \kappa_\lambda(\Delta_j)^2
    ] 
    (\bar{r}_n^\circ + \bar{s}_n^\circ + q_n), \\
    & \overset{(\mathrm{i})}{\leq} D^{-2} \tsum_{i,j} h(i,j) \Exp[
         \kappa_\lambda(\Delta_j)
    ] 
    (\bar{r}_n^\circ + \bar{s}_n^\circ + q_n), \\
    & \leq D^{-1} K_{\max} (\bar{r}_n^\circ + \bar{s}_n^\circ + q_n) 
    \overset{\mathrm{(ii)}}{=} o(\bar\sigma_n^2),
\end{aligned}
\end{equation*}
where $(\mathrm{i})$ follows from that $\kappa_\lambda(\Delta_j) \leq 1$ almost surely, and $(\mathrm{ii})$ follows from that $\sqrt{K_{\max}/D} = O(\bar\sigma_n)$. 

For the second term,
\begin{equation*}
\begin{aligned}
    & D^{-2} \Exp\left[\abs*{
        \tsum_{i,j} h(i,j) {W}_{\theta,i} \Delta_{\nu,j}
    }\right] \\
    & \leq D^{-2} \Exp\left[ \begin{aligned}
        \sum_{i,j} h(i,j) \kappa_\lambda(\Delta_i)
        \abs*{
            \xi_i(\bar\mu, \bar\pi) - \tilde\theta_{i^*}(t^*_{N_{i^*}}) 
        }
        \abs{\Delta_{\nu,j}}
    \end{aligned} \right] \\
    & \overset{(\mathrm{i})}{\leq} D^{-2} \sqrt{\Exp\left[ \begin{aligned}
        \sum_{i,j} h(i,j) \kappa_\lambda(\Delta_i)^2
        \left\{
            \xi_i(\bar\mu, \bar\pi) - \tilde\theta_{i^*}(t^*_{N_{i^*}}) 
        \right\}^2
    \end{aligned} \right]} 
    \sqrt{\Exp\left[ \begin{aligned}
        \sum_{i,j} h(i,j) \Delta_{\nu,j}^2
    \end{aligned} \right]} \\
    & \overset{(\mathrm{ii})}{\lesssim} \sqrt{K_{\max} / D} 
    \sqrt{n K_{\max}} \{ \begin{aligned}
        \bar\sigma_n (\bar{r}_n^H + \bar{s}_n^H + q_n^H) 
        + \bar{s}_n^\circ \bar{r}_n^H + \bar{r}_n^\circ \bar{s}_n^H 
        + (\bar{r}_n^\circ + \bar{s}_n^\circ) q_n^H
    \end{aligned}\}
    \\
    & \overset{(\mathrm{iii})}{=} o(\bar\sigma_n^2),
\end{aligned}
\end{equation*}
where $(\mathrm{i})$ follows from Hölder's inequality, $(\mathrm{ii})$ from \cref{eq:delta_nu}; and $(\mathrm{iii})$ from that $\sqrt{n K_{\max}}(\bar{r}_n^H, \bar{s}_n^H, q_n^H) = O(\bar{r}_n^\circ, \bar{s}_n^\circ, q_n^\circ)$, that $\bar{r}_n^\circ \bar{s}_n^\circ = o(\bar\sigma_n)$ and that $\max\{ \sqrt{K_{\max}/D}, \bar{r}^\circ_n s^\circ_n, r^\circ_n \bar{s}^\circ_n, q_n \} = O( \bar\sigma_n )$.

For the third term,
\begin{equation*}
\begin{aligned}
    & D^{-2} \Exp\left[\abs*{
        \tsum_{i,j} h(i,j) {W}_{\nu,i} \Delta_{\theta,j}
    }\right] \\
    & \leq D^{-2} \Exp\left[ \begin{aligned}
        \sum_{i,j} h(i,j) \kappa_\lambda(\Delta_j) 
        {W}_{\nu,i}
        \abs*{
            \left( \hat\xi_j(\hat\mu, \hat\pi) - \hat\theta_{i^*}(t^*_{N_{i^*}}) \right) -
            \left( \xi_j(\bar\mu, \bar\pi) - \tilde\theta_{i^*}(t^*_{N_{i^*}}) \right)
        }
    \end{aligned} \right] \\
    & \overset{(\mathrm{i})}{\lesssim} D^{-2} \sqrt{\Exp\left[ \begin{aligned}
        \sum_{i,j} h(i,j) {W}_{\nu,i}^2
    \end{aligned} \right]} \\
    & \quad \times \sqrt{\Exp\left[ \begin{aligned}
        \sum_{i,j} h(i,j) \kappa_\lambda(\Delta_j)^2
        \left\{
            \left( \hat\xi_j(\hat\mu, \hat\pi) - \hat\theta_{i^*}(t^*_{N_{i^*}}) \right) -
            \left( \xi_j(\bar\mu, \bar\pi) - \tilde\theta_{i^*}(t^*_{N_{i^*}}) \right)
        \right\}^2
    \end{aligned} \right]} \\
    & \overset{(\mathrm{ii})}{\lesssim} D^{-1} \sqrt{n K_{\max}}
    (\bar{r}_n^H s_n^\circ + r_n^\circ \bar{s}_n^H + q_n^H)
    \sqrt{\Exp[
        \tsum_{i,j} h(i,j) \kappa_\lambda(\Delta_j)^2
    ]}
    (\bar{r}_n^\circ + \bar{s}_n^\circ + q_n)\\
    & \lesssim  \sqrt{nK_{\max}} (\bar{r}_n^H s_n^\circ + r_n^\circ \bar{s}_n^H + q_n^H) \sqrt{K_{\max}/D} (\bar{r}_n^\circ + \bar{s}_n^\circ + q_n)
    \overset{(\mathrm{iii})}{=} o(\bar\sigma_n^2),
\end{aligned}
\end{equation*}
where $(\mathrm{i})$ follows from Hölder's inequality; $(\mathrm{ii})$ from \cref{eq:W_nu_i_bound,eq:delta_theta}; and $(\mathrm{iii)}$ from that $\sqrt{n K_{\max}}(\bar{r}_n^H, \bar{s}_n^H, q_n^H) = O(\bar{r}_n^\circ, \bar{s}_n^\circ, q_n^\circ)$, that $\bar{r}_n^\circ \bar{s}_n^\circ = o(\bar\sigma_n)$ and that $\max\{ \sqrt{K_{\max}/D}, \bar{r}^\circ_n s^\circ_n, r^\circ_n \bar{s}^\circ_n, q_n \} = O( \bar\sigma_n )$.

For the fourth term,
\begin{equation*}
\begin{aligned}
    & D^{-2} \Exp\left[\abs*{
        \tsum_{i,j} h(i,j) {W}_{\nu,i} \Delta_{\nu,j}
    }\right] \\
    & \overset{(\mathrm{i})}{\leq} D^{-2} \sqrt{\Exp\left[ \begin{aligned}
        \sum_{i,j} h(i,j) {W}_{\nu,i}^2
    \end{aligned} \right]} 
    \sqrt{\Exp\left[ \begin{aligned}
        \sum_{i,j} h(i,j) \Delta_{\nu,j}^2
    \end{aligned} \right]}\\
    & \overset{(\mathrm{ii})}{\lesssim} \sqrt{n K_{\max}}
    (\bar{r}_n^H s_n^\circ + r_n^\circ \bar{s}_n^H + q_n^H) 
    \sqrt{n K_{\max}} \{ \begin{aligned}
        \bar\sigma_n (\bar{r}_n^H + \bar{s}_n^H + q_n^H) 
        + \bar{s}_n^\circ \bar{r}_n^H + \bar{r}_n^\circ \bar{s}_n^H 
        + (\bar{r}_n^\circ + \bar{s}_n^\circ) q_n^H
    \end{aligned}\} \\
    & \overset{(\mathrm{iii})}{=} o(\bar\sigma_n^2),
\end{aligned}
\end{equation*}
where $(\mathrm{i})$ follows from Hölder's inequality; $(\mathrm{ii})$ from \cref{eq:W_nu_i_bound,eq:delta_nu}; and $(\mathrm{iii)}$ from that $\sqrt{n K_{\max}}(\bar{r}_n^H, \bar{s}_n^H, q_n^H) = O(\bar{r}_n^\circ, \bar{s}_n^\circ, q_n^\circ)$, that $\bar{r}_n^\circ \bar{s}_n^\circ = o(\bar\sigma_n)$ and that $\max\{ \sqrt{K_{\max}/D}, \bar{r}^\circ_n s^\circ_n, r^\circ_n \bar{s}^\circ_n, q_n \} = O( \bar\sigma_n )$.

The rest terms are second-order terms, which are easy to show to be of size $o(\sigma_n^2)$ using similar arguments as for the first four terms. Combining the results on the individual terms back to \cref{eq:empirical_oracle_var_decomp}, we show that $\hat\sigma_n - \check\sigma_n = o(\sigma_n)$ and conclude the proof for the consistency of the empirical sandwich variance estimator.

\subsection{Structural Equation Model Examples} \label{app:sem_example}

A common structural equation model for causal inference under network interference (see \citet{van2014causal,ogburn2022causal}) takes the form
\begin{equation*}
\begin{aligned}
    X_i & = f_X\!\left(\vareps_{X_i}\right), \\
    T_i & = f_T\!\left(s_X(X_{N_i}), \vareps_{T_i}\right), \\
    Y_i & = f_Y\!\left(s_T(T_{N_i}), s_X(X_{N_i}), \vareps_{Y_i}\right),
\end{aligned}
\end{equation*}
where $s_X$ and $s_T$ are $d_X$- and $d_T$-dimensional summary mappings of $X_{N_i}$ and $T_{N_i}$. Typical choices include
$s_X(X_{N_i}) = \bigl(X_i, \mathrm{Avg}(X_{N_i\setminus\{i\}})\bigr)$ or $\bigl(X_i, \mathrm{Sum}(X_{N_i\setminus\{i\}})\bigr)$, with analogous choices for $s_T$.
For the exogenous noises, we assume $\vareps_{X_i}$, $\vareps_{T_i}$, $\vareps_{Y_i}$ are i.i.d.

Assuming that covariates are i.i.d., and treatment is assigned independently across nodes given covariates, as formalized in \cref{eq:propensity_ind}, we then consider a nonparametric kernel smoothing estimator for
\[
\mu(T_{N_i},X_{N_i},\mathcal{G}_{N_i})
= \Exp\!\left[Y_i \mid T_{N_i}, X_{N_i}, \mathcal{G}_{N_i}\right]
\quad\text{and}\quad
\pi_i^\circ(X_{N_i},\mathcal{G}_{N_i})
= \Exp\!\left[T_i \mid X_{N_i}, \mathcal{G}_{N_i}\right],
\]
using the empirical product measure for the covariate distribution. 
Then the nuisance estimation rates in this toy example yield
\[
r_n^\circ
= O\!\left(n^{-\frac{1}{2+d_T+d_X}}\right),
\qquad
s_n^\circ
= O\!\left(K_{\max}\, n^{-\frac{1}{2+d_X}}\right),
\qquad
q_n
= O\!\left(K_{\max}^{\frac{1}{2}} n^{-\frac{1}{2}}\right),
\]
where $K_{\max}$ denotes the maximal dependence degree and $r_n^\circ$, $s_n^\circ$ and $q_n$ refers to the convergence rates of outcome regression, propensity score and covariate distribution as defined in \cref{app:pf_consistency}. 

The additional condition $\max\{r^\circ_n \, s^\circ_n, q_n\} = o( \sigma_n )$ in \cref{thm:asymptotic_normality} requires the variance of KECENI to be dominated by the kernel smoothing step contribution:
\begin{align*}
    \Var\!\left[
        \hat{D}^{-1}\sum_{i=1}^n \kappa_\lambda(\Delta_i)\xi_i(\bar\mu,\bar\pi)
    \,\middle|\,\mathcal{G} \right] 
    = \Exp\!\left[ \left(
        \hat{D}^{-1}\sum_{i=1}^n \kappa_\lambda(\Delta_i)\xi_i(\bar\mu,\bar\pi)
        -\tilde\theta_{i^*}(t^*_{N_{i^*}})
    \right)^2 \,\middle|\, \mathcal{G}
    \right],
\end{align*}
which scales as $\sqrt{K_{\max}/D}$ under the above dependence structure. Under a stable degree distribution and $\Delta_i = \left\lVert s_T(t^*_{N_{i^*}}) - s_T(T_{N_i}) \right\rVert$,
$D$ typically scales as $n\lambda_n^{d_T}$. Putting these pieces together, the additional condition $\max\{r^\circ_n \, s^\circ_n, q_n\} = o( \sigma_n )$ holds when
\begin{equation*}
    \max\left\{
        \sqrt{\frac{K_{\max}}{n}}, 
        \frac{K_{\max}}{ n^{\frac{1}{2+d_X} + \frac{1}{2+d_T+d_X}}}
    \right\}
    = o\!\left( \sqrt{\frac{K_{\max}}{n \lambda_n^{d_T}}}\right).
\end{equation*}
This simplifies to the requirement
\begin{equation*}
    K_{\max}^{\frac{1}{2}} 
    = o\!\left( n^{\frac{1}{2+d_X} + \frac{1}{2+d_T+d_X} - \frac{1}{2}} \lambda_n^{-\frac{d_T}{2}}\right).
\end{equation*}
With the bandwidth choice that is optimal for prediction, $\lambda_n \asymp (K_{\max}/n)^{\frac{1}{2+d_T}}$, this condition reduces to
\begin{equation*}
    K_{\max}
    = o\!\left(
    n^{\left(\frac{1}{2+d_X} + \frac{1}{2+d_T+d_X} - \frac{1}{2+d_T}\right)\big/\left(1 - \frac{1}{2 + d_T}\right)}
    \right),
\end{equation*}
which imposes a polynomial upper bound on the growth of $K_{\max}$. This is compatible, for example, with degree distributions whose tails are well approximated by a power law with an exponential cutoff or by a lognormal tail, both widely used as descriptive models for empirical networks \citep{broido2019scale,clauset2009power,mitzenmacher2004brief}.

%% file: app/c_supp_simulation.tex
\section{Supplementary Simulation Results} \label{app:supp_simulation}

\subsection{Simulation Setting and Estimation Details for Section~\ref{sec:simulation_node_wise}}
\label{subsec:settings}
\paragraph{Setting.}

\begin{figure}[tbp]
    \centering
    \begin{subfigure}[t]{0.36\linewidth}
        \captionsetup{margin={5mm, 0mm}}
        \centering
        \subcaption{}
        \includegraphics[height=0.88\linewidth]{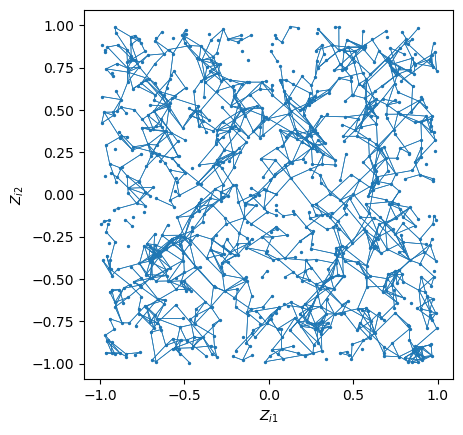}
        \label{fig:latent_linear}
    \end{subfigure}
    \qquad\qquad
    \begin{subfigure}[t]{0.48\linewidth}
        \captionsetup{margin={5mm, 0mm}}
        \centering
        \subcaption{}
        \includegraphics[height=0.66\linewidth]{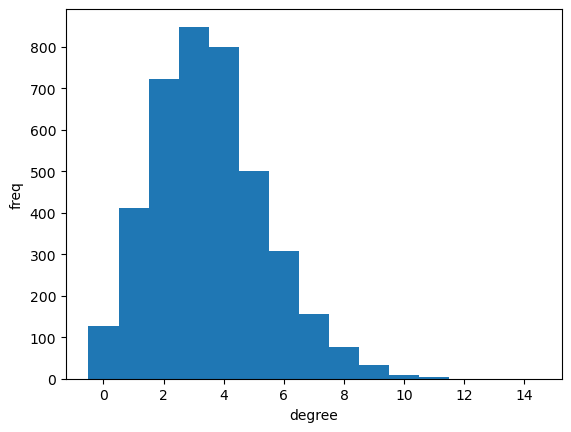}
        \label{fig:hist_degree_linear}
    \end{subfigure}
    \caption{(a) The $1000$ node latent positions and edges (b) the histogram of node degrees.}
    \label{fig:result_network}
\end{figure}

We sample \(\mathcal{G}\) with $n = 1000$ units once from a latent variable network model, where the probability of an edge between a pair of nodes is determined by the distance between their latent positions.  We draw the latent positions $Z_i$ i.i.d.\ from the uniform distribution on $[-1, 1]^2$ and model \(\mathcal{G}\) as a simple undirected network with edges drawn independently with probability 
\begin{equation} \label{eq:adj_prob}
    \Pr[(i,j) \in \mathcal{E}] = \rho \exp\left( - e^{\beta \cdot \norm{Z_i - Z_j}_\infty} \right) , 
\end{equation}
for all $i < j$.   The parameters \(\rho\) and \(\beta\) govern the network structure, with \(\rho\) controlling the overall edge density and \(\beta\) controlling the influence of the latent positions.   We set \(\rho = 2\) and \(\beta = 10\) in this simulation. 
\cref{fig:latent_linear} shows the sampled latent positions and the corresponding adjacency structure among the units, while \cref{fig:hist_degree_linear} presents the histogram of the degree distribution for those units.

With the network fixed, we sample covariate feature vectors \(X_i \in \mathbb{R}^3\), treatment variables \(T_i \in \{0, 1\}\), and outcome variables \(Y_i \in \mathbb{R}\) for nodes \(i \in \{1, \dots, n\}\). 
All covariates are sampled as i.i.d.\ standard normals. Given the covariates, treatments are independently assigned as 
\begin{equation} \label{eq:propensity_true}
\begin{aligned}
    \Pr[T_i = 1| X_{N_i} = x_{N_i}, \mathcal{G}_{N_i}]
    = \mathrm{expit}\left(
        \beta_\pi^{(0)}
        + \inner{\beta_\pi^{(1)}, x_i}
        + \inner*{\beta_\pi^{(2)}, \mathrm{Avg}(x_{N_i \setminus \{i\}})}
    \right)
\end{aligned}
\end{equation}
for some $\beta_\pi^{(1)}, \beta_\pi^{(2)} \in \reals^3$, where $\mathrm{expit}$ is the logistic sigmoid function, $\mathrm{expit}(x) \equiv \frac{e^x}{1+e^x}$, and $\mathrm{Avg}(x_{N_i \setminus \{i\}}) \equiv \mathbf{1}\{\abs{N_i} > 1\} (\frac{1}{\abs{N_i} - 1} \tsum_{j \in N_i \setminus \{i\}} x_j)$. 
Finally, incorporating two-hop neighborhood covariates as described in \cref{rem:two_hop_neighborhoods}, we define the outcomes by 
\begin{equation} \label{eq:outcome_true}
\begin{aligned}
    & \Exp[Y_i | T_{N_i} = t_{N_i}, X_{N_i^{(2)}} = x_{N_i^{(2)}}, \mathcal{G}_{N_i^{(2)}}] \\
    & = \beta_\mu^{(0)} 
    + \beta_\mu^{(1)} (t_i - 0.5) 
    + \beta_\mu^{(2)} \mathrm{Avg}(t_{N_i \setminus \{i\}} - 0.5) 
    + \inner*{\beta_\mu^{(3)}, x_i} 
    + \inner*{\beta_\mu^{(4)}, \mathrm{Avg}(x_{N_i^{(2)} \setminus \{i\}})},
\end{aligned}
\end{equation}
with an independent standard Gaussian random error, for some $\beta_\mu^{(1)}, \beta_\mu^{(2)} \in \reals$, $\beta_\mu^{(3)}, \beta_\mu^{(4)} \in \reals^3$.
We set $\beta_\pi^{(0)} = 0$, $\beta_\pi^{(1)} = (0.5, 0.5, 0.5)$, $\beta_\pi^{(2)} = (0, 0, 0)$, $\beta_\mu^{(0)} = 0$, $\beta_\mu^{(1)} = \beta_\mu^{(2)} = 2$, and $\beta_\mu^{(3)} = \beta_\mu^{(4)} = (-1.55, -1.55, -1.55)$. The treatment assignment and outcome are designed so that $\Exp[Y_i | T_i = 1] \approx \Exp[Y_i | T_i = 0]$. In order to recover the correct treatment effect, one should account for the confounding by the covariate vector.


We select the node $i^*$ with the smallest $\ell_\infty$ norm of the latent position among the observed nodes $\{1, \dots, n\}$ and apply KECENI to estimate its counterfactual mean outcome, $\theta_{i^*}(t^*_{N_{i^*}})$, under a given treatment allocation \(t^*_{N_{i^*}} \in \{0, 1\}^{\abs{N_{i^*}}}\). This choice of target node \(i^*\) ensures that network edge effects are avoided. The selected node has two adjacent units, and we consider two contrasting treatment scenarios: \(t^{(0)}_{N_{i^*}}\), where no neighbors are treated (\(t^{(0)}_i = 0\) for all \(i \in N_{i^*}\)), and \(t^{(1)}_{N_{i^*}}\), where all neighbors are treated (\(t^{(1)}_i = 1\) for all \(i \in N_{i^*}\)). 
According to the true outcome model in \cref{eq:outcome_true}, the expected outcomes under these two scenarios are \(\theta_{i^*}(t^{(0)}_{N_{i^*}}) = -2\) and \(\theta_{i^*}(t^{(1)}_{N_{i^*}}) = 2\), respectively.

\paragraph{Estimation.}

We estimate the propensity scores conditionally on covariates and the connectivity of two-hop neighborhoods as described in \cref{rem:two_hop_neighborhoods}. 
The outcome regression $\mu(T_{N_i}, X_{N_i^{(2)}}, \mathcal{G}_{N_i^{(2)}})$ is estimated by linear regression under \cref{eq:outcome_true}.
The propensity score model assumes that intervention assignments are conditionally independent across nodes, given local covariates and network structure as formulated in \cref{eq:propensity_ind}, 
where the node-level propensity $\pi^\circ(T_i | X_{N_i}, \mathcal{G}_{N_i})$ is estimated by logistic regression under \cref{eq:propensity_true}.
To estimate the covariate distribution, we use the empirical product measure $\hat{P}^\otimes_{X_{[n]}|\mathcal{G}}$ as defined in \cref{eq:empirical_product_measure}.

For the dissimilarity metric, we use 
\begin{equation} \label{eq:delta_avg}
    \Delta_i \equiv \norm*{(T_i, \mathrm{Avg}(T_{N_i \setminus \{i\}} - 0.5)) 
    - (t^*_{i^*}, \mathrm{Avg}(t^*_{N_{i^*} \setminus \{i^*\}} - 0.5))}_1.
\end{equation}

\subsection{Dependency on Sample Size} \label{app:simulation_n_node}

We use the same simulation dataset as in \cref{sec:simulation_node_wise} to further investigate the accuracy of KECENI in relation to the sample size \(n\), representing the number of observed units or nodes in the network. We apply KECENI to simulation datasets with seven different sample sizes: \(n \in \{250, 500, 750, 1000, 2000, 3000, 4000\}\). To isolate the effect of sample size, we generate a network \(\mathcal{G}\) for each \(n\) while maintaining consistent network sparsity and connectivity of the target node. We first sample \(Z_i \in \mathbb{R}^2\) for \(i \in \{1, \dots, 4000\}\) from a uniform distribution on \([-2, 2]^2\) and then generate a preliminary network \(\mathcal{G}^{\mathrm{pre}}\) from the latent variable network model with adjacency probability as given in \cref{eq:adj_prob}. For each \(n\), we define \(\mathcal{G}\) as a subgraph of \(\mathcal{G}^{\mathrm{pre}}\), selecting nodes with the smallest \(n\) values of \(\norm{Z_i}_\infty\). We then proceed with the same feature generation and estimation procedures as in \cref{sec:simulation_node_wise}.

\begin{figure}[tbp]
    \centering
    \begin{subfigure}[t]{0.4\linewidth}
        \captionsetup{margin={5mm, 0mm}}
        \centering
        \subcaption{}
        \includegraphics[width=1\textwidth]{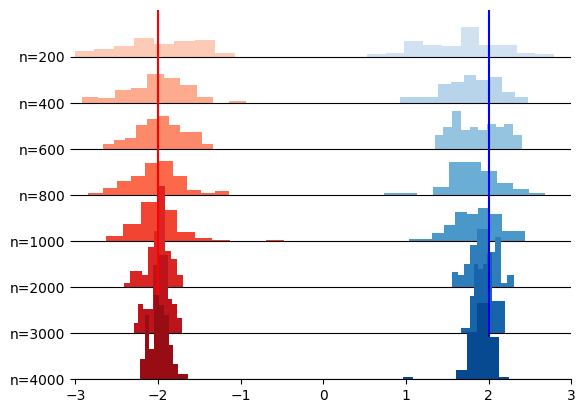}
        \label{fig:hist_T1_T2_n_node}
    \end{subfigure}
    \qquad
    \begin{subfigure}[t]{0.4\linewidth}
        \captionsetup{margin={5mm, 0mm}}
        \centering
        \subcaption{}
        \includegraphics[width=1\textwidth]{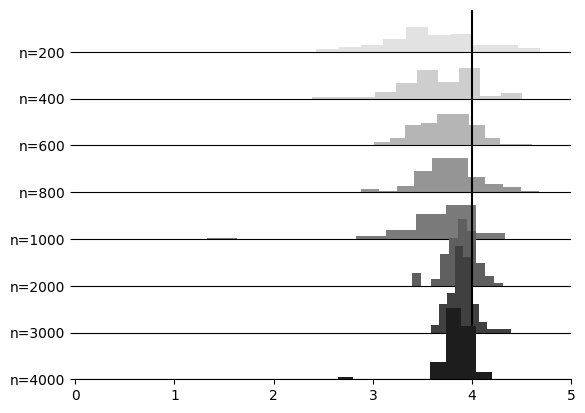}
        \label{fig:hist_Td_n_node}
    \end{subfigure}
    
    \begin{subfigure}[t]{0.4\linewidth}
        \captionsetup{margin={5mm, 0mm}}
        \centering
        \subcaption{}
        \includegraphics[width=1\textwidth]{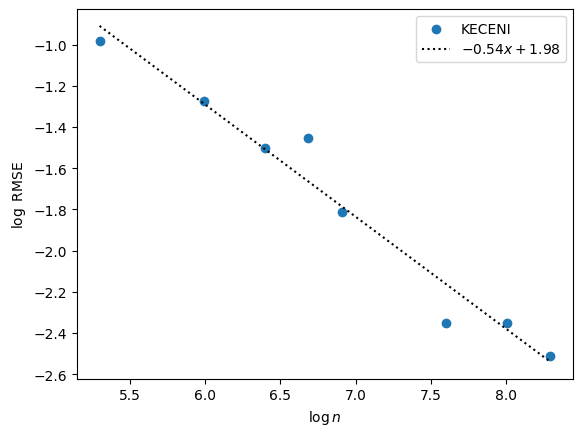}
        \label{fig:rmse_n_node}
    \end{subfigure}
    \caption{(a,b) Histogram of the estimated counterfactual means and the treatment effect by KECENI in different sample sizes. (c) Plot of their root mean squared errors against sample sizes in log scale.}
    \label{fig:result_n_node}
\end{figure}

\cref{fig:result_n_node} summarizes the results of 80 simulation runs for each sample size \(n\). \cref{fig:hist_T1_T2_n_node,fig:hist_Td_n_node} show the distributions of estimated counterfactual means and treatment effects for different values of \(n\). As \(n\) increases, both the bias and variance of the KECENI estimates decrease, leading to more accurate treatment effect estimates. To estimate the convergence rate, we plot the logarithms of the root-mean-square errors (RMSE) against \(\log n\) in \cref{fig:rmse_n_node}, which show an excellent linear fit with a slope of \(-0.30\). This rate, though slower than the parametric rate of \(n^{-1/2}\), is expected due to the nonparametric nature of the kernel smoothing estimation. This result demonstrates the consistency of KECENI under the given simulation setting.

For average treatment effects, this bias has been theoretically derived and studied through simulations by \citet{forastiere2021identification}. Our simulation study extends the literature by demonstrating a similar bias in the estimation of counterfactual means for specific node types.

\subsection{Results of Classical Methods (Ignoring Network Interference)} \label{app:simulation_nonet}

We use the same simulation dataset as in \cref{sec:simulation_node_wise} to demonstrate the bias resulting from ignoring interference. 
Consider the classical augmented inverse propensity weighting method (AIPW; \citealp{bang2005doubly}), which imposes the stable unit treatment value assumption (SUTVA). Under SUTVA, each unit is assumed to be independent and identically distributed, making the individual treatment effect equivalent to the average treatment effect. 

\begin{figure}[tbp]
    \centering
    \begin{subfigure}[t]{0.4\linewidth}
        \captionsetup{margin={5mm, 0mm}}
        \centering
        \subcaption{}
        \includegraphics[height=0.65\textwidth]{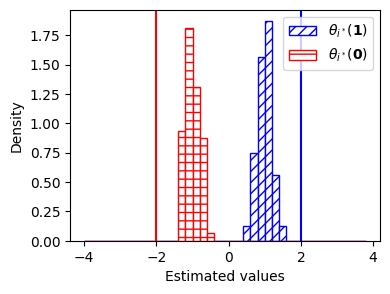}
        \label{fig:hist_T1_T2_aipw}
    \end{subfigure}
    \qquad
    \begin{subfigure}[t]{0.4\linewidth}
        \captionsetup{margin={5mm, 0mm}}
        \centering
        \subcaption{} 
        \includegraphics[height=0.65\textwidth]{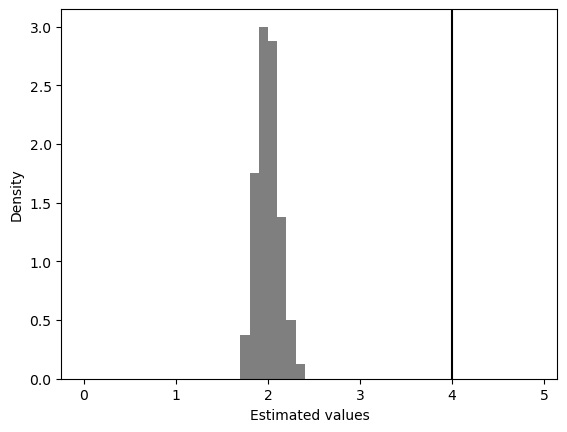}
        \label{fig:hist_Td_aipw}
    \end{subfigure}
    \caption{Histograms of (a) the AIPW estimates for the counterfactual means under two treatment scenarios and (b) the estimated treatment effects.}
    \label{fig:result_aipw}
\end{figure}

\cref{fig:result_aipw} shows the distribution of the AIPW estimates applied to the same 80 datasets as in \cref{sec:simulation_node_wise}. The mean of the treatment effect estimates is $2.001$, indicating a significant bias compared to the true value $4$. This mean value is close to the direct treatment effect, defined as
\begin{equation*}
    \Exp[Y_{i^*}(t_{i^*} = 1, t_{N_{i^*} \setminus \{i^*\}} = T_{N_{i^*} \setminus \{i^*\}}) | \mathcal{G}] - \Exp[Y_{i^*}(t_{i^*} = 0, t_{N_{i^*} \setminus \{i^*\}} = T_{N_{i^*} \setminus \{i^*\}}) | \mathcal{G}],
\end{equation*}
with a slight abuse of notation.
This result aligns with \citet{savje2021average}, who found that commonly used estimators for treatment effects without interference are consistent for the direct effect under certain experimental designs. Their Proposition 2 established this for the inverse propensity weighting estimator \citep{horvitz1952generalization} under randomized experimental designs and degree restrictions. Our simulation study suggests a possible extension to doubly robust estimators, such as AIPW. Identifying specific conditions under which this argument holds for doubly robust estimators is a potential area for future research.

\subsection{Cross-validation on Choice of Tuning Hyperparameters} \label{app:simulation_cv}

We use the same simulation dataset as in \cref{sec:simulation_node_wise} to demonstrate data-driven choice of tuning hyperparmeters using the leave-neighborhood-out cross-validation in \cref{app:cross_validation}, exclusively offered by the node-wise counterfactual mean framework of KECENI.

We used cross validation to evaluate the generalization performance of KECENI across candidate hyperparameter combinations $(\lambda, \kappa, \Delta_i)$. For the bandwidth $\lambda$, we considered the grid $\{1/2, 1/3, \dots, 1/20\}$. For the kernel $\kappa$, we considered the exponential, box, triangular, and Gaussian kernels so as to include both bounded support and infinite support cases. For the dissimilarity metric $\Delta_i$, we considered three options: the metric induced by the correct exposure mapping (\cref{eq:delta_avg}), the metric induced by an incorrect exposure mapping, and the Gromov Wasserstein based metric (\cref{eq:delta_gw}). Specifically, under the incorrect exposure mapping, we defined
\begin{equation*}
    g(T_{N_i}) = \left( T_i, ~\frac{\sum_{j \in N_i \setminus \{i\}} \abs{N_i \cap N_j} T_j}{\sum_{j \in N_i \setminus \{i\}} \abs{N_i \cap N_j}} \right),
\end{equation*}
and took the corresponding dissimilarity metric to be $\Delta_i \equiv \norm*{g(T_{N_i}) - g(t^*_{N_{i^*}})}_1$.

\begin{figure}[tbp]
    \centering
    \begin{subfigure}[t]{0.4\linewidth}
        \captionsetup{margin={5mm, 0mm}}
        \centering
        \subcaption{}
        \includegraphics[height=0.65\textwidth]{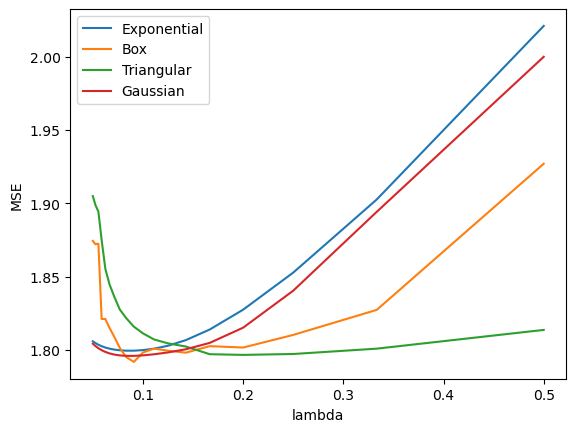}
        \label{fig:mse_cv_kernel}
    \end{subfigure}
    \qquad
    \begin{subfigure}[t]{0.4\linewidth}
        \captionsetup{margin={5mm, 0mm}}
        \centering
        \subcaption{}
        \includegraphics[height=0.65\textwidth]{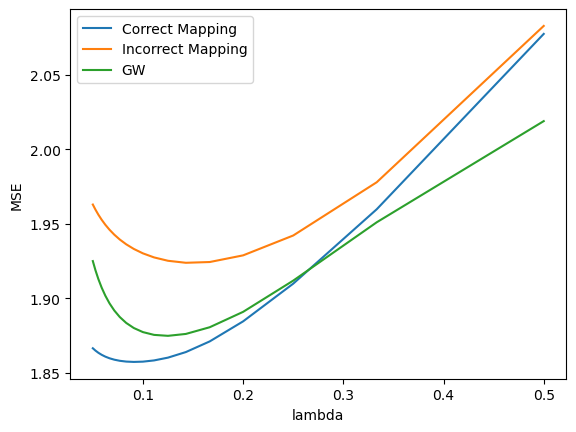}
        \label{fig:mse_cv_Ds}
    \end{subfigure}
    
    \begin{subfigure}[t]{0.4\linewidth}
        \captionsetup{margin={5mm, 0mm}}
        \centering
        \subcaption{}
        \includegraphics[height=0.65\textwidth]{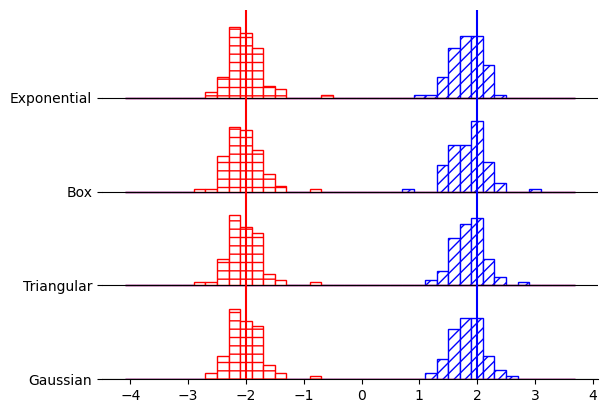}
        \label{fig:hist_T1_T2_kernel}
    \end{subfigure}
    \qquad
    \begin{subfigure}[t]{0.4\linewidth}
        \captionsetup{margin={5mm, 0mm}}
        \centering
        \subcaption{}
        \includegraphics[height=0.65\textwidth]{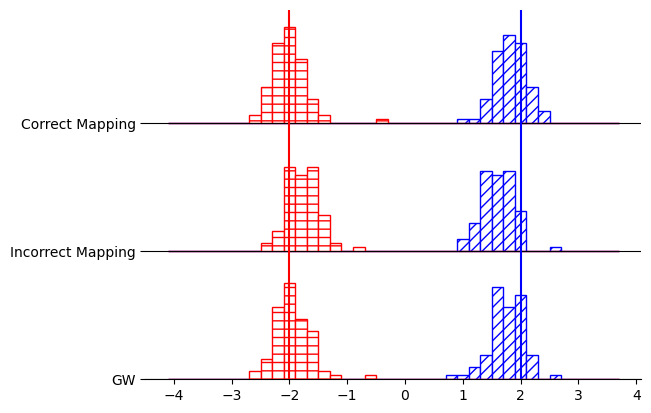}
        \label{fig:hist_Td_Ds}
    \end{subfigure}
    
    \begin{subfigure}[t]{0.4\linewidth}
        \captionsetup{margin={5mm, 0mm}}
        \centering
        \subcaption{}
        \includegraphics[height=0.65\textwidth]{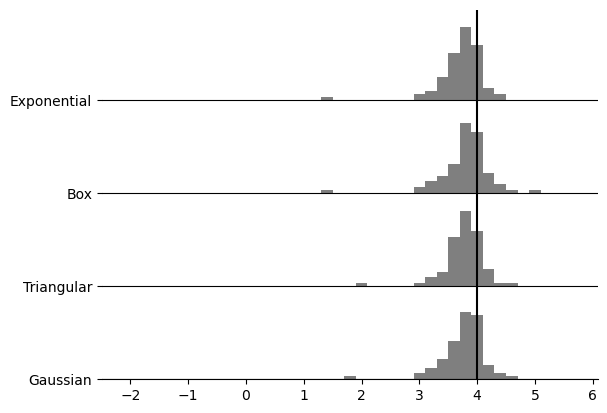}
        \label{fig:hist_T1_T2_Ds}
    \end{subfigure}
    \qquad
    \begin{subfigure}[t]{0.4\linewidth}
        \captionsetup{margin={5mm, 0mm}}
        \centering
        \subcaption{}
        \includegraphics[height=0.65\textwidth]{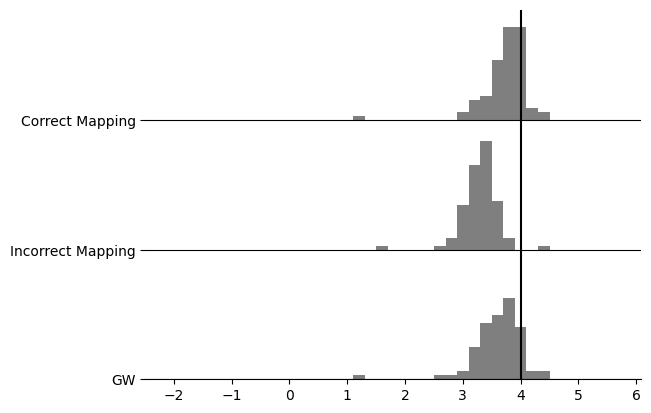}
        \label{fig:hist_Td_kernel}
    \end{subfigure}
    \caption{Leave-neighborhood-out cross-validated MSEs for candidate sets of $(\kappa, \lambda, \Delta_i)$ (a) when $\Delta_i$'s were fixed given by the correct summary statistics and (b) when $\kappa$ was fixed at the exponential kernel. Histograms of the KECENI estimates for the counterfactual means under two treatment scenarios and for the corresponding treatment effects (c,e) over candidate $\kappa$'s and (d,f) over candidate $\Delta_i$'s, with cross-validated $\lambda$.}
    \label{fig:result_cv}
\end{figure}

\cref{fig:mse_cv_kernel,fig:mse_cv_Ds} reports the cross validated MSE across different combinations of tuning hyperparameters. In \cref{fig:mse_cv_kernel}, we vary $\lambda$ and $\kappa$ while fixing $\Delta_i$ at \cref{eq:delta_avg}. In \cref{fig:mse_cv_Ds}, we vary $\lambda$ and $\Delta_i$ while fixing $\kappa$ to the exponential kernel. In both settings, the choice of $\lambda$ has a substantial effect on the MSE of KECENI. By contrast, the choice of kernel has little, if any, effect on finite sample performance relative to the other tuning parameters. This observation in cross-validation maintains in both counterfactual mean estimation and treatment effect estimation. The distribution of the KECENI estimates in \cref{fig:hist_T1_T2_kernel,fig:hist_Td_kernel} shows that once the bandwidth is selected by cross validation, KECENI behaves similarly across kernel functions. This agrees with the standard intuition from smoothing based nonparametric estimation, where performance is typically driven more by bandwidth selection than by the precise kernel shape.

For dissimilarity metrics, the cross-validation errors were smallest for the correct mapping, and the corresponding estimators performed best on the subsequent node-wise counterfactual mean estimation task, as shown in \cref{fig:hist_T1_T2_Ds,fig:hist_Td_Ds}. Moreover, cross-validation preferred the Gromov Wasserstein based metric over the incorrect exposure mapping, illustrating that the procedure can guide practitioners away from poor specifications when the correct mapping is unknown.

\subsection{Simulation setting and estimation details for Section~\ref{sec:simulation_dr}}

\paragraph{Setting.} 

We use the same $80$ datasets with $n = 1000$ units and the same target node $i^*$ as in the previous simulation.  
%
%
The objective is to estimate the direct treatment effect $i^*$. We set $t^{(0)}_{N_{i^*}}$ so that $t^{(0)}_{i^*} = 0$ and $|N_{i^*}|^{-1} \sum_{i \in N_{i^*}} T_i = 0.5$. The second treatment scenario $t^{(1)}_{N_{i^*}}$ had the same treatment assignment as $t^{(0)}_{N_{i^*}}$ except that $t^{(1)}_{i^*} = 1$. The expected outcomes for these scenarios were \(\theta_{i^*}(t^{(0)}_{N_{i^*}}) = -1\) and \(\theta_{i^*}(t^{(1)}_{N_{i^*}}) = 1\), respectively. 

\paragraph{Estimation.} 

We investigate the sensitivity of the G-computation and KECENI to model misspecification. We quantify the extent of misspecification in the propensity score and outcome regression models by $\alpha_\pi$ and $\alpha_\mu$, respectively. For outcome regression, we fit a linear regression model
\begin{equation*}
\begin{aligned}
    \mu(t_{N_i}, x_{N_i^{(2)}}, \mathcal{G}_{N_i^{(2)}})
    & = \beta_\mu^{(0)} 
    + \beta_\mu^{(1)} (t_i - 0.5) 
    + \beta_\mu^{(2)} \mathrm{Avg}(t_{N_i \setminus \{i\}} - 0.5) \\
    & \quad + \inner*{\beta_\mu^{(3)}, (1-\alpha_\mu) x_i + \alpha_\mu x_i^2}
    + \inner*{\beta_\mu^{(4)}, \mathrm{Avg}((1-\alpha_\mu) x_{N_i^{(2)} \setminus \{i\}} + \alpha_\mu x_{N_i^{(2)} \setminus \{i\}}^2)}.
\end{aligned}
\end{equation*}
The propensity score model follows \cref{eq:propensity_ind}, where each node-level probability is estimated by logistic regression 
\begin{equation*}
\begin{aligned}
    & \pi^\circ(1| x_{N_i}, \mathcal{G}_{N_i}) \\
    & = \mathrm{expit}\left(
        \beta_\pi^{(0)}
        + \inner*{\beta_\pi^{(1)}, (1-\alpha_\pi) x_i + \alpha_\pi x_i^2}
        + \inner*{\beta_\pi^{(2)}, \mathrm{Avg}((1-\alpha_\pi) x_{N_i \setminus \{i\}} + \alpha_\pi x_{N_i \setminus \{i\}}^2))}
    \right). \\
\end{aligned}
\end{equation*}
These models are correctly specified when $\alpha_\pi = \alpha_\mu = 0$ and become increasingly misspecified as $\alpha_\pi$ and $\alpha_\mu$ approach $1$, reaching substantial misspecification at $\alpha_\pi = \alpha_\mu = 1$. We quantified the accuracy of G-computation and KECENI by RMSE while varying both  $\alpha_\pi$ and $\alpha_\mu$ from 0 to 1 with a step size of 0.1.

\subsection{Simulation Setting and Estimation Details for Section~\ref{sec:simulation_ate}}

\paragraph{Setting.}

The simulation consisted of $40$ repeated runs, using the same network model $\mathcal{G}$ as that in the previous subsection with $n = 4000$ units.
For the compatibility with existing methods, we work with binary two-dimensional covariate vectors $X_i = (X_i^{(1)}, X_i^{(2)}) \in \{0, 1\}^2$, treatment variables $T_i \in \{0, 1\}$ and outcome variables $Y_i \in \{0, 1\}$. Each covariate was independently sampled from the Bernoulli distribution with probability $0.5$.  Treatments were randomly assigned as 
\begin{equation*}
    \Pr[T_i = 1| X_{N_i} = x_{N_i}, \mathcal{G}_{N_i}]
    = \mathrm{expit}\left(
        \beta_\pi \mathrm{Avg}((x_{N_i \setminus \{i\}}^{(1)} - 0.5)(x_{N_i \setminus \{i\}}^{(2)} - 0.5))
    \right),
\end{equation*}
independently across $i$.
Outcomes $Y_i \in \{0,1\}$ were generated as
\begin{equation} \label{eq:outcome_ate}
\begin{aligned}
    & \Pr[Y_i=1 | T_{N_i} = t_{N_i}, X_{N_i^{(2)}} = x_{N_i^{(2)}}, \mathcal{G}_{N_i^{(2)}}] \\
    & = \mathrm{expit}\left(
        \beta_\mu^{(1)} (t_i - 0.5) 
        + \beta_\mu^{(2)} \mathrm{Avg}((x_{N_i \setminus \{i\}}^{(1)} - 0.5)(x_{N_i \setminus \{i\}}^{(2)} - 0.5))
    \right),
\end{aligned}
\end{equation}
independently across $i$. We set $\beta_\pi = 5$, $\beta_\mu^{(1)} = 1$ and $\beta_\mu^{(2)} = -7$. The treatment assignment and outcome model were designed so that $\Exp[Y_i | T_i = 1] \approx \Exp[Y_i | T_i = 0]$. In order to recover the correct treatment effect, one has to  account for the confounding by the covariate vector. 

We focused on the task of estimating the average counterfactual means under two treatment assignments and their difference. 
We considered the none-treated and all-treated scenarios: $t_i^{(0)} = 0, t_i^{(1)} = 1, \forall i$, respectively. The average counterfactual means for these scenarios were \( n^{-1} \sum_{i=1}^n \theta_{i}(t^{(0)}_{N_{i}}) \approx 0.406\) and \( n^{-1}\sum_{i=1}^n\theta_{i}(t^{(1)}_{N_{i}}) \approx 0.594\). The average treatment effect was approximately $0.188$.

\paragraph{Estimation Details and Results.}

We evaluate the performance of three competing methods (AUTOGNET, TMLENET and KECENI) when prior knowledge about confounding through the interaction term $(X_{N_i \setminus \{i\}}^{(1)} - 0.5)(X_{N_i \setminus \{i\}}^{(2)} - 0.5)$ is unavailable. In this setting, a natural strategy for AUTOGNET is to construct structural equation models using summary statistics $\mathrm{Avg}(X_{N_i \setminus \{i\}})$. The resulting models for the treatment and outcome mechanisms are specified as follows: 
\begin{equation*}
\begin{aligned}
    & \Pr[T_i = 1| T_{N_i \setminus \{i\}} = t_{N_i \setminus \{i\}}, X_{N_i} = x_{N_i}, \mathcal{G}_{N_i}] \\
    & = \mathrm{expit}\left(
        \beta_\pi^{(0)} 
        + \beta_\pi^{(1)} \mathrm{Avg}(t_{N_i \setminus \{i\}})
        + \inner*{\beta_\pi^{(2)}, x_i}
        + \inner*{\beta_\pi^{(3)}, \mathrm{Avg}(x_{N_i \setminus \{i\}})}
    \right); \\
    & \Pr[Y_i=1 | Y_{N_i \setminus \{i\}} = y_{N_i \setminus \{i\}}, T_{N_i} = t_{N_i}, X_{N_i^{(2)}} = x_{N_i^{(2)}}, \mathcal{G}_{N_i^{(2)}}] \\
    & = \mathrm{expit}\left(
        \beta_\mu^{(0)}
        + \beta_\mu^{(1)} \mathrm{Avg}(y_{N_i \setminus \{i\}})
        + \beta_\mu^{(2)} t_i
        + \beta_\mu^{(3)} \mathrm{Avg}(t_{N_i \setminus \{i\}})
        + \inner*{\beta_\mu^{(4)}, x_i}
        + \inner*{\beta_\mu^{(5)}, \mathrm{Avg}(x_{N_i \setminus \{i\}})}
    \right),
\end{aligned}
\end{equation*}
where $\beta_\pi^{(0)}, \beta_\pi^{(1)} \in \reals$, $\beta_\pi^{(2)}, \beta_\pi^{(3)} \in \reals^3$, $\beta_\mu^{(0)}, \beta_\mu^{(1)}, \beta_\mu^{(2)}, \beta_\mu^{(3)} \in \reals$, $\beta_\mu^{(4)}, \beta_\mu^{(5)} \in \reals^3$. These models omit the interaction term and therefore misrepresent the true data-generating process.
TMLENET adopts a similar approach and also results in misspecified models for the outcome regression and propensity score models omitting the interaction term.

We implement KECENI assuming that treatment is assigned independently across nodes given covariates, as formalized in \cref{eq:propensity_ind}. 
We further assume that the node-level propensity score \(\pi^\circ(t_i | x_{N_i}, \mathcal{G}_{N_i})\) and outcome regression \(\mu(t_{N_i}, x_{N_i^{(2)}}, \mathcal{G}_{N_i^{(2)}})\) change smoothly with the distribution of \(t_j\) and \(x_j\) over \(j \in N_i \setminus \{i\}\). Specifically, we assume:
\begin{equation} \label{eq:nuisance_fgw}
\begin{aligned}
    & \abs{\mu(t_{N_i}, x_{N_i^{(2)}}, \mathcal{G}_{N_i^{(2)}}) - \mu(t'_{N_{i^*}}, x'_{N_{i^*}^{(2)}}, \mathcal{G}_{N_{i^*}^{(2)}})} \\
    & \leq L_\mu \left(\norm{(t_i, x_i) - (t'_{i^*}, x'_{i^*})}_1 + W_{1,1}\left(\hat\Pr\{(t_j, x_j)\}_{j \in N_i \setminus \{i\}}, \hat\Pr\{(t'_j, x'_j)\}_{j \in N_{i^*} \setminus \{i^*\}}\right)\right); \\
    & \abs{\pi^\circ(x_{N_i}) - \pi^\circ(x'_{N_{i^*}})} \leq L_\pi \left(\norm{x_i - x'_{i^*}}_1 + W_{1,1}\left(\hat\Pr\{x_j\}_{j \in N_i \setminus \{i\}}, \hat\Pr\{x'_j\}_{j \in N_{i^*} \setminus \{i^*\}}\right)\right),
\end{aligned}
\end{equation}
for some \(L_\mu, L_\pi > 0\) and any given \(t_{N_i}\), \(t'_{N_{i^*}}\), \(x_{N_i}\), and \(x'_{N_{i^*}}\), where \(\hat\Pr\) is the empirical probability measure on the respective set, and \(W_{1,1}(\cdot, \cdot)\) denotes the Wasserstein 1-distance with respect to the \(\ell_1\) metric. We then use kernel smoothing estimators to fit \(\mu\) and \(\pi^\circ\), yielding the estimates \(\hat\mu\) and \(\hat\pi\). The Wasserstein distance is also used to define the dissimilarity metric $\Delta_i$ which relates the observed treatment assignments and network characteristics to those of the target node, 
\begin{equation}
    \Delta_i \equiv \abs{T_i - t^*_{i^*}} 
    + W_{1,1}\left(\hat\Pr\{T_j\}_{j \in N_i \setminus \{i\}}, \hat\Pr\{t^*_j\}_{j \in N_{i^*} \setminus \{i^*\}}\right).
\end{equation}
Finally, \(\hat{P}_{X_{[n]}|\mathcal{G}}\) is taken to be the empirical product measure in \cref{eq:empirical_product_measure}.

\subsection{Comparison with Existing Methods under Known Parametric Model or Summary Statistics} \label{app:comparison_known}

Under the simulation setting of \cref{sec:simulation_ate}, 
we evaluate the performance of AUTOGNET, TMLENET, and KECENI when the prior knowledge about confounding through the interaction term is available. In this setting, we construct structural equation models for AUTOGNET using summary statistics $\mathrm{Avg}(w(X_{N_i \setminus \{i\}}))$ instead of $\mathrm{Avg}(X_{N_i \setminus \{i\}})$, where $w(X_{N_i \setminus \{i\}}) \equiv (X_{N_i \setminus \{i\}}^{(1)} - 0.5)(X_{N_i \setminus \{i\}}^{(2)} - 0.5)$, unlikely to the scenario in \cref{sec:simulation_ate}. The resulting models for the treatment and outcome mechanisms, specified as 
\begin{equation*}
\begin{aligned}
    & \Pr[T_i = 1| T_{N_i \setminus \{i\}} = t_{N_i \setminus \{i\}}, X_{N_i} = x_{N_i}, \mathcal{G}_{N_i}] \\
    & = \mathrm{expit}\left(\begin{aligned}
        \beta_\pi^{(0)} 
        + \beta_\pi^{(1)} \mathrm{Avg}(t_{N_i \setminus \{i\}})
        + \beta_\pi^{(2)} w(x_i)
        + \beta_\pi^{(3)} \mathrm{Avg}(w(x_{N_i \setminus \{i\}}))
    \end{aligned}\right); \\
    & \Pr[Y_i=1 | Y_{N_i \setminus \{i\}} = y_{N_i \setminus \{i\}}, T_{N_i} = t_{N_i}, X_{N_i^{(2)}} = x_{N_i^{(2)}}, \mathcal{G}_{N_i^{(2)}}] \\
    & = \mathrm{expit}\left( \begin{aligned}
        \beta_\mu^{(0)}
        + \beta_\mu^{(1)} \mathrm{Avg}(y_{N_i \setminus \{i\}})
        + \beta_\mu^{(2)} t_i
        + \beta_\mu^{(3)} \mathrm{Avg}(t_{N_i \setminus \{i\}})
        + \beta_\mu^{(4)} w(x_i)
        + \beta_\mu^{(5)} \mathrm{Avg}(w(x_{N_i \setminus \{i\}}))
    \end{aligned}\right),
\end{aligned}
\end{equation*}
where $\beta_\pi^{(0)}, \beta_\pi^{(1)}, \beta_\pi^{(2)}, \beta_\pi^{(3)} \in \reals$, and $\beta_\mu^{(0)}, \beta_\mu^{(1)}, \beta_\mu^{(2)}, \beta_\mu^{(3)}, \beta_\mu^{(4)}, \beta_\mu^{(5)} \in \reals$, correctly specify the true data-generating process.
TMLENET adopts a similar approach and also results in correctly specified models for the outcome regression and propensity score models.

%
For KECENI, we use a kernel smoother approach with the distance metric
\[
\Delta_i \equiv \norm*{(T_i, \mathrm{Avg}(T_{N_i \setminus \{i\}} - 0.5)) - (t^*_{i^*}, \mathrm{Avg}(t^*_{N_{i^*} \setminus \{i^*\}} - 0.5))}_1
\]
and the covariate distribution estimate \( \hat{P}_{X_{[n]}|\mathcal{G}} \equiv \prod_{i=1}^n \hat{P}_{X_i|\mathcal{G}} \), where \( \hat{P}_{X_i|\mathcal{G}} \) is the empirical distribution of observed covariates.
We considered two approaches for estimating propensity scores and outcome regressions in KECENI. In the first, we assumed known parametric forms:
\begin{equation*}
\begin{aligned}
    & \pi^\circ(1| x_{N_i}, \mathcal{G}_{N_i}) \\
    & = \mathrm{expit}\left(\begin{aligned}
        \beta_\pi^{(0)} 
        + \beta_\pi^{(1)} w(x_i)
        + \beta_\pi^{(2)} \mathrm{Avg}(w(x_{N_i \setminus \{i\}}))
    \end{aligned}\right); \\
\end{aligned}
\end{equation*}
\begin{equation*}
\begin{aligned}
    & \mu(t_{N_i}, x_{N_i^{(2)}}, \mathcal{G}_{N_i^{(2)}}) \\
    & = \mathrm{expit}\left( \begin{aligned}
        \beta_\mu^{(0)}
        + \beta_\mu^{(1)} t_i
        + \beta_\mu^{(2)} \mathrm{Avg}(t_{N_i \setminus \{i\}})
        + \beta_\mu^{(3)} w(x_i)
        + \beta_\mu^{(4)} \mathrm{Avg}(w(x_{N_i \setminus \{i\}}))
    \end{aligned}\right),
\end{aligned}
\end{equation*}
where $\beta_\pi^{(0)}, \beta_\pi^{(1)}, \beta_\pi^{(2)} \in \reals$, and $\beta_\mu^{(0)}, \beta_\mu^{(1)}, \beta_\mu^{(2)}, \beta_\mu^{(3)}, \beta_\mu^{(4)} \in \reals$,
yielding parametric estimates $\hat\mu$ and $\hat\pi$. We compare this approach with AUTOGNET as the both methods assume known parametric models regarding the treatment and outcome processes. In the second approach, we assume that the node-level propensity score \(\pi^\circ(t_i | x_{N_i}, \mathcal{G}_{N_i})\) and outcome regression \(\mu(t_{N_i}, x_{N_i^{(2)}}, \mathcal{G}_{N_i^{(2)}})\) change smoothly with summary statistics $(w(x_i), \mathrm{Avg}(w(x_{N_i \setminus \{i\}})))$ and $(t_i, \mathrm{Avg}(t_{N_i \setminus \{i\}} - 0.5), w(x_i), \mathrm{Avg}(w(x_{N_i \setminus \{i\}})))$, respectively. Specifically, we assume:
\[
\begin{aligned}
    & \abs{\mu(t_{N_i}, x_{N_i^{(2)}}, \mathcal{G}_{N_i^{(2)}}) - \mu(t'_{N_{i^*}}, x'_{N_{i^*}^{(2)}}, \mathcal{G}_{N_{i^*}^{(2)}})} \\
    & \leq L_\mu \norm*{ \begin{aligned}
        & (t_i, \mathrm{Avg}(t_{N_i \setminus \{i\}} - 0.5), w(x_i), \mathrm{Avg}(w(x_{N_i \setminus \{i\}}))) \\
        & - (t'_{i^*}, \mathrm{Avg}(t'_{N_{i^*} \setminus \{i^*\}} - 0.5), (w(x'_{i^*}), \mathrm{Avg}(w(x'_{N_{i^*} \setminus \{i^*\}})))
    \end{aligned} }_1;
\end{aligned}
\]
\[
\begin{aligned}
    & \abs{\pi^\circ(1| x_{N_i}, \mathcal{G}_{N_i}) - \pi^\circ(1| x'_{N_{i^*}}, \mathcal{G}_{N_{i^*}})} \\
    & \leq L_\pi \norm{
        (w(x_i), \mathrm{Avg}(w(x_{N_i \setminus \{i\}}))) 
        - (w(x'_{i^*}), \mathrm{Avg}(w(x'_{N_{i^*} \setminus \{i^*\}})))
    }_1,
\end{aligned}
\]
for some \(L_\mu, L_\pi > 0\) and any given \(t_{N_i}\), \(t'_{N_{i^*}}\), \(x_{N_i^{(2)}}\), and \(x'_{N_{i^*}^{(2)}}\). We then use kernel smoothing estimators to fit \(\mu\) and \(\pi^\circ\), yielding the estimates \(\hat\mu\) and \(\hat\pi\).
%
We compare this approach to TMLENET.

\begin{figure}[t]
    \centering
    \begin{subfigure}[t]{0.4\linewidth}
        \captionsetup{margin={5mm, 0mm}}
        \centering
        \subcaption{}
        \includegraphics[width=0.9\textwidth]{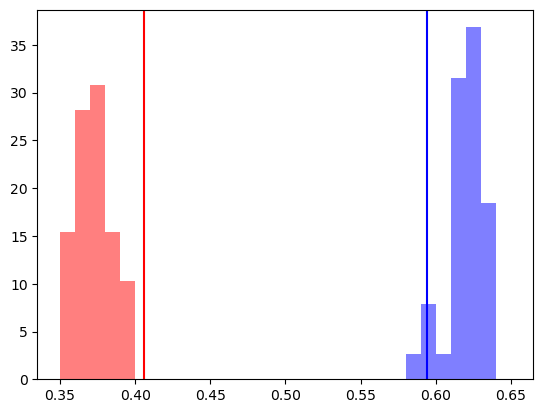}
        \label{fig:result_W_autognet}
    \end{subfigure}
    \qquad
    \begin{subfigure}[t]{0.4\linewidth}
        \captionsetup{margin={5mm, 0mm}}
        \centering
        \subcaption{}
        \includegraphics[width=0.9\textwidth]{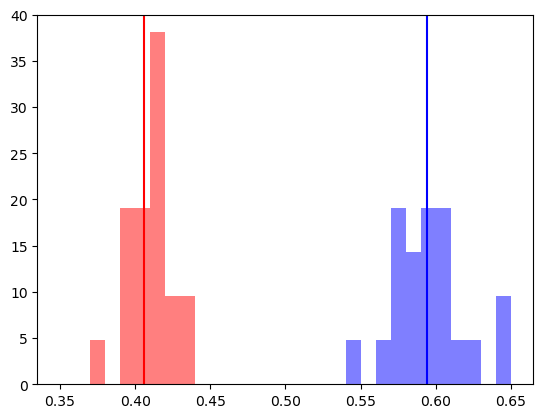}
        \label{fig:result_W_tmlenet}
    \end{subfigure}
    
    \begin{subfigure}[t]{0.4\linewidth}
        \captionsetup{margin={5mm, 0mm}}
        \centering
        \subcaption{}
        \includegraphics[width=0.9\textwidth]{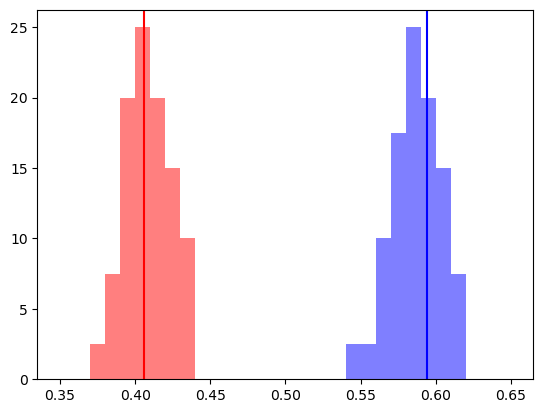}
        \label{fig:result_W_linear}
    \end{subfigure}
    \qquad
    \begin{subfigure}[t]{0.4\linewidth}
        \captionsetup{margin={5mm, 0mm}}
        \centering
        \subcaption{}
        \includegraphics[width=0.9\textwidth]{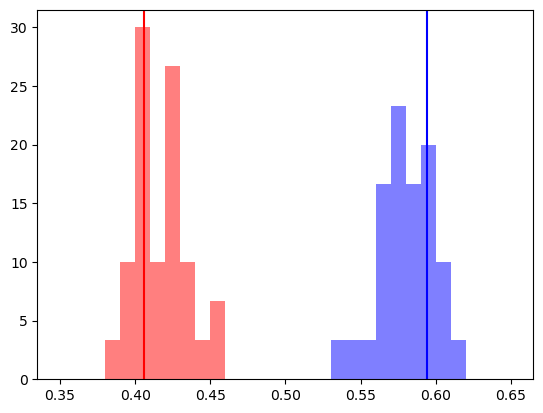}
        \label{fig:result_W_kernel}
    \end{subfigure}
    \caption{\textbf{Histogram of counterfactual mean estimates under known parametric model or summary statistics by (a) AUTOGNET, (b) TMLENET, (c) parametric KECENI, and (d) semi-parametric KECENI using known summary statistics.}}
    \label{fig:result_ate_W}
\end{figure}

\cref{fig:result_ate_W} presents the estimation results of the four methods. When the structural equation models are correctly specified, AUTOGNET and TMLENET accurately estimate the average counterfactual means and treatment effects (\cref{fig:result_W_autognet,fig:result_W_tmlenet}). The corresponding implementations of KECENI also produce accurate estimates (\cref{fig:result_W_linear,fig:result_W_kernel}). The root-mean-square errors (RMSE) for the methods are shown in the top rows of \cref{tab:result_ate}, where KECENI outperform or match the performance of its competitors.

\begin{table}[]
    \centering
    \begin{tabular}{lc}
        \toprule
        \textbf{Method} & \textbf{RMSE}  \\
        \midrule
        AUTOGNET, correct structural equation model & 0.065 \\
        TMLENET, known summary statistics & 0.035 \\
        KECENI, parametric & 0.022 \\
        KECENI, non-parametric, known summary statistics & 0.032 \\
        \midrule 
        AUTOGNET, incorrect structural equation model & 0.183 \\
        TMLENET, incorrect summary statistics & 0.184 \\
        KECENI, fully non-parametric & 0.063 \\
        \bottomrule
    \end{tabular}
    \caption{Root-mean-square errors (RMSE) of the competing methods: AUTOGNET, TMLENET, KECENI.}
    \label{tab:result_ate}
\end{table}

%% file: app/d_supp_real_data.tex
\section{Supplementary Indian Rurla Social Networks Analysis Details} \label{app:supp_real_data}

\paragraph{Interference Network.} 
The interference network have the same node set as the observed multilayer network (i.e., the subsampled villagers), and the edge set is formed by the union of the following  network layers:  `those who visit the respondent's home', `those whose homes the respondent visits', `kin in the village', `nonrelatives with whom the respondent socializes', `those to whom the respondent would lend money', `those from whom the respondent would borrow material goods', and `those with whom the respondent prays'.

\paragraph{Estimation details.}
We assume that the node-level propensity score \(\pi^\circ(t_i | x_{N_i}, \mathcal{G}_{N_i})\) and outcome regression \(\mu(t_{N_i}, x_{N_i^{(2)}}, \mathcal{G}_{N_i^{(2)}})\) change smoothly with summary statistics $(x_i, \mathrm{Avg}(x_{N_i \setminus \{i\}}))$ and $(t_i, \mathrm{Avg}(t_{N_i \setminus \{i\}} - 0.5), x_i, \mathrm{Avg}(x_{N_i^{(2)} \setminus \{i\}}))$, respectively. Specifically, we assume:
\[
\begin{aligned}
    & \abs{\mu(t_{N_i}, x_{N_i^{(2)}}, \mathcal{G}_{N_i^{(2)}}) - \mu(t'_{N_{i^*}}, x'_{N_{i^*}^{(2)}}, \mathcal{G}_{N_{i^*}^{(2)}})} \\
    & \leq L_\mu \norm{
        (t_i, \mathrm{Avg}(t_{N_i \setminus \{i\}} - 0.5), x_i, \mathrm{Avg}(x_{N_i^{(2)} \setminus \{i\}})) 
        - (t'_{i^*}, \mathrm{Avg}(t'_{N_{i^*} \setminus \{i^*\}} - 0.5), x'_{i^*}, \mathrm{Avg}(x'_{N_{i^*}^{(2)} \setminus \{i^*\}}))
    }_1;
\end{aligned}
\]
\[
    \abs{\pi^\circ(1| x_{N_i}, \mathcal{G}_{N_i}) - \pi^\circ(1| x'_{N_{i^*}}, \mathcal{G}_{N_{i^*}})} 
    \leq L_\pi \norm{
        (x_i, \mathrm{Avg}(x_{N_i \setminus \{i\}})) 
        - (x'_{i^*}, \mathrm{Avg}(x'_{N_{i^*} \setminus \{i^*\}}))
    }_1,
\]
for some \(L_\mu, L_\pi > 0\) and any given \(t_{N_i}\), \(t'_{N_{i^*}}\), \(x_{N_i^{(2)}}\), and \(x'_{N_{i^*}^{(2)}}\). We then use kernel smoothing estimators to fit \(\mu\) and \(\pi^\circ\), yielding the estimates \(\hat\mu\) and \(\hat\pi\).
%
For the dissimilarity metric, we used $\Delta_i 
    \equiv \norm{(T_i, \mathrm{Avg}(T_{N_i \setminus \{i\}} - 0.5)) - (t^*_{i^*}, \mathrm{Avg}(t^*_{N_{i^*} \setminus \{i^*\}} - 0.5))}_1$
and the empirical product measure in \cref{eq:empirical_product_measure} for the covariate distribution estimate.

%% file: 2_ref-main.bib
@article{forastiere2021identification,
  title={Identification and estimation of treatment and interference effects in observational studies on networks},
  author={Forastiere, Laura and Airoldi, Edoardo M and Mealli, Fabrizia},
  journal={Journal of the American Statistical Association},
  volume={116},
  number={534},
  pages={901--918},
  year={2021},
  publisher={Taylor \& Francis}
}

@article{tchetgen2021auto,
  title={Auto-{G}-computation of causal effects on a network},
  author={Tchetgen Tchetgen, Eric J and Fulcher, Isabel R and Shpitser, Ilya},
  journal={Journal of the American Statistical Association},
  volume={116},
  number={534},
  pages={833--844},
  year={2021},
  publisher={Taylor \& Francis}
}

@article{ogburn2022causal,
  title={Causal inference for social network data},
  author={Ogburn, Elizabeth L and Sofrygin, Oleg and Diaz, Ivan and Van der Laan, Mark J},
  journal={Journal of the American Statistical Association},
  volume={119},
  number={545},
  pages={597--611},
  year={2024},
  publisher={Taylor \& Francis}
}

@article{li2022random,
  title={Random graph asymptotics for treatment effect estimation under network interference},
  author={Li, Shuangning and Wager, Stefan},
  journal={Annals of Statistics},
  volume={50},
  number={4},
  pages={2334--2358},
  year={2022},
  publisher={Institute of Mathematical Statistics}
}

@article{van2014causal,
  title={Causal inference for a population of causally connected units},
  author={van der Laan, Mark J},
  journal={Journal of Causal Inference},
  volume={2},
  number={1},
  pages={13--74},
  year={2014},
  publisher={De Gruyter}
}

@article{kennedy2017non,
  title={Non-parametric methods for doubly robust estimation of continuous treatment effects},
  author={Kennedy, Edward H and Ma, Zongming and McHugh, Matthew D and Small, Dylan S},
  journal={Journal of the Royal Statistical Society Series B: Statistical Methodology},
  volume={79},
  number={4},
  pages={1229--1245},
  year={2017},
  publisher={Oxford University Press}
}

@article{vayer2020fused,
  title={Fused Gromov-Wasserstein distance for structured objects},
  author={Vayer, Titouan and Chapel, Laetitia and Flamary, R{\'e}mi and Tavenard, Romain and Courty, Nicolas},
  journal={Algorithms},
  volume={13},
  number={9},
  pages={212},
  year={2020},
  publisher={MDPI}
}

@unpublished{kojevnikov2021bootstrap,
  title={The bootstrap for network dependent processes},
  author={Kojevnikov, Denis},
  journal={arXiv preprint arXiv:2101.12312},
  note={\emph{arXiv preprint}},
  year={2021}
}

@article{ross2011fundamentals,
  title={Fundamentals of {S}tein’s method},
  author={Ross, Nathan},
  journal={Probability Surveys},
  volume={8},
  pages={210--293},
  year={2011}
}

@book{vaart1996weak,
  title={Weak Convergence and Empirical Processes: With Applications to Statistics},
  author={van der Vaart, Aad W. and Wellner, Jon A.},
  series={Springer Series in Statistics},
  year={1996},
  publisher={Springer},
  isbn={978-3-031-29040-4}
}

@book{vershynin2018high,
  title={High-dimensional probability: An introduction with applications in data science},
  author={Vershynin, Roman},
  volume={47},
  year={2018},
  publisher={Cambridge university press}
}

@article{bang2005doubly,
  title={Doubly robust estimation in missing data and causal inference models},
  author={Bang, Heejung and Robins, James M},
  journal={Biometrics},
  volume={61},
  number={4},
  pages={962--973},
  year={2005},
  publisher={Oxford University Press}
}

@article{savje2021average,
  title={Average treatment effects in the presence of unknown interference},
  author={S{\"a}vje, Fredrik and Aronow, Peter and Hudgens, Michael},
  journal={Annals of Statistics},
  volume={49},
  number={2},
  pages={673},
  year={2021},
  publisher={NIH Public Access}
}

@article{horvitz1952generalization,
  title={A generalization of sampling without replacement from a finite universe},
  author={Horvitz, Daniel G and Thompson, Donovan J},
  journal={Journal of the American Statistical Association},
  volume={47},
  number={260},
  pages={663--685},
  year={1952},
  publisher={Taylor \& Francis}
}

@article{robins1986new,
  title={A new approach to causal inference in mortality studies with a sustained exposure period—application to control of the healthy worker survivor effect},
  author={Robins, James},
  journal={Mathematical Modelling},
  volume={7},
  number={9-12},
  pages={1393--1512},
  year={1986},
  publisher={Elsevier}
}

@article{leung2022causal,
  title={Causal inference under approximate neighborhood interference},
  author={Leung, Michael P},
  journal={Econometrica},
  volume={90},
  number={1},
  pages={267--293},
  year={2022},
  publisher={Wiley Online Library}
}

@article{aronow2017estimating,
  title={Estimating average causal effects under general interference, with application to a social network experiment},
  author={Aronow, Peter M and Samii, Cyrus},
  journal={Annals of Applied Statistics},
  volume={11},
  number={4},
  pages={1912--1947},
  year={2017},
  publisher={Institute of Mathematical Statistics}
}

@article{liu2016inverse,
  title={On inverse probability-weighted estimators in the presence of interference},
  author={Liu, Lan and Hudgens, Michael G and Becker-Dreps, Sylvia},
  journal={Biometrika},
  volume={103},
  number={4},
  pages={829--842},
  year={2016},
  publisher={Oxford University Press}
}

@unpublished{khatami2024graph,
  title={Graph Neural Network based Double Machine Learning Estimator of Network Causal Effects},
  author={Khatami, Seyedeh Baharan and Parikh, Harsh and Chen, Haowei and Roy, Sudeepa and Salimi, Babak},
  journal={arXiv preprint arXiv:2403.11332},
  note={\emph{arXiv preprint}},
  year={2024}
}

@unpublished{leung2022graph,
  title={Graph Neural Networks for Causal Inference Under Network Confounding},
  author={Leung, Michael P and Loupos, Pantelis},
  journal={arXiv preprint arXiv:2211.07823},
  note={\emph{arXiv preprint}},
  year={2022}
}

@article{auerbach2021local,
  title={The local approach to causal inference under network interference},
  author={Auerbach, Eric and Guo, Hongchang and Tabord-Meehan, Max},
  journal={Quantitative Economics},
  volume={17},
  number={1},
  pages={173--199},
  year={2026},
  publisher={Wiley Online Library}
}

@article{shalizi2011homophily,
  title={Homophily and contagion are generically confounded in observational social network studies},
  author={Shalizi, Cosma Rohilla and Thomas, Andrew C},
  journal={Sociological Methods \& Research},
  volume={40},
  number={2},
  pages={211--239},
  year={2011},
  publisher={Sage Publications Sage CA: Los Angeles, CA}
}

@article{kojevnikov2021limit,
  title={Limit theorems for network dependent random variables},
  author={Kojevnikov, Denis and Marmer, Vadim and Song, Kyungchul},
  journal={Journal of Econometrics},
  volume={222},
  number={2},
  pages={882--908},
  year={2021},
  publisher={Elsevier}
}

@article{banerjee2013diffusion,
  title={The diffusion of microfinance},
  author={Banerjee, Abhijit and Chandrasekhar, Arun G and Duflo, Esther and Jackson, Matthew O},
  journal={Science},
  volume={341},
  number={6144},
  pages={1236498},
  year={2013},
  publisher={American Association for the Advancement of Science}
}

@book{van2011targeted,
  title={Targeted Learning: Causal Inference for Observational and Experimental Data},
  author={van der Laan, Mark J and Rose, Sherri and others},
  volume={4},
  year={2011},
  publisher={Springer}
}

@article{jiang2022new,
  title={A new central limit theorem for the augmented IPW estimator: Variance inflation, cross-fit covariance and beyond},
  author={Jiang, Kuanhao and Mukherjee, Rajarshi and Sen, Subhabrata and Sur, Pragya},
  journal={The Annals of Statistics},
  volume={53},
  number={2},
  pages={647--675},
  year={2025},
  publisher={Institute of Mathematical Statistics}
}

@article{funk2011doubly,
  title={Doubly robust estimation of causal effects},
  author={Funk, Michele Jonsson and Westreich, Daniel and Wiesen, Chris and St{\"u}rmer, Til and Brookhart, M Alan and Davidian, Marie},
  journal={American journal of epidemiology},
  volume={173},
  number={7},
  pages={761--767},
  year={2011},
  publisher={Oxford University Press}
}

@article{shook2025double,
  title={Double robust variance estimation with parametric working models},
  author={Shook-Sa, Bonnie E and Zivich, Paul N and Lee, Chanhwa and Xue, Keyi and Ross, Rachael K and Edwards, Jessie K and Stringer, Jeffrey SA and Cole, Stephen R},
  journal={Biometrics},
  volume={81},
  number={2},
  pages={ujaf054},
  year={2025},
  publisher={Oxford University Press}
}

@article{arcones1993limit,
  title={Limit Theorems for {$U$}-Processes},
  author={Arcones, Miguel A and Gine, Evarist},
  journal={Annals of Probability},
  volume={21},
  number={3},
  pages={1494--1542},
  year={1993},
  publisher={Institute of Mathematical Statistics}
}

@article{stefanski2002calculus,
  title={The calculus of {$M$}-estimation},
  author={Stefanski, Leonard A and Boos, Dennis D},
  journal={The American Statistician},
  volume={56},
  number={1},
  pages={29--38},
  year={2002},
  publisher={Taylor \& Francis}
}

@unpublished{peng2019asymptotic,
  title={Asymptotic distributions and rates of convergence for random forests via generalized {$U$}-statistics},
  author={Peng, Wei and Coleman, Tim and Mentch, Lucas},
  journal={arXiv preprint arXiv:1905.10651},
  note={\emph{arXiv preprint}},
  year={2019}
}

@article{munoz2012population,
  title={Population intervention causal effects based on stochastic interventions},
  author={Mu{\~n}oz, Iv{\'a}n D{\'\i}az and van der Laan, Mark},
  journal={Biometrics},
  volume={68},
  number={2},
  pages={541--549},
  year={2012},
  publisher={Oxford University Press}
}

@article{buchanan2022spillover,
  title={Spillover benefit of pre-exposure prophylaxis for HIV prevention: evaluating the importance of effect modification using an agent-based model},
  author={Buchanan, Ashley L and Park, Carolyn J and Bessey, Sam and Goedel, William C and Murray, Eleanor J and Friedman, Samuel R and Halloran, M Elizabeth and Katenka, Natallia V and Marshall, Brandon DL},
  journal={Epidemiology \& Infection},
  volume={150},
  pages={e192},
  year={2022},
  publisher={Cambridge University Press}
}

@article{bramoulle2009identification,
  title={Identification of peer effects through social networks},
  author={Bramoull{\'e}, Yann and Djebbari, Habiba and Fortin, Bernard},
  journal={Journal of econometrics},
  volume={150},
  number={1},
  pages={41--55},
  year={2009},
  publisher={Elsevier}
}

@article{lomi2011some,
  title={Why are some more peer than others? Evidence from a longitudinal study of social networks and individual academic performance},
  author={Lomi, Alessandro and Snijders, Tom AB and Steglich, Christian EG and Torl{\'o}, Vanina Jasmine},
  journal={Social science research},
  volume={40},
  number={6},
  pages={1506--1520},
  year={2011},
  publisher={Elsevier}
}

@article{bhadra2025causal,
  title={Causal Inference Under Network Interference},
  author={Bhadra, Subhankar and Schweinberger, Michael},
  journal={arXiv preprint arXiv:2508.06808},
  year={2025}
}

@article{broido2019scale,
  title={Scale-free networks are rare},
  author={Broido, Anna D and Clauset, Aaron},
  journal={Nature communications},
  volume={10},
  number={1},
  pages={1017},
  year={2019},
  publisher={Nature Publishing Group UK London}
}

@article{clauset2009power,
  title={Power-law distributions in empirical data},
  author={Clauset, Aaron and Shalizi, Cosma Rohilla and Newman, Mark EJ},
  journal={SIAM review},
  volume={51},
  number={4},
  pages={661--703},
  year={2009},
  publisher={SIAM}
}

@article{mitzenmacher2004brief,
  title={A brief history of generative models for power law and lognormal distributions},
  author={Mitzenmacher, Michael},
  journal={Internet mathematics},
  volume={1},
  number={2},
  pages={226--251},
  year={2004},
  publisher={Taylor \& Francis}
}

@book{wasserman2006all,
  title={All of nonparametric statistics},
  author={Wasserman, Larry},
  year={2006},
  publisher={Springer}
}

@article{takatsu2025debiased,
  title={Debiased inference for a covariate-adjusted regression function},
  author={Takatsu, Kenta and Westling, Ted},
  journal={Journal of the Royal Statistical Society Series B: Statistical Methodology},
  volume={87},
  number={1},
  pages={33--55},
  year={2025},
  publisher={Oxford University Press UK}
}

@article{calonico2018effect,
  title={On the effect of bias estimation on coverage accuracy in nonparametric inference},
  author={Calonico, Sebastian and Cattaneo, Matias D and Farrell, Max H},
  journal={Journal of the American Statistical Association},
  volume={113},
  number={522},
  pages={767--779},
  year={2018},
  publisher={Taylor \& Francis}
}

@article{kandiros2024conflict,
  title={The Conflict Graph Design: Estimating Causal Effects under Arbitrary Neighborhood Interference},
  author={Kandiros, Vardis and Pipis, Charilaos and Daskalakis, Constantinos and Harshaw, Christopher},
  journal={arXiv preprint arXiv:2411.10908},
  year={2024}
}

@incollection{kennedy2016semiparametric,
  title={Semiparametric theory and empirical processes in causal inference},
  author={Kennedy, Edward H},
  booktitle={Statistical causal inferences and their applications in public health research},
  pages={141--167},
  year={2016},
  publisher={Springer}
}


%% file: 2_ref-supp.bib
@unpublished{shook2024double,
  title={Double Robust Variance Estimation},
  author={Shook-Sa, Bonnie E and Zivich, Paul N and Lee, Chanhwa and Xue, Keyi and Ross, Rachael K and Edwards, Jessie K and Stringer, Jeffrey SA and Cole, Stephen R},
  journal={arXiv preprint arXiv:2404.16166},
  note={\emph{arXiv preprint}},
  year={2024}
}
